\documentclass[thmsa,11pt,sw20elba]{article}%
\usepackage{amssymb}
\usepackage{amsfonts}
\usepackage{amsmath}
\usepackage{geometry}
\usepackage[pdftex]{hyperref}
\usepackage[pdftex]{graphicx}
\usepackage{threeparttable}
\usepackage{color}%
\def\dsum\limits{\sum\limits}
\def\dprod\limits{\prod\limits}
\begin{document}

\author{Victor Aguirregabiria\thanks{We have benefited from comments and conversations
with Bo Honor\'{e}, Kyoo il Kim, Roger Koenker, Bob Miller, Pedro Mira, Ismael
Mourifi\'{e}, Whitney Newey, Elie Tamer, Yuanyuan Wan, Jeff Wooldridge, and
from participants in seminars at Harvard-MIT, Michigan State, Princeton,
Toronto, the conference at UCL on \textit{Implementation of Structural Dynamic
Models}, and the conference at Vanderbilt on \textit{Identification in
Econometrics}.}\\University of Toronto and CEPR
\and Jiaying Gu$^{\ast}$\\University of Toronto
\and Yao Luo$^{\ast}$\\University of Toronto}
\title{\textbf{Sufficient Statistics for Unobserved Heterogeneity}\\\textbf{in Structural Dynamic Logit Models }}
\date{May 8th, 2018}
\maketitle

\begin{abstract}
We study the identification and estimation of structural parameters in dynamic
panel data logit models where decisions are forward-looking and the joint
distribution of unobserved heterogeneity and observable state variables is
nonparametric, i.e., fixed-effects model. We consider models with two
endogenous state variables: the lagged decision variable, and the time
duration in the last choice. This class of models includes as particular cases
important economic applications such as models of market entry-exit,
occupational choice, machine replacement, inventory and investment decisions,
or dynamic demand of differentiated products. The identification of structural
parameters requires a sufficient statistic that controls for unobserved
heterogeneity not only in current utility but also in the continuation value
of the forward-looking decision problem. We obtain the minimal sufficient
statistic and prove identification of some structural parameters using a
conditional likelihood approach. We apply this estimator to a machine
replacement model.

\bigskip


\bigskip


\bigskip

\noindent Victor Aguirregabiria. Address: 150 St. George Street. Toronto, ON,
M5S 3G7, Canada.

\noindent E-mail: \texttt{victor.aguirregabiria@utoronto.ca}

\medskip

\noindent Jiaying Gu. Address: 150 St. George Street. Toronto, ON, M5S 3G7, Canada.

\noindent E-mail: \texttt{jiaying.gu@utoronto.ca}

\medskip

\noindent Yao Luo. Address: 150 St. George Street. Toronto, ON, M5S 3G7, Canada.

\noindent E-mail: \texttt{yao.luo@utoronto.ca}

\end{abstract}

\thispagestyle{empty}\setcounter{page}{0}\baselineskip22pt

\newpage

\section{Introduction}

Persistent unobserved heterogeneity is pervasive in empirical applications
using panel data of individuals, households, or firms. An important challenge
in these applications consists of distinguishing between \textit{true
dynamics} due to state dependence and \textit{spurious dynamics} due to
unobserved heterogeneity (Heckman, 1981). The identification of \textit{true
dynamics}, when persistent unobserved heterogeneity is present, should deal
with two key econometric issues: the \textit{incidental parameters problem},
and the \textit{initial conditions problem}. The first one establishes that a
simple dummy-variables estimator, that treats each individual unobservable as
a parameter to be estimated jointly with the parameters of interest, is
inconsistent in most nonlinear panel data models when $T$ is fixed (Neyman and
Scott, 1948, Lancaster, 2000). Given this issue, it would seem reasonable to
consider a nonparametric (or a flexible) joint distribution of the unobserved
heterogeneity and the observables variables, and construct a likelihood
function that is integrated over unobservables. In this context, the
\textit{initial conditions problem }establishes that the joint distribution of
the unobserved heterogeneity and the initial values of the observable
variables is not nonparametrically identified, but the misspecification of
this joint distribution can generate important biases in the estimation of the
parameters of interest (Heckman, 1981, Chamberlain, 1985, among others).

There are two general approaches to deal with this issue: random effects and
fixed effects models/methods. Random-effects models impose restrictions on the
distribution of unobserved heterogeneity (e.g., parametric, finite mixture),
and on the joint distribution of these unobservables and the initial
conditions of the observable explanatory variables. Under these restrictions,
the parameters of interest and the distribution of the unobserved
heterogeneity are jointly estimated. In contrast, fixed-effects methods focus
on the estimation of the parameters of interest and they do not try to
identify the distribution of the unobserved heterogeneity. These methods are
more robust because they are fully nonparametric in the specification of the
joint distribution of unobserved heterogeneity and exogenous or predetermined
explanatory variables.\footnote{See Arellano and Honor\'{e} (2001), and
Arellano and Bonhomme (2012, 2017) for recent surveys on the econometrics of
nonlinear panel data models.}

A fixed effect method, pioneered by Andersen (1970) and extended by
Chamberlain (1980), is based on the derivation of sufficient statistics for
the incidental parameters (fixed effects) and the maximization of a likelihood
function conditional on these sufficient statistics. This paper deals with
this fixed effects - sufficient statistics - conditional maximum likelihood
approach (FE-CML hereinafter). We study the applicability of this approach to
structural dynamic discrete choice models where agents are
forward-looking.\footnote{Among the class of fixed-effects estimators in short
panels, the dummy-variables estimator is the simplest of these methods.
However, as mentioned above, this estimator is inconsistent in most nonlinear
panel data models when $T$ is fixed. Two-step bias reduction methods, both
analytical and simulation-based, have been proposed to correct for the
asymptotic bias of these dummy-variables fixed-effect estimators (e.g., Hahn
and Newey, 2004, Browning and Carro, 2010, and Hahn and Kuersteiner, 2011,
among others). Other fixed-effects estimators are based on a transformation of
the model that eliminates the fixed effects, e.g., Manski's maximum score
estimator (Manski, 1987). However, in nonlinear models, these estimators
require strict exogeneity of the explanatory variables, ruling out nonlinear
dynamic models. Bonhomme (2012) presents a \textit{functional differencing
approach} that includes as particular cases different fixed effects estimators
in the literature.}

There is a wide class of nonlinear panel data models where the FE-CML approach
cannot identify the structural parameters.\footnote{In this paper, the
concepts of identification and consistent estimation, as $N$ goes not infinity
and $T$ is fixed, are used as synonymous.} In general, a sufficient statistic
of the incidental parameters always exists.\footnote{For instance, we could
define as sufficient statistic the complete choice history of an individual.
Obviously, the conditional likelihood function based on this sufficient
statistic does not depend neither on incidental nor on structural parameters.
Though this is an extreme example, it illustrates that the key identification
problem is not finding a sufficient statistic for the incidental parameters
but showing that there are sufficient statistics for which the conditional
likelihood still depends on the structural parameters.} The identification
problem appears when the minimal sufficient statistic is such that the
likelihood conditional on this statistic does not depend on the structural
parameters. For instance, in the context of binary choice models, Chamberlain
(1993, 2010) shows that a necessary and sufficient condition for (point)
identification under the FE-CML\ approach is that the distribution of the
time-varying unobservable is logistic.\footnote{Chamberlain (1993, 2010)
considers the model where the time-varying unobservables are independently and
identically distributed. Magnac (2004) studies a two-period model where the
two time-varying unobservables have a general joint distribution. Honor\`{e}
and Tamer (2006) study partial identification of the dynamic Probit model and
derive sharp bounds on parameters.} Similarly, identification is not possible
in discrete choice models where unobserved heterogeneity appears in the slope
parameters, interacting with predetermined explanatory variables
\footnote{Browining and Carro (2014) study the identification of this type of
dynamic binary choice model with \textit{maximal heterogeneity} in short
panels. The fixed-effects model (nonparametric specification of the unobserved
heterogeneity) is not identified. They consider a finite mixture specification
of the heterogeneous parameters. This is in the same spirit as Kasahara and
Shimotsu (2009), though these other authors consider a nonparametric Markov
chain with finite mixture unobserved heterogeneity.} This has important
implications for structural dynamic discrete choice models. In these models,
an agent's optimal decision depends not only on her current utility but also
on the continuation value function, which is an endogenous object. In general,
unobserved heterogeneity enters non-additively in the continuation value
function and interacts with the observable state variables, even when this
unobserved heterogeneity is additively separable in the one-period utility
function. This interaction between the unobserved heterogeneity and the
endogenous state variables implies that structural parameters are not
identified in the fixed-effects model.

For non-structural (i.e., myopic) dynamic logit models with unobserved
heterogeneity only in the intercept, Chamberlain (1985) and Honor\'{e} and
Kyriazidou (2000) have shown that the FE-CML approach can identify the
parameters of interest.\footnote{In the models of these papers, the only
endogenous (predetermined) explanatory variable is the lagged decision. For
instance, time duration in the last choice is not an explanatory variable. In
our model, we include both lagged decision and duration as state variables.}
In contrast, all the methods and applications for structural dynamic discrete
choice models have considered random-effects models with a finite mixture
distribution, e.g., Keane and Wolpin (1997), Aguirregabiria and Mira (2007),
Kasahara and Shimotsu (2009), Arcidiacono and Miller (2011), among many
others. This random-effects approach imposes important restrictions: the
number of points in the support of the unobserved heterogeneity is finite and
is typically reduced to a small number of points; furthermore, the joint
distribution of the unobserved heterogeneity and the initial conditions of the
observable state variables is restricted.

In this paper, we revisit the applicability of FE-CML methods to the
identification and estimation of structural dynamic discrete choice models. We
follow the sufficient statistics approach to study the identification of
payoff function parameters in structural dynamic logit models with a
fixed-effects specification of the time-invariant unobserved heterogeneity. We
consider multinomial models with two types of endogenous state variables: the
lagged value of the decision variable, and the time duration in the last
choice. The main challenge for the identification of this model comes from the
fact that unobserved heterogeneity enters not only in current utility but also
in the continuation value of the forward-looking decision problem. In general,
this continuation value is a nonlinear function of unobserved heterogeneity
and state variables.\footnote{In fact, before solving the model, we do not
know how unobserved heterogeneity and state variables enter this continuation
value function. Therefore, for fixed-effects estimation, it is as if we had a
nonparametric specification of this function.} Therefore, identification
requires a sufficient statistic that controls for this continuation value but
implies a conditional likelihood that still depends on the structural
parameters that capture true state dependence. We derive the minimal
sufficient statistic and show that some structural parameters are identified.
The forward-looking model where the only state variable is the lagged decision
is identified under the same conditions as the myopic version of the model.
Instead, with duration dependence, there are some parameters identified in the
myopic model but not in the forward-looking model.

Based on our identification results, we consider a conditional maximum
likelihood estimator, and a test for the validity of a correlated random
effects specification. We apply this estimator and the test to the bus engine
model Rust (1987) using both simulated and actual data.

In most empirical applications of structural models, the researcher is not
only interested in the value of the structural parameters but also on the
estimation of marginal effects and counterfactual experiments. The
identification of marginal effects and counterfactuals requires the
identification of the distribution of the observed heterogeneity. Point
identification requires imposing restrictions on the joint distribution of
unobserved heterogeneity and the initial conditions of the state variables.
Alternatively, the researcher may prefer not to impose these restrictions and
then set-identify the distribution of the unobservables and the marginal
effects (Chernozhukov, Fernandez-Val, Hahn, and Newey, 2013). We discuss this
problem is section 3.4.

This paper contributes to the literature on structural dynamic discrete choice
models. The structure of the payoff function and of the endogenous state
variables that we consider in this paper includes as particular cases
important economic applications in the literature of dynamic discrete choice
structural models, such as models of market entry and exit either binary
(Roberts and Tybout, 1997, Aguirregabiria and Mira, 2007) or multinomial
(Sweeting, 2013; Caliendo et al, 2015); occupational choice models (Miller,
1984; Keane and Wolpin, 1997); machine replacement models (Rust, 1987; Das,
1992; Kennet, 1993; and Kasahara, 2009); inventory and investment decision
models (Aguirregabiria 1999; Ryan, 2013; Kalouptsidi, 2014); demand of
differentiated products with consumer brand switching costs (Erdem, Keane, and
Sun, 2008) or storable products (Erdem, Imai, and Keane, 2003; Hendel and
Nevo, 2006); and dynamic pricing models with menu costs (Willis, 2006), or
with duration dependence due to inflation or other forms of depreciation
(Slade, 1998; Aguirregabiria, 1999; Kano, 2013); among others.\footnote{Note
that most of the empirical applications cited above in this paragraph do not
allow for time-invariant unobserved heterogeneity. This is still a common
restriction in empirical applications of dynamic structural models, and it is
mostly justified by computational convenience. The exceptions, within the
cited papers, are Keane and Wolpin (1997), Erdem, Imai, and Keane (2003),
Willis (2006), Aguirregabiria and Mira (2007), and Erdem, Keane, and Sun
(2008).} Our paper also contributes to the literature on nonlinear dynamic
panel data models by providing new identification results of fixed effects
dynamic logit models with duration dependence (Frederiksen, Honor\'{e}, and
Hu, 2007).

The rest of the paper is organized as follows. Section 2 describes the class
of models that we study in this paper. Section 3 presents our identification
results. Section 4 deals with estimation and inference. In section 5, we
illustrate our identification results in the context of a bus replacement
model. Section 6 summarizes and concludes. Proofs of Lemmas and Propositions
are in the Appendix. Also in the Appendix, we show that our identification
results extend to an extended version of our model where the endogenous state
variables have a stochastic transition rule.

\section{Model}

Time is discrete and indexed by $t$ that belongs to $\{1,2,...,\infty
\}$.\footnote{The time horizon of the decision problem is infinite.} Agents
are indexed by $i$. Every period $t$, agent $i$ chooses a value of the
discrete variable $y_{it}\in\mathcal{Y}=\{0,1,...,J\}$ to maximize her
expected and discounted intertemporal utility $\mathbb{E}_{t}\left[
%TCIMACRO{\tsum \nolimits_{j=0}^{\infty}}%
%BeginExpansion
{\textstyle\sum\nolimits_{j=0}^{\infty}}
%EndExpansion
\delta_{i}^{j}\text{ }\Pi_{i,t+j}(y_{i,t+j})\right]  $, where $\delta_{i}%
\in(0,1)$ is agent $i$'s time discount factor, and $\Pi_{it}(y)$ is her
one-period utility if she chooses action $y$. This utility is a function of
four types of state variables which are known to the agent at period $t$:%
\begin{equation}
\Pi_{it}(y)=\alpha\left(  y,\boldsymbol{\eta}_{i},\mathbf{z}_{it}\right)
+\beta\left(  y,\mathbf{x}_{it},\mathbf{z}_{it}\right)  +\varepsilon
_{it}(y)\text{.} \label{one-period utility general specification}%
\end{equation}
$\mathbf{z}_{it}$ and $\mathbf{x}_{it}$ are observable to the researcher, and
$\mathbf{\varepsilon}_{it}$ and $\boldsymbol{\eta}_{i}$ are unobservable. The
vector $\mathbf{z}_{it}$ contains exogenous state variables and it follows a
Markov process with transition probability function $f_{\mathbf{z}}%
(\mathbf{z}_{i,t+1}|\mathbf{z}_{it})$. The vector $\mathbf{x}_{it}$ contains
endogenous state variables. We describe below the nature of these endogenous
state variables and their transition rules. Both $\mathbf{z}_{it}$ and
$\mathbf{x}_{it}$ have discrete supports $\mathcal{Z}$ and $\mathcal{X}$,
respectively. The unobservable variables $\{\varepsilon_{it}(y):y\in
\mathcal{Y}\}$ are $i.i.d.$ over $(i,t,y)$ with an extreme value type I
distribution. The vector $\boldsymbol{\eta}_{i}$ represents time-invariant
unobserved heterogeneity from the point of view of the researcher. Let
$\mathbf{\theta}_{i}\equiv(\boldsymbol{\eta}_{i},\delta_{i})$ represent the
unobserved heterogeneity from individual $i$. The probability distribution of
$\mathbf{\theta}_{i}$ conditional on the history of observable state variables
$\{\mathbf{z}_{it},\mathbf{x}_{it}:t=1,2,...\}$ is unrestricted and
nonparametrically specified, i.e., fixed effects model. Functions
$\alpha\left(  y,\boldsymbol{\eta},\mathbf{z}\right)  $ and $\beta\left(
y,\mathbf{x},\mathbf{z}\right)  $ are nonparametrically specified but they are bounded.

Our specification of the utility function represents a general semiparametric
fixed-effect logit model. It builds on \textit{Rust model} (Rust, 1987, 1994)
and generalizes it in two directions. First, Rust assumes that all the
unobservables satisfy the conditions of \textit{additive separability} and
\textit{conditional independence}, and they have an extreme value
distribution. While our time-varying unobservables $\varepsilon_{it}(y)$
satisfy these conditions, our time-invariant unobserved heterogeneity
interacts, in an unrestricted way, with the exogenous state variables and the
choice, and they do not satisfy the conditional independence assumption.
Second, we allow for unobserved heterogeneity in the discount factor.

The assumption of additive separability between $\boldsymbol{\eta}_{i}$ and
the endogenous state variables in $\mathbf{x}_{it}$ is key for the
identification results in this paper. This condition does not imply that the
conditional-choice value functions, that describe the solution of the dynamic
model, are additive separability between $\boldsymbol{\eta}_{i}$ and
$\mathbf{x}_{it}$. In general, the solution of the dynamic programming problem
implies a value function that is not additively separable in $\boldsymbol{\eta
}_{i}$ and $\mathbf{x}_{it}$ even when the utility function is additive in
these variables.

The model includes two types of endogenous state variables that correspond to
two different types of state dependence, $\mathbf{x}_{it}=(y_{i,t-1},d_{it})$:
(a) dependence on the the lagged decision variable, $y_{i,t-1}$; and (b)
\textit{duration dependence}, where $d_{it}\in\{1,2,...,\infty\}$ is the
number of periods since the last change in choice. The lagged decision has the
obvious transition rule. The transition rule for the duration variable is
$d_{i,t+1}=$ $1\left\{  y_{it}=y_{i,t-1}\right\}  $ $d_{it}+1$, where $1\{.\}$
is the indicator function.\footnote{Note that these endogenous state variables
follow deterministic transition rules. In the Appendix, we present a version
of the model that allows for stochastic transition rules for the endogenous
state variables.}

The term $\beta\left(  y,\mathbf{x}_{it},\mathbf{z}_{it}\right)  $ in the
payoff function captures the dynamics, or structural state dependence, in the
model. We distinguish in this function two additive components that correspond
to the two forms of state dependence in the model:%
\begin{equation}
\beta\left(  y,\mathbf{x}_{it},\mathbf{z}_{it}\right)  =1\{y=y_{i,t-1}\}\text{
}\beta_{d}\left(  y,d_{it},\mathbf{z}_{it}\right)  +1\{y\neq y_{i,t-1}\}\text{
}\beta_{y}\left(  y,y_{i,t-1},\mathbf{z}_{it}\right)
\label{two components beta}%
\end{equation}
Function $\beta_{d}\left(  y,d_{it},\mathbf{z}_{it}\right)  $ captures
duration dependence. For instance, in an occupational choice model, this term
captures the return on earnings of job experience in the current occupation.
Function $\beta_{y}\left(  y,y_{i,t-1},\mathbf{z}_{it}\right)  $ represents
switching costs. In an occupational choice model, this term represents the
cost of switching from occupation $y_{i,t-1}$ to occupation \thinspace$y$. The
additive separability between switching costs and "returns to experience" is
not without loss of generality. For instance, the cost of switching occupation
could depend on experience in the current job not only through the loss of the
returns of experience, i.e., $\beta_{y}(.)$ could depend on $d_{it}$. However,
this additive separability facilitates our analysis of identification and the
model is still more general than previous fixed-effects discrete choice models.

We impose a restriction on the structural function $\beta_{d}\left(
y,d,\mathbf{z}_{it}\right)  $ that plays a role in our identification results
for this function. We assume that there is not duration dependence in choice
alternative $y=0$, i.e., $\beta_{d}\left(  0,d,\mathbf{z}_{it}\right)  =0$ for
any value of $d$. Also, but without loss of generality, we set $\beta
_{y}(y,y,\mathbf{z}_{it})=0$, i.e., the switching cost of no-switching is
zero.\footnote{Given the payoff function in equation
(\ref{two components beta}), the parameter $\beta_{y}(y,y)$ is completely
irrelevant for an individual's optimal decision. When $y_{it}=y_{i,t-1}=y$, we
have that $\beta\left(  y,\mathbf{x}_{it}\right)  =\beta_{d}\left(
y,d_{it}\right)  +0$ such that the term $\beta_{y}\left(  y,y\right)  $ never
enters in the relevant payoff function. Therefore, $\beta_{y}\left(
y,y\right)  $ can be normalized to zero without loss of generality.}
Assumption 1 summarizes our basic conditions on the model. For the rest of the
paper, we assume that this assumption holds.

\medskip

\noindent\textit{ASSUMPTION 1. (A) The time horizon is infinite and }%
$\delta_{i}\in(0,1)$\textit{. (B) The utility function has the form given by
equations (\ref{one-period utility general specification}) and
(\ref{two components beta}), and functions }$\alpha\left(  y,\eta
,\mathbf{z}\right)  $\textit{, }$\beta_{d}\left(  y,d,\mathbf{z}\right)
$\textit{, and }$\beta_{y}(y,y_{-1},\mathbf{z})$ \textit{are bounded. (C)
}$\beta_{y}(y,y,\mathbf{z})=0$\textit{, }$\beta_{d}\left(  0,d,\mathbf{z}%
\right)  =0$\textit{. (D) }$\{\varepsilon_{it}(y):y\in\mathcal{Y}%
\}$\textit{\ are }$i.i.d.$\textit{\ over }$(i,t,y)$\textit{\ with a extreme
value type I distribution.} \textit{(E) }$\mathbf{z}_{it}$\textit{\ has
discrete and finite support }$\mathcal{Z}$\textit{\ and follows a
time-homogeneous Markov process. (F) The probability distribution of
}$\mathbf{\theta}_{i}\equiv(\boldsymbol{\eta}_{i},\delta_{i})$
\textit{conditional on} $\{\mathbf{z}_{it},\mathbf{x}_{it}:t=1,2,...\}$%
\textit{\ is nonparametrically specified and completely unrestricted.}%
$\qquad\blacksquare$

\medskip

Since the model does not have duration dependence when at choice alternative
$0$, it is convenient for notation to make duration equal to zero at state
$y_{t-1}=0$. In other words, we consider the following modification in the
transition rule for duration:%
\begin{equation}
d_{i,t+1}=\left\{
\begin{array}
[c]{ccc}%
1\left\{  y_{it}=y_{i,t-1}\right\}  \text{ }d_{it}+1 & \text{if} & y_{it}>0\\
0 & \text{if} & y_{it}=0
\end{array}
\right.  \label{modified transition for duration}%
\end{equation}

For our identification results in forward-looking models with duration
dependence, we also impose the following assumption.

\medskip

\noindent\textit{ASSUMPTION 2. For any }$y\in\mathcal{Y}$ \textit{there is a
finite value of duration, }$d_{y}^{\ast}<\infty$\textit{, such that the
marginal return of duration is zero for values greater that} $d_{y}^{\ast}%
$:\footnote{The assumption of no duration dependence in choice alternative
$y=0$ is equivalent to assuming $d_{0}^{\ast}=1$.}%
\begin{equation}
\beta_{d}\left(  y,d,\mathbf{z}\right)  =\beta_{d}\left(  y,d_{y}^{\ast
},\mathbf{z}\right)  \text{ \ \textit{for any} }d\geq d_{y}^{\ast}%
\qquad\blacksquare\label{assumption on d*}%
\end{equation}

\medskip

For the moment, we assume that the researcher knows the values of $d_{y}%
^{\ast}$. In section 4, we show that these values $\{d_{y}^{\ast}\}$ are
identified from the data.

The following are some examples of models within the class defined by
Assumption 1.

\medskip

\noindent\textit{(a) Market entry-exit models. }In its simpler version, there
is only one market, and the choice variable is binary and represents a firm's
decision of being active in the market ($y_{it}=1$) or not ($y_{it}=0$), e.g.,
Dunne et al. (2013). The only endogenous state variable is the lagged
decision, $y_{i,t-1}$. The parameter $-\beta_{y}\left(  1,0,\mathbf{z}\right)
$ represents the cost of entry in the market. Similarly, the parameter
$-\beta_{y}\left(  0,1,\mathbf{z}\right)  $ represents the cost of exit from
the market. An extension of the basic entry model includes as an endogenous
state variable the number of periods of experience since last entry in the
market, $d_{it}$, which follows the transition rule $d_{i,t+1}=d_{it}+1$ if
$y_{it}=1$ and $d_{i,t+1}=0$ if $y_{it}=0$. The parameter $\beta_{d}\left(
1,d,\mathbf{z}\right)  $ represents the effect of market experience on the
firm's profit (Roberts and Tybout, 1997). The model can be extended to $J$
markets (Sweeting, 2013; Caliendo et al, 2015). The two endogenous state
variables are the index of the market where the firm was active at the
previous period ($y_{i,t-1}$) and the number of periods of experience in the
current market ($d_{it}$). The parameter $\beta_{y}\left(  y,y_{-1}%
,\mathbf{z}\right)  $ represents the cost of switching from market $y_{-1}$ to
market $y$. There is not duration dependence if a firm is not active in any
market (if $y=0$), and the marginal return to experience in market $y$ is zero
after $d_{y}^{\ast}$ periods in the market.

\medskip

\noindent\textit{(b) Occupational choice models (Miller, 1984; Keane and
Wolpin, 1997). }A worker chooses between $J$ occupations and the choice
alternative of not working ($y=0$). There are costs of switching occupations
such that a worker's occupation at previous period, $y_{it-1}$, is a state
variable of the model. There is (passive) learning that increases productivity
in the current occupation. There is not duration dependence if the worker is unemployed.

\medskip

\noindent\textit{(c) Machine replacement models (Rust, 1987; Das, 1992;
Kennet, 1993; and Kasahara, 2009). }The choice variable is binary and it
represents the decision of keeping a machine ($y_{it}=1$) or replacing it
($y_{it}=0$). The only endogenous state variable is the number of periods
since the last replacement, $d_{it}$, i.e., the machine age. The evolution of
the machine age is $d_{i,t+1}=d_{it}+1$ if $y_{it}=1$ and $d_{i,t+1}=0$ if
$y_{it}=0$. The parameter $\beta_{d}\left(  1,d,\mathbf{z}\right)  $
represents the effect of age on the firm's profit, e.g., productivity declines
and maintenance costs increase with age.\footnote{In some versions of this
model, such as Rust (1987), the endogenous state variable represents
cumulative usage of the machine and it can follow a stochastic transition
rule. We consider this stochastic version of the model in the Appendix.} More
generally, the class of models in this paper includes binary choice models of
investment in capital, inventory, or capacity (Aguirregabiria 1999; Ryan,
2013; Kalouptsidi, 2014), as long as the depreciation of the stock is deterministic.

\medskip

\noindent\textit{(d) Dynamic demand of differentiated products (Erdem, Imai,
and Keane, 2003; Hendel and Nevo, 2006). }A differentiated product has $J$
varieties and a consumer chooses which one, if any, to purchase (no purchase
is represented by $y=0$). Brand switching costs imply that the brand in the
last purchase is a state variable (Erdem, Keane, and Sun, 2008). For storable
products, the duration since last purchase, $d_{it}$, represents (or proxies)
the consumer's level of inventory that is an endogenous state variable.
Function $\beta_{d}\left(  y,d,\mathbf{z}\right)  $ captures the effect of
inventory on the consumer's utility, and function $\beta_{y}\left(
y,y_{-d},,\mathbf{z}\right)  $ represents brand switching costs.

\medskip

\noindent\textit{(e) Menu costs models of pricing (Slade, 1998;
Aguirregabiria, 1999; Willis, 2006; Kano, 2013). }A firm sells a product and
chooses its price to maximize intertemporal profits. The firm's profit has two
components: a variable profit that depends on the real price (in logarithms),
$r_{it}$; and a fixed menu cost that is paid only if the firm changes its
nominal price. There is a constant inflation rate, $\pi$, that erodes the real
price. Every period, the firm decides whether to keep its nominal price
($y_{it}=1$) or to adjust it ($y_{it}=0$) such that current real price becomes
$r^{\ast}$. The evolution of log-real-price is: $r_{it+1}=r_{it}-\pi$ if
$y_{it}=1$, and $r_{it+1}=r^{\ast}-\pi$ if $y_{it}=0$. If $d_{it}$ represents
the time duration since the last nominal price change, we can represent the
transition rule of the real price as follows: $(r_{it+1}-r^{\ast})/\pi
=d_{it}+1$ if $y_{it}=1$, and $(r_{it+1}-r^{\ast})/\pi=0$ if $y_{it}=0$. This
model has a similar structure as the machine replacement models described
above.$\qquad\blacksquare$

\medskip

We now derive the optimal decision rule and the conditional choice
probabilities in this model. Agent $i$ chooses $y_{it}$ to maximize its
expected and discounted intertemporal utility.\ Given the infinite horizon and
the time-homogeneous utility and transition probability functions, Blackwell's
Theorem establishes that the value function and the optimal decision rule are
time-invariant (Blackwell, 1965). Let $V_{\mathbf{\theta}_{i}}\left(
y_{t},d_{t},\mathbf{z}_{t}\right)  $ be the integrated (or smoothed) value
function for agent type $\mathbf{\theta}_{i}$, as defined by Rust
(1994).\footnote{The integrated value function is defined as the integral of
the value function over the distribution of the i.i.d. unobservable state
variables $\varepsilon$.} The optimal choice at period $t$ can be represented
as:%
\begin{equation}%
\begin{array}
[c]{ccl}%
y_{it} & = & \arg\max\limits_{y\in\mathcal{Y}}\left\{  \text{ }\alpha\left(
y,\boldsymbol{\eta}_{i},\mathbf{z}_{it}\right)  +\beta\left(  y,\mathbf{x}%
_{it},\mathbf{z}_{it}\right)  +\varepsilon_{it}(y)+\delta_{i}\text{
}\mathbb{E}\left[  V_{\mathbf{\theta}_{i}}\left(  y,d_{i,t+1},\mathbf{z}%
_{i,t+1}\right)  \text{ }|\text{ }y\text{, }\mathbf{x}_{it},\mathbf{z}%
_{it}\right]  \text{ }\right\}
\end{array}
\label{optimal decision rule}%
\end{equation}
Note that $d_{i,t+1}$ is a deterministic function of $(y$, $\mathbf{x}_{it})$,
i.e., conditional on $y_{it}=y$. Therefore, we can represent the continuation
value $\mathbb{E}[V_{\mathbf{\theta}_{i}}\left(  y,d_{i,t+1},\mathbf{z}%
_{i,t+1}\right)  $ $|$ $y$, $\mathbf{x}_{it},\mathbf{z}_{it}]$ using a
function $v_{\mathbf{\theta}_{i}}(y,d_{t+1}[y,\mathbf{x}_{it}]),\mathbf{z}%
_{it})$ with $d_{t+1}[y,\mathbf{x}_{it}]=0$ if $y=0$ and $d_{t+1}%
[y,\mathbf{x}_{it}]=$ $1\{y=y_{it-1}\}d_{it}+1$ if $y>0$. The extreme value
type 1 distribution of the unobservables $\varepsilon$ implies that the
\textit{conditional choice probability} (CCP) function has the following form:%
\begin{equation}%
\begin{array}
[c]{ccl}%
P_{\mathbf{\theta}_{i}}\left(  y\text{ }|\text{ }\mathbf{x}_{it}%
,\mathbf{z}_{it}\right)  & = & \dfrac{\exp\left\{  \text{ }\alpha\left(
y,\boldsymbol{\eta}_{i},\mathbf{z}_{it}\right)  +\beta\left(  y,\mathbf{x}%
_{it},\mathbf{z}_{it}\right)  +v_{\mathbf{\theta}_{i}}(y,d_{t+1}%
[y,\mathbf{x}_{it}],\mathbf{z}_{it})\text{ }\right\}  }{\sum\limits_{j\in
\mathcal{Y}}\exp\left\{  \text{ }\alpha\left(  j,\boldsymbol{\eta}%
_{i},\mathbf{z}_{it}\right)  +\beta\left(  j,\mathbf{x}_{it},\mathbf{z}%
_{it}\right)  +v_{\mathbf{\theta}_{i}}(j,d_{t+1}[j,\mathbf{x}_{it}%
],\mathbf{z}_{it})\text{ }\right\}  }%
\end{array}
\label{general CCP function}%
\end{equation}

The continuation value function $v_{\mathbf{\theta}_{i}}$ has two properties
which play an important role in our identification results. These properties
establish conditions under which the continuation values do not depend on
current endogenous state variables, $(y_{i,t-1}.d_{it})$.

\medskip

\noindent\textit{Property 1. }In a model without duration dependence (i.e.,
$\beta_{d}=0$), the continuation value function becomes $v_{\mathbf{\theta
}_{i}}(y,\mathbf{z}_{it})$ that does not depend on the state variable,
$y_{it-1}$.

\medskip

\noindent\textit{Property 2. }In a model with duration dependence, the
continuation $v_{\mathbf{\theta}_{i}}(y,d_{t+1}[y,\mathbf{x}_{it}%
],\mathbf{z}_{it})$ is equal to $v_{\mathbf{\theta}_{i}}(y,d_{y}^{\ast
},\mathbf{z}_{it})$ for any duration $d_{t+1}[y,\mathbf{x}_{it}]\geq
d_{y}^{\ast}$.

\section{Identification}

\subsection{Preliminaries}

The researcher has a panel dataset of $N$ individuals over $T$ periods of
time, $\{y_{it},$ $\mathbf{x}_{it}$ $,$ $\mathbf{z}_{it}:i=1,2,...,N$ $;$
$t=1,2,...,T\}$. We consider microeconometric applications where $N$ is large
and $T$ is small. More precisely, our identification results and the
asymptotic properties of the proposed estimator assume that $N$ goes to
infinity and $T$ is small and fixed.\footnote{Note that $T$ represents the
number of periods with data on the decision variable and the state variables
for all the individuals. The set of observable state variables includes the
endogenous state variables $y_{i,t-1}$ and $d_{it}$. Knowing the values of
these state variables at the initial period $t=1$ (i.e., $y_{i0}$ and $d_{i1}%
$) may require data on the individual's choices for periods before $t=1$.
Therefore, the time dimension $T$ may not correspond to the actual time
dimension of the required panel dataset.} We are interested in the
identification of the functions $\beta_{y}$ and $\beta_{d}$ that represent the
dependence of utility with respect to the endogenous state variables.

For the rest of this section, we omit the individual subindex $i$ in most of
the expressions, and instead we include $\mathbf{\theta}$ as an argument (or
subindex) in those functions that depend on the time-invariant unobserved
heterogeneity, i.e., $\alpha_{\mathbf{\theta}}\left(  y,\mathbf{z}\right)  $
and $v_{\mathbf{\theta}}\left(  \mathbf{x},\mathbf{z}\right)  $. We use
$\mathbf{\beta}$ to represent the vector of structural parameters that define
the functions $\beta_{y}$ and $\beta_{d}$.\footnote{Since $\left(
y_{t},\mathbf{x}_{t},\mathbf{z}_{t}\right)  $ has finite support, we can
represent the structural functions $\beta_{y}\left(  y_{t},y_{t-1}%
,\mathbf{z}_{t}\right)  $ and $\beta_{d}\left(  y_{t},d_{t},\mathbf{z}%
_{t}\right)  $ using a finite vector of parameters.}

As in Honor\'{e} and Kyriazidou (2000), our sufficient statistics include the
condition that the exogenous state variables, $\mathbf{z}$, remains constant
over several periods. For notational simplicity, we omit $\mathbf{z}$ as an
argument in most of the expressions for the rest of this section. In section
4, we explain how to deal with this condition in the implementation of the
conditional maximum likelihood estimator.

Let $\widetilde{\mathbf{y}}=\{y_{1},y_{2},...,y_{T}\}$ be an individual's
observed history of choices and exogenous state variables, respectively. The
model implies that:%
\begin{equation}
\mathbb{P}\left(  \widetilde{\mathbf{y}}\text{ }|\text{ }\mathbf{x}%
_{1},\mathbf{\theta},\mathbf{\beta}\right)  =\dprod\limits_{t=1}^{T}%
\dfrac{\exp\left\{  \text{ }\alpha_{\mathbf{\theta}}\left(  y_{t}\right)
+\beta\left(  y_{t},\mathbf{x}_{t}\right)  +v_{\mathbf{\theta}}\left(
y_{t},d_{t+1}[y_{t},\mathbf{x}_{t}]\right)  \text{ }\right\}  }{\sum
\limits_{j\in\mathcal{Y}}\exp\left\{  \text{ }\alpha_{\mathbf{\theta}}\left(
j\right)  +\beta\left(  j,\mathbf{x}_{t}\right)  +v_{\mathbf{\theta}}\left(
j,d_{t+1}[j,\mathbf{x}_{t}]\right)  \text{ }\right\}  }
\label{prob choice history}%
\end{equation}
Our identification results, for different versions of the model, have the
following common features. First, we show that the log-probability function
$\ln\mathbb{P}\left(  \widetilde{\mathbf{y}}\text{ }|\text{ }\mathbf{x}%
_{1},\mathbf{\theta},\mathbf{\beta}\right)  $ has the following structure:%
\begin{equation}
\ln\mathbb{P}\left(  \widetilde{\mathbf{y}}\text{ }|\text{ }\mathbf{x}%
_{1},\mathbf{\theta},\mathbf{\beta}\right)  =U(\widetilde{\mathbf{y}%
},\mathbf{x}_{1})^{\prime}g_{\mathbf{\theta}}+S(\widetilde{\mathbf{y}%
},\mathbf{x}_{1})^{\prime}\mathbf{\beta}^{\ast}
\label{stylized representation logP}%
\end{equation}
where $U(\widetilde{\mathbf{y}},\mathbf{x}_{1})$ and $S(\widetilde{\mathbf{y}%
},\mathbf{x}_{1})$ are vectors of statistics (i.e., deterministic functions of
the history $(\widetilde{\mathbf{y}},\mathbf{x}_{1})$), $g_{\mathbf{\theta}}$
is a vector of functions of $\mathbf{\theta}$, and $\mathbf{\beta}^{\ast}$ is
a vector of linear combinations of the original vector of structural
parameters $\mathbf{\beta}$. This representation is such that each of the
vectors, $U$, $g_{\mathbf{\theta}}$, $S$, and $\mathbf{\beta}^{\ast}$, has
elements which are linearly independent.\footnote{Suppose that $S$ and
$\mathbf{\beta}$ are $K\times1$ vectors, and only $K^{\ast}<K$ elements in $S$
are linearly independent. Then, $S=[S_{a},S_{b}]$ where $S_{a}$ contains
$K^{\ast}$ linearly independent elements, and $S_{b}=\mathbf{A}$ $S_{a}$ where
$\mathbf{A}$ is a $(K-K^{\ast})\times K^{\ast}$ matrix. This implies that
$S^{\prime}\mathbf{\beta}=$ $S_{a}^{\prime}\mathbf{\beta}^{\ast}$ with
$\mathbf{\beta}^{\ast}=[\mathbf{I}$ $\vdots$ $\mathbf{A}]^{\prime
}\mathbf{\beta}$, such that $S_{a}$ and $\mathbf{\beta}^{\ast}$ are vectors
with linearly independent elements.} Based on this representation of the
log-probability of a choice history, we establish the following results. For
notational simplicity, we use $U$ and $S$ to represent $U(\widetilde
{\mathbf{y}},\mathbf{x}_{1})$ and $S(\widetilde{\mathbf{y}},\mathbf{x}_{1})$, respectively.

\medskip

\noindent\textit{(i) Sufficiency.} $U$ is a sufficient statistic for
$\mathbf{\theta}$, i.e., for any $(\widetilde{\mathbf{y}},\mathbf{x}_{1})$ and
$\mathbf{\theta}$, $\mathbb{P}\left(  \widetilde{\mathbf{y}}|\mathbf{x}%
_{1},\mathbf{\theta},U\right)  $ does not depend on $\mathbf{\theta}$. By
definition, $\ln\mathbb{P}\left(  \widetilde{\mathbf{y}}|\mathbf{x}%
_{1},\mathbf{\theta},U\right)  $ is equal to $\ln\mathbb{P}\left(
\widetilde{\mathbf{y}}|\mathbf{x}_{1},\mathbf{\theta}\right)  -$
$\ln\mathbb{P}\left(  U|\mathbf{x}_{1},\mathbf{\theta}\right)  $, and taking
into account the form of the log-probability in equation
(\ref{stylized representation logP}), we have:%
\begin{equation}%
\begin{array}
[c]{ccl}%
\ln\mathbb{P}\left(  \widetilde{\mathbf{y}}|\mathbf{x}_{1},U,\mathbf{\beta
}^{\ast}\right)  & = & U^{\prime}g_{\mathbf{\theta}}+S^{\prime}\mathbf{\beta
}^{\ast}-\ln\left(
%TCIMACRO{\tsum \limits_{j:U(j)=U}}%
%BeginExpansion
{\textstyle\sum\limits_{j:U(j)=U}}
%EndExpansion
\exp\left\{  U(j)^{\prime}g(\mathbf{\theta})+S(j)^{\prime}\mathbf{\beta}%
^{\ast}\right\}  \right) \\
&  & \\
& = & S^{\prime}\mathbf{\beta}^{\ast}-\ln\left(
%TCIMACRO{\tsum \limits_{j:U(j)=U}}%
%BeginExpansion
{\textstyle\sum\limits_{j:U(j)=U}}
%EndExpansion
\exp\left\{  S(j)^{\prime}\mathbf{\beta}^{\ast}\right\}  \right)
\end{array}
\label{sufficiency general.}%
\end{equation}
where $%
%TCIMACRO{\tsum \nolimits_{j:U(j)=U}}%
%BeginExpansion
{\textstyle\sum\nolimits_{j:U(j)=U}}
%EndExpansion
$ represents the sum over all the histories that imply the same value $U$ of
the vector of sufficient statistics. Equation (\ref{sufficiency general.})
shows that the structure of the log-probability in
(\ref{stylized representation logP}) implies that $U$ is a sufficient
statistic for $\mathbf{\theta}$.

\medskip

\noindent\textit{(ii) Minimal sufficiency.} $U$ is a \textit{minimal
sufficient statistic}, i.e., it does not contain redundant information. More
formally, let $\mathbf{U}$ be a matrix where each row corresponds to a value
of the choice history $(\widetilde{\mathbf{y}},\mathbf{x}_{1})$. Then, $U$ is
\textit{minimal }if and only if matrix $\mathbf{U}$ is full-column rank.

\medskip

\noindent\textit{(iii) Identification. }Define the conditional log-likelihood
function, in the population, $\ell\left(  \mathbf{\beta}^{\ast}\right)
\equiv$ $\mathbb{E}_{(\widetilde{\mathbf{y}},\mathbf{x}_{1})}\left[
\ln\mathbb{P}\left(  \widetilde{\mathbf{y}}|\mathbf{x}_{1},U,\mathbf{\beta
}^{\ast}\right)  \right]  $. The vector of parameters $\mathbf{\beta}^{\ast}$
is point identified if the population likelihood is uniquely maximized at the
true value of $\mathbf{\beta}^{\ast}$. Lemma 1 establishes a necessary and
sufficient condition for identification that is simple to verify. Let $K$ be
the dimension of the vector of statistics $S$ and of the vector of
parameters\textit{ }$\mathbf{\beta}^{\ast}$.

\medskip

\noindent\textit{LEMMA 1. Given }$K+1$ \textit{histories }$(\widetilde
{\mathbf{y}},\mathbf{x}_{1})$\textit{, say }$\{A_{j}:j=0,1,2...,K\}$\textit{,
define a }$K\times K$ \textit{matrix }$\mathbf{S}$ \textit{such that every row
}$j$ \textit{is associated to a history and contains the vector of statistics
}$S(A_{j})^{\prime}-S(A_{0})^{\prime}$\textit{. The vector of parameters
}$\mathbf{\beta}^{\ast}$ \textit{is identified if and only if there exist
}$K+1$ \textit{histories with the same value of the statistic }$U$ \textit{and
a non-singular matrix }$\mathbf{S}$\textit{.}\qquad$\blacksquare$

\medskip

\noindent\textit{Corollary: If }$K=1$\textit{, parameter }$\beta^{\ast}%
$\textit{ is identified iff there are two histories, }$A$ \textit{and }$B$,
\textit{such that }$U(A)=U(B)$ \textit{and }$S(A)\neq S(B)$.

\medskip

The derivation of these sufficient statistics should deal with two issues that
do not appear in the previous literature on FE-CMLE of non-structural (or
myopic) nonlinear panel data models. First, we consider models with duration
dependence. Second, we should take into account that unobserved heterogeneity
enters in the continuation value function, $v_{\mathbf{\theta}}$. This implies
that the sufficient statistic $U$ should control not only for $\alpha
_{\mathbf{\theta}}\left(  y_{t}\right)  $ but also for the continuation values
$v_{\mathbf{\theta}}\left(  y_{t},d_{t+1}\right)  $. This is challenging
because, in general, these continuation values depend on the endogenous state
variables. We cannot fully control for (or condition on) the value of the
state variables because the identification condition (iii) would not hold. We
show that there are states where the continuation value does not depend on
current state variables once we condition on current choices.

The presentation of our identification results tries to emphasize both the
links and extensions with previous results in the literature. For this reason,
we start presenting identification results for the binary choice model, that
is the model more extensively studied in the literature of nonlinear dynamic
panel data. For this binary choice model, we present new identification
results for the myopic model with duration dependence and for the
forward-looking model with and without duration dependence. Then, we present
our identification results for multinomial models.

\medskip

\noindent\textit{Some useful statistics. }We show below that, in our model,
the log-probability of a choice history, $\mathbb{P}\left(  \widetilde
{\mathbf{y}}\text{ }|\text{ }y_{0},d_{1},\mathbf{\theta},\mathbf{\beta
}\right)  $, can be written in terms of several sets of statistics or
functions of $(y_{0},d_{1},\widetilde{\mathbf{y}})$: the initial and final
choices, $\{y_{0},y_{T}\}$; the initial and final durations, $\{d_{1}%
,d_{T+1}\}$; and the statistics that we denote as \textit{hits},
\textit{dyads}, \textit{histogram of states, }and \textit{histogram of
choice-states}. We now define these statistics. Note that each of these
statistics for a single history $(y_{0},d_{1},\widetilde{\mathbf{y}})$.

\textit{Hit statistics.} For any choice alternative $y\in\mathcal{Y}$, the
\textit{hit }statistic $T^{(y)}$ represents the number of times that
alternative $y$ is visited (or \textit{hit}) during the choice history
$\widetilde{\mathbf{y}}$, i.e., $T^{(y)}\equiv\sum_{t=1}^{T}1\{y_{t}=y\}$.

\textit{Dyad statistics}. For $y_{-1}$ and $y$ in $\mathcal{Y}$, the
\textit{dyad }statistic $D^{(y_{-1},y)}$ is the number of times that the
sequence $(y_{-1},y)$ is observed at two consecutive periods in the choice
history $(y_{0},\widetilde{\mathbf{y}})$, i.e., $D^{(y_{-1},y)}\equiv$
$\sum_{t=1}^{T}1\{y_{t-1}=y_{-1}$, $y_{t}=y\}$.

\textit{Histogram of states}. Given a history $(y_{0},d_{1},\widetilde
{\mathbf{y}})$, the statistic $H^{(y)}(d)$ (for $y\in\mathcal{Y}$ and $d\geq
0$) is the number of times that we observe state $(y_{t-1},d_{t})=(y,d)$,
i.e., $H^{(y)}(d)=\sum_{t=1}^{T}1\{y_{t-1}=y$, $d_{t}=d\}$.

\textit{Extended histogram of states}. For any $y\in\mathcal{Y}$ and $d\geq0$,
the statistic $X^{(y)}(d)$ represents the number of times that we observe
state $(y_{t-1},d_{t})=(y,d)$ and the individual decides to continue one more
period in choice $y$. By definition, $X^{(y)}(d)=\sum_{t=1}^{T}1\{y_{t-1}%
=y_{t}=y$, $d_{t}=d\}$.

\textit{Difference between final and initial states. }For any $y\in
\mathcal{Y}$ and $d\geq0$, the statistic $\Delta^{(y)}(d)$ is defined as
$1\{y_{T}=y$, $d_{T+1}=d\}-1\{y_{0}=y,$ $d_{1}=d\}$. When the difference
applies only to the choice variable, we represent it as $\Delta^{(y)}%
\equiv1\{y_{T}=y\}-1\{y_{0}=y\}$.

Table 1 summarizes our definition of statistics. The following Lemma 2
establishes several properties of these statistics that we apply in our derivations.

\medskip

\noindent\textit{LEMMA 2. For any history }$(y_{0},d_{1},\widetilde
{\mathbf{y}})$\textit{ and value }$y>0$ \textit{the following properties
apply: (i) }$H^{(y)}(0)=0$\textit{; (ii) }$X^{(y)}(0)=0$\textit{; (iii) }%
$\sum_{d\geq1}H^{(y)}(d)=$\textit{ }$T^{(y)}-\Delta^{(y)}$\textit{; (iv)
}$\sum_{d\geq1}X^{(y)}(d)=D^{(y,y)}$\textit{; (v) for }$d\geq1$, $X^{(y)}(d)=$
$H^{(y)}(d+1)+\Delta^{(y)}(d+1)$\textit{; (vi) }$\sum_{d\geq1}\Delta
^{(y)}(d)=$ $\Delta^{(y)}$\textit{; and (vii) for }$y\geq1$\textit{,}
$\sum_{y_{-1}\neq y}D^{(y_{-1},y)}=$ $H^{(y)}(1)+\Delta^{(y)}$\textit{.}%
\qquad$\blacksquare$

\medskip

\begin{center}%
\begin{tabular}
[c]{r|c}\hline\hline
\multicolumn{2}{c}{\textbf{Table 1}}\\
\multicolumn{2}{c}{\textbf{Definition of statistics for a choice history
}$\mathbf{\{}y_{0},d_{1}$ $|$ $\widetilde{\mathbf{y}}\mathbf{\}}$}\\\hline
& \\
\textit{Name: Symbol} & \multicolumn{1}{|l}{\textit{Definition}}\\\hline
\multicolumn{1}{c|}{} & \multicolumn{1}{|l}{}\\
\textit{Hits: }$T^{(y)}$ & \multicolumn{1}{|l}{$\sum_{t=1}^{T}1\{y_{t}=y\}$}\\
& \multicolumn{1}{|l}{}\\
\textit{Dyad: }$D^{(y_{-1},y)}$ & \multicolumn{1}{|l}{$\sum_{t=1}%
^{T}1\{y_{t-1}=y_{-1}$, $y_{t}=y\}$}\\
& \multicolumn{1}{|l}{}\\
\textit{Histogram of states: }$H^{(y)}(d)$ & \multicolumn{1}{|l}{$\sum
_{t=1}^{T}1\{y_{t-1}=y$, $d_{t}=d\}$}\\
& \multicolumn{1}{|l}{}\\
\textit{Extended histogram of states: }$X^{(y)}(d)$ &
\multicolumn{1}{|l}{$\sum_{t=1}^{T}1\{y_{t-1}=y_{t}=y$, $d_{t}=d\}$}\\
& \\
\textit{Diff. final-initial states: }$\Delta^{(y)}(d)$ &
\multicolumn{1}{|l}{$1\{y_{T}=y$,$d_{T+1}=d\}-1\{y_{0}=y,d_{1}=d\}$}\\
$\Delta^{(y)}$ & \multicolumn{1}{|l}{$1\{y_{T}=y\}-1\{y_{0}=y\}$}\\
& \multicolumn{1}{|l}{}\\\hline\hline
\end{tabular}

\end{center}

\subsection{Binary choice models}

Consider the binary choice version of the model characterized by Assumption 1.
The optimal decision rule in this model is:%
\begin{equation}%
\begin{array}
[c]{ccl}%
y_{t} & = & 1\left\{
\begin{array}
[c]{l}%
\alpha_{\mathbf{\theta}}(1)-\alpha_{\mathbf{\theta}}(0)+\beta(1,y_{t-1}%
,d_{t})-\beta(0,y_{t-1},d_{t})\\
\\
+v_{\mathbf{\theta}}\left(  1,d_{t}+1\right)  -v_{\mathbf{\theta}}\left(
0\right)  +\varepsilon_{t}(1)-\varepsilon_{t}(0)\geq0
\end{array}
\right\}
\end{array}
\label{binary choice model optimal decision}%
\end{equation}
where for choice $y=0$ we use $v_{\mathbf{\theta}}\left(  0\right)  $ instead
$v_{\mathbf{\theta}}\left(  0,0\right)  $ to emphasize that there is not
duration dependence when the state is $y=0$. We now present identification
results for different versions of this model, starting with the myopic model
without duration dependence that has been studied by Chamberlain (1985) and
Honor\'{e} and Kyriazidou (2000).

\subsubsection{Myopic dynamic model without duration dependence}

Consider the model in equation (\ref{binary choice model optimal decision})
under the restrictions of myopic behavior (i.e., $\delta=0$) and no duration
dependence (i.e., $\beta_{d}(y,d)=0$). These restrictions imply that the
continuation values, $v_{\mathbf{\theta}}\left(  1,d_{t}+1\right)  $ and
$v_{\mathbf{\theta}}\left(  0\right)  $, become zero, and the term
$\beta(1,y_{t-1},d_{t})-\beta(0,y_{t-1},d_{t})$ becomes equal to $\beta
_{y}(1,0)-y_{t-1}$ $\left[  \beta_{y}(1,0)+\beta_{y}(0,1)\right]  $. We can
present this model using the more standard representation,%
\begin{equation}
y_{t}=1\left\{  \widetilde{\alpha}_{\mathbf{\theta}}+\widetilde{\beta}%
_{y}\text{ }y_{t-1}+\widetilde{\varepsilon}_{t}\geq0\right\}
\label{binary myopic no duration}%
\end{equation}
with $\widetilde{\alpha}_{\mathbf{\theta}}\equiv\alpha_{\mathbf{\theta}%
}(1)-\alpha_{\mathbf{\theta}}(0)+\beta_{y}(1,0)$, $\widetilde{\beta}_{y}%
\equiv-\beta_{y}(1,0)-\beta_{y}(0,1)$, and $\widetilde{\varepsilon}_{t}%
\equiv\varepsilon_{t}(1)-\varepsilon_{t}(0)$. In a model of market entry-exit,
the parameter $\widetilde{\beta}_{y}$ represents the sum of the costs of entry
and exit, or equivalently the sunk cost of entry. This is an important
structural parameter.

Define function $\sigma_{\mathbf{\theta}}(y_{t-1})\equiv-\ln\left(
1+\exp\left\{  \widetilde{\alpha}_{\mathbf{\theta}}+\widetilde{\beta}%
_{y}y_{t-1}\right\}  \right)  $. The log-probability of the choice history
$\widetilde{\mathbf{y}}$ conditional on $(y_{0},\mathbf{\theta})$ is:%
\begin{equation}%
\begin{array}
[c]{ccl}%
\ln\mathbb{P}\left(  \widetilde{\mathbf{y}}\text{ }|\text{ }y_{0}%
,\mathbf{\theta}\right)  & = & \dsum\limits_{t=1}^{T}y_{t}\left[
\widetilde{\alpha}_{\mathbf{\theta}}+\widetilde{\beta}_{y}y_{t-1}\right]
+(1-y_{t-1})\text{ }\sigma_{\mathbf{\theta}}(0)+y_{t-1}\text{ }\sigma
_{\mathbf{\theta}}(1)
\end{array}
\label{Binary myopic no duration: log prob}%
\end{equation}
Proposition 1 establishes (i) the sufficient statistic, (ii) minimal
sufficiency, and (iii) identification for this model.

\medskip

\noindent\textit{PROPOSITION 1. In the myopic binary choice model without
duration dependence the log-probability of a choice history} \textit{has the
form}%
\begin{equation}%
\begin{array}
[c]{ccl}%
\ln\mathbb{P}\left(  \widetilde{\mathbf{y}}\text{ }|\text{ }y_{0}%
,\mathbf{\theta}\right)  & = & T^{(1)}\,g_{\mathbf{\theta},1}+\Delta
^{(1)}\,g_{\mathbf{\theta},2}+\widetilde{\beta}_{y}\text{ }D^{(1,1)}%
\end{array}
\label{Prop 1 binary myopic nodura}%
\end{equation}
\textit{with }$g_{\mathbf{\theta},1}\equiv\widetilde{\alpha}_{\mathbf{\theta}%
}+\sigma_{\mathbf{\theta}}(1)-\sigma_{\mathbf{\theta}}(0)$\textit{, and
}$g_{\mathbf{\theta},2}\equiv\sigma_{\mathbf{\theta}}(0)-\sigma
_{\mathbf{\theta}}(1)$\textit{, such that }$U=\{T^{(1)},$ $\Delta^{(1)}%
\}$\textit{, }$S=D^{(1,1)}$\textit{, and }$\mathbf{\beta}^{\ast}%
=\widetilde{\beta}_{y}$. \textit{We have that: (i) }$U=\{T^{(1)},$
$\Delta^{(1)}\}$ \textit{is a sufficient statistic; (ii) }$T^{(1)}$
\textit{and} $\Delta^{(1)}$ \textit{are linearly independent such that }%
$U$\textit{ is a minimal sufficient statistic; and (iii) for }$T\geq3$
\textit{there is a pair of histories }$\{y_{0}|\widetilde{\mathbf{y}}%
\}$\textit{, say }$A$ \textit{and }$B$\textit{, with }$U(A)=U(B)$ \textit{and
}$S(A)\neq S(B)$ \textit{such that the parameter }$\widetilde{\beta}_{y}$
\textit{is identified as} $\left[  \ln\mathbb{P}\left(  A|U\right)
-\ln\mathbb{P}\left(  B|U\right)  \right]  /$\textit{ }$\left[  D_{A}%
^{(1,1)}-D_{B}^{(1,1)}\right]  $\textit{. For instance, with }$T=3$\textit{,
}$A=\{0|0,1,1\}$ \textit{and }$B=\{0|1,0,1\}$.\qquad$\blacksquare$

\medskip

This Proposition 1 is almost identical to the identification result in
Chamberlain (1985). Chamberlain shows that the vector of statistics
$\{T^{(1)},y_{0}$, $y_{T}\}$ is sufficient for $\mathbf{\theta}$, and
conditional on this vector the parameter $\widetilde{\beta}_{y}$ is
identified. Our Proposition 1 shows that Chamberlain's sufficient statistic is
not minimal and the minimal statistic is $\{T^{(1)},y_{T}-y_{0}\}$. However,
it turns out that, in this binary choice model, the extra variation left by
the minimal sufficient statistic does not help in the identification of
$\widetilde{\beta}_{y}$, so the two CMLEs are equivalent.

\subsubsection{Forward-looking dynamic model without duration dependence}

Consider a forward-looking version of the model in equation
(\ref{binary choice model optimal decision}) but still without duration
dependence. Since the model is of forward-looking behavior, now we have the
continuation values $v_{\mathbf{\theta}}\left(  1,d_{t}+1\right)
-v_{\mathbf{\theta}}\left(  0\right)  $. However, there is not duration
dependence, and the only state variable is $y_{t-1}$. Therefore, for this
version of the model we have that $v_{\mathbf{\theta}}\left(  1,d_{t}%
+1\right)  -v_{\mathbf{\theta}}\left(  0\right)  $ becomes $v_{\mathbf{\theta
}}(1)-v_{\mathbf{\theta}}(0)\equiv$ $\widetilde{v}_{\mathbf{\theta}}$, i.e.,
continuation values depend on current choices but not on the current state
variable $y_{t-1}$. We can represent this model as,%
\begin{equation}
y_{t}=1\{\widetilde{\alpha}_{\mathbf{\theta}}+\widetilde{v}_{\mathbf{\theta}%
}+\widetilde{\beta}_{y}y_{t-1}+\widetilde{\varepsilon}_{t}\geq0\}
\label{Bynary forward-looking no duration}%
\end{equation}
The only difference between this model and the myopic model is that now the
fixed effect has two components: $\widetilde{\alpha}_{\mathbf{\theta}}$ that
comes from current profit, and $\widetilde{v}_{\mathbf{\theta}}$ that comes
from the continuation values. However, from the point of view of
identification and estimation, the two models are observationally equivalent.

A key feature of this model, that determines the observational equivalence
with the myopic model, is the property that the state variable at period $t+1$
depends on the choice at period $t$ but not on the state variable at period
$t$, i.e., $x_{t+1}=y_{t}$.

Proposition 2 establishes this equivalence.

\medskip

\noindent\textit{PROPOSITION 2. In the forward-looking binary choice model
without duration dependence the log-probability of a choice history}
\textit{has the form}%
\begin{equation}%
\begin{array}
[c]{ccl}%
\ln\mathbb{P}\left(  \widetilde{\mathbf{y}}\text{ }|\text{ }y_{0}%
,\mathbf{\theta}\right)  & = & T^{(1)}\,g_{\mathbf{\theta},1}+\Delta
^{(1)}\,g_{\mathbf{\theta},2}+\widetilde{\beta}_{y}\text{ }D^{(1,1)}%
\end{array}
\label{Prop 2 binary myopic nodura}%
\end{equation}
\textit{with }$g_{\mathbf{\theta},1}\equiv\widetilde{\alpha}_{\mathbf{\theta}%
}+\widetilde{v}_{\mathbf{\theta}}+\sigma_{\mathbf{\theta}}(1)-\sigma
_{\mathbf{\theta}}(0)$\textit{, and }$g_{\mathbf{\theta},2}\equiv
\sigma_{\mathbf{\theta}}(0)-\sigma_{\mathbf{\theta}}(1)$\textit{, such that
}$U=\{T^{(1)},$ $\Delta^{(1)}\}$\textit{, }$S=D^{(1,1)}$\textit{, and
}$\mathbf{\beta}^{\ast}=\widetilde{\beta}_{y}$. \textit{We have that: (i)
}$U=\{T^{(1)},$ $\Delta^{(1)}\}$ \textit{is a sufficient statistic; (ii)
}$T^{(1)}$ \textit{and} $\Delta^{(1)}$ \textit{are linearly independent such
that }$U$\textit{ is a minimal sufficient statistic; and (iii) for }$T\geq3$
\textit{there is a pair of histories }$\{y_{0}|\widetilde{\mathbf{y}}%
\}$\textit{, say }$A$ \textit{and }$B$\textit{, with }$U(A)=U(B)$ \textit{and
}$S(A)\neq S(B)$ \textit{such that the parameter }$\widetilde{\beta}_{y}$
\textit{is identified as} $\left[  \ln\mathbb{P}\left(  A|U\right)
-\ln\mathbb{P}\left(  B|U\right)  \right]  /$\textit{ }$\left[  D_{A}%
^{(1,1)}-D_{B}^{(1,1)}\right]  $\textit{.}\qquad$\blacksquare$

\subsubsection{Myopic dynamic model with duration dependence}

The continuation values are zero, and the term $\beta(1,y_{t-1},d_{t}%
)-\beta(0,y_{t-1},d_{t})$ is equal to $(1-y_{t-1})$ $\beta_{y}(1,0)+$
$y_{t-1}$ $\beta_{d}(1,d_{t})-$ $y_{t-1}\beta_{y}(0,1)$, and it can be
represented as $\beta_{y}(1,0)+\widetilde{\beta}_{y}$ $y_{t-1}+\beta
_{d}(1,d_{t})$ $y_{t-1}$. Therefore, we can present this model as%
\begin{equation}
y_{t}=1\{\widetilde{\alpha}_{\mathbf{\theta}}+\widetilde{\beta}_{y}%
y_{t-1}+\beta_{d}(1,d_{t})\text{ }y_{t-1}+\widetilde{\varepsilon}_{t}\geq0\}
\label{Binary myopic with duration}%
\end{equation}
For this model, the log-probability of the choice history $\widetilde
{\mathbf{y}}$ conditional on $(y_{0},d_{1},\mathbf{\theta})$ is:%
\begin{equation}%
\begin{array}
[c]{ccl}%
\ln\mathbb{P}\left(  \widetilde{\mathbf{y}}\text{ }|\text{ }y_{0}%
,d_{1},\mathbf{\theta}\right)  & = & \dsum\limits_{t=1}^{T}y_{t}\left[
\widetilde{\alpha}_{\mathbf{\theta}}+\widetilde{\beta}_{y}\text{ }%
y_{t-1}+\beta_{d}(1,d_{t})\text{ }y_{t-1}\right]  +\sigma_{\mathbf{\theta}%
}(y_{t-1},d_{t})
\end{array}
\label{log prob for binary myopic duration}%
\end{equation}
where $\sigma_{\mathbf{\theta}}(y_{t-1},d_{t})\equiv-\ln\left(  1+\exp\left\{
\widetilde{\alpha}_{\mathbf{\theta}}+\widetilde{\beta}_{y}\text{ }%
y_{t-1}+\beta_{d}(1,d_{t})\text{ }y_{t-1}\right\}  \right)  $. In order to
emphasize that $\sigma_{\mathbf{\theta}}(y_{t-1},d_{t})$ does not depend on
$d_{t}$ when $y_{t-1}=0$, we use the notation $\sigma_{\mathbf{\theta}}(0)$ to
represent $\sigma_{\mathbf{\theta}}(0,0)$.

Proposition 3 establishes the minimal sufficient statistic and identification
of structural parameters in this model.

\medskip

\noindent\textit{PROPOSITION 3. In the myopic binary choice model with
duration dependence under Assumption 1, the log-probability of a choice
history} \textit{has the form}%
\begin{equation}%
\begin{array}
[c]{ccl}%
\ln\mathbb{P}\left(  \widetilde{\mathbf{y}}|y_{0},d_{1}\right)  & = &
\sum\limits_{d\geq1}H^{(1)}(d)\text{ }g_{\mathbf{\theta},1}(d)+\Delta
^{(1)}\text{ }g_{\mathbf{\theta},2}+\sum\limits_{d\geq1}\Delta^{(1)}(d)\text{
}\gamma(d-1)
\end{array}
\label{Prop 3 log prob}%
\end{equation}
\textit{with }$g_{\mathbf{\theta},1}(d)\equiv\widetilde{\alpha}%
_{\mathbf{\theta}}+\sigma_{\mathbf{\theta}}(1,d)-\sigma_{\mathbf{\theta}%
}(0)+\gamma(d-1)$\textit{, }$g_{\mathbf{\theta},2}\equiv\widetilde{\alpha
}_{\mathbf{\theta}}$\textit{, }$\gamma(d)\equiv\widetilde{\beta}_{y}+\beta
_{d}(1,d)$\textit{, and }$\gamma(0)=0$\textit{, such that }$U=\{H^{(1)}%
(d):d\geq1,$ $\Delta^{(1)}\}$\textit{, }$S=$ $\{\Delta^{(1)}(d):d\geq1\}$,
\textit{and }$\mathbf{\beta}^{\ast}=$ $\{\gamma(d):d\geq1\}$. \textit{Then, we
have that: (i) }$U$ \textit{is a sufficient statistic. (ii) The elements in
the vector }$U$ \textit{are linearly independent such that }$U$\textit{ is a
minimal sufficient statistic. (iii) Conditional on }$U$\textit{, the
statistics }$\{\Delta^{(1)}(d):d\geq1\}$ \textit{have variation and the
structural parameters }$\{\gamma(d):1\leq d\leq T-2\}$ \textit{are identified,
i.e., for any }$1\leq d\leq T-2$\textit{, there is a pair of histories, }$A$
\textit{and }$B$\textit{, such that }$U(A)=U(B)$ \textit{and }$\gamma(d)=$
$\ln\mathbb{P}\left(  A|U\right)  -\ln\mathbb{P}\left(  B|U\right)  $%
.\qquad$\blacksquare$

\medskip

\noindent\textit{Proof. }The derivation of equation (\ref{Prop 3 log prob}) is
in the Appendix. Proof of (iii). For any duration $n$, with $1\leq n\leq T-2$,
define a sub-history $\{y_{0},d_{1}$ $|$ $y^{n+2}\}$, and consider the
sub-histories $A=\left\{  0,0\text{ }|\text{ }0,\mathbf{1}_{n+1}\right\}  $
and $B=\left\{  0,0\text{ }|\text{ }\mathbf{1}_{n},0,1\right\}  $, where
$\mathbf{1}_{n}$ represents a sequence of $n$ consecutive $1^{\prime}$s. The
corresponding histories of durations $\{d_{t}:$ $t=1,...,n+2\}$ are: for $A$,
$\left\{  0,0,1,...,n\right\}  $; and for $B$, $\left\{  0,1,...,n,0\right\}
$. It is clear that the histogram of durations is the same under the two
histories: $H_{A}^{(1)}(d)=H_{B}^{(1)}(d)=1$ for any $1\leq d\leq n$, and
$H_{A}^{(1)}(d)=H_{B}^{(1)}(d)=0$ for $d\geq n+1$. Also, $\Delta_{A}%
^{(1)}=y_{n+2,A}-y_{0,A}=1$ and $\Delta_{B}^{(1)}=y_{n+2,B}-y_{0,B}=1$.
Therefore, we conclude that $U(A)=U(B)$. For the statistics associated to the
structural parameters: $d_{1,A}=0$ and $d_{n+3,A}=n+1$, such that $\Delta
_{A}^{(1)}(n+1)=1$ and $\Delta_{A}^{(1)}(d)=0$ for any $d\neq n+1$;
$d_{1,B}=0$ and $d_{n+3,B}=1$, such that $\Delta_{B}^{(1)}(d)=0$ for any
$d\geq2$. Therefore, $\ln\mathbb{P}\left(  A|U\right)  -\ln\mathbb{P}\left(
B|U\right)  =$ $[\Delta_{A}^{(1)}(n+1)-\Delta_{B}^{(1)}(n+1)]$ $\gamma(n)=$
$\gamma(n)$, and this structural parameter is identified.\qquad$\blacksquare$

\medskip

For this model, the vector of sufficient statistics include the histogram of
durations, $\{H^{(1)}(d):d\geq1\}$. Conditional on these statistics, the
identification of the structural parameter $\gamma(d)$ comes from the
difference between the final and the initial value of duration, $\Delta
^{(1)}(d+1)=1\{d_{T+1}=d+1\}-$ $1\{d_{1}=d+1\}$. The identification result in
Proposition 3 for the myopic model with duration dependence does not depend on
Assumption 2.

In this binary choice model, the parameters $\widetilde{\beta}_{y}$ and
$\beta_{d}(1,n)$ cannot be separately identified. However, given the
parameters $\{\gamma(d):1\leq d\leq T-2\}$, we can identify the marginal
returns to experience $\beta_{d}(1,d)-\beta_{d}(1,d-1)$ as $\gamma
(d)-\gamma(d-1)$ for any value $d$ between $2$ and $T-2$.\footnote{In this
binary choice model with both switching costs and duration dependence, it is
not possible to separately identify the switching cost parameter
$\widetilde{\beta}_{y}$ and the level of the return to experience $\beta
_{d}(1,d)$. This result resembles the under-identification of the
autoregressive of the order two model studied by Chamberlain (1985). In that
model, we have $y_{it}=1\{\widetilde{\alpha}_{i}+\beta_{1}$ $y_{i,t-1}%
+\beta_{2}$ $y_{i,t-2}+\widetilde{\varepsilon}_{it}\geq0\}$. Chamberlain shows
that the parameter $\beta_{2}$ is identified but the parameter $\beta_{1}$ is
not.}

\subsubsection{Forward-looking dynamic model with duration dependence}

Now, the optimal decision rule includes the difference of continuation values
$v_{\mathbf{\theta}}\left(  1,d_{t}+1\right)  -v_{\mathbf{\theta}}\left(
0\right)  $. Therefore, the model is:%
\begin{equation}
y_{t}=1\left\{  \widetilde{\alpha}_{\mathbf{\theta}}+\widetilde{\beta}%
_{y}\text{ }y_{t-1}+\beta_{d}(1,d_{t})\text{ }y_{t-1}+v_{\mathbf{\theta}%
}\left(  1,d_{t}+1\right)  +\widetilde{\varepsilon}_{t}\geq0\right\}
\label{binary forward and duration model}%
\end{equation}
where now $\widetilde{\alpha}_{\mathbf{\theta}}\equiv\alpha_{\mathbf{\theta}%
}(1)-\alpha_{\mathbf{\theta}}(0)+\beta_{y}(1,0)-v_{\mathbf{\theta}}\left(
0\right)  $. For this model, the log-probability of the choice history
$\widetilde{\mathbf{y}}$ conditional on $(y_{0},d_{1},\mathbf{\theta})$ is:%
\begin{equation}%
\begin{array}
[c]{ccl}%
\ln\mathbb{P}\left(  \widetilde{\mathbf{y}}\text{ }|\text{ }y_{0}%
,d_{1},\mathbf{\theta}\right)  & = & \dsum\limits_{t=1}^{T}y_{t}\left[
\widetilde{\alpha}_{\mathbf{\theta}}+\widetilde{\beta}_{y}y_{t-1}+\beta
_{d}(1,d_{t})y_{t-1}+v_{\mathbf{\theta}}\left(  1,d_{t}+1\right)  \right]
+\sigma_{\mathbf{\theta}}(y_{t-1},d_{t})
\end{array}
\label{binary foward duration log prob}%
\end{equation}
with $\sigma_{\mathbf{\theta}}(y_{t-1},d_{t})\equiv$ $-\ln(1+$ $\exp
\{\widetilde{\alpha}_{\mathbf{\theta}}+$ $\widetilde{\beta}_{y}y_{t-1}+$
$\beta_{d}(1,d_{t})y_{t-1}+$ $v_{\mathbf{\theta}}\left(  1,d_{t}+1\right)
\})$. Comparing equation (\ref{binary foward duration log prob}) with
(\ref{log prob for binary myopic duration}) we can see the forward looking
model has the additional term $\sum_{t=1}^{T}y_{t}$ $v_{\mathbf{\theta}%
}\left(  1,d_{t}+1\right)  $.

Proposition 4 establishes that under Assumption 1 (and without Assumption 2)
there is not identification of any structural parameter.

\medskip

\noindent\textit{PROPOSITION 4. In the forward-looking binary choice model
with duration dependence under Assumption 1, the log-probability of a choice
history has the following form}%
\begin{equation}%
\begin{array}
[c]{ccl}%
\ln\mathbb{P}\left(  \widetilde{\mathbf{y}}|y_{0},d_{1},\mathbf{\theta}\right)
& = & \sum\limits_{d\geq1}H^{(1)}(d)\text{ }g_{\mathbf{\theta},1}%
(d)+\sum\limits_{d\geq1}\Delta^{(1)}(d)\text{ }g_{\mathbf{\theta},2}(d)
\end{array}
\label{Prop 4 log prob}%
\end{equation}
\textit{with }$g_{\mathbf{\theta},1}(d)\equiv\widetilde{\alpha}%
_{\mathbf{\theta}}+\sigma_{\mathbf{\theta}}(1,d)-\sigma_{\mathbf{\theta}%
}(0)+\gamma(d-1)+v_{\mathbf{\theta}}\left(  1,d\right)  $\textit{,
}$g_{\mathbf{\theta},2}(d)\equiv\widetilde{\alpha}_{\mathbf{\theta}%
}+v_{\mathbf{\theta}}\left(  1,d\right)  +\gamma(d-1)$\textit{, }%
$\gamma(d)\equiv\widetilde{\beta}_{y}+\beta_{d}(1,d)$\textit{, and }%
$\gamma(0)=0$\textit{, such that }$S=$ $\{\Delta^{(1)}(d):d\geq1\}$
\textit{and} $U=\{H^{(1)}(d):d\geq1,$ $\Delta^{(1)}(d):d\geq1\}$. \textit{The
minimal sufficient statistic }$U$ \textit{includes the whole vector }%
$S$\textit{, and therefore, the structural parameters} $\gamma(d)$\textit{ are
not identified.}\qquad$\blacksquare$

\medskip

In terms of the minimal sufficient statistic, the difference between this
forward-looking model and its myopic counterpart is that now we need to
control for the difference between final and initial duration, $\Delta
^{(1)}(d+1)$. These additional statistics are also the only statistics
associated with the structural parameter $\gamma(d)$. Therefore, after
controlling for the vector of sufficient statistics $U$, there is not
variation left that can identify structural parameters in this model.

The under-identification result in Proposition 4 applies to the model under
Assumption 1 but without Assumption 2. Under Assumption 2, continuation values
are such that $v_{\mathbf{\theta}}\left(  1,d\right)  =v_{\mathbf{\theta}%
}\left(  1,d^{\ast}\right)  $ for any $d\geq d^{\ast}$. This property provides
identification of some structural parameters. Proposition 5 establishes this result.

\medskip

\noindent\textit{PROPOSITION 5. In the forward-looking binary choice model
with duration dependence under Assumptions 1 and 2, the log-probability of a
choice history has the following form}%
\begin{equation}%
\begin{array}
[c]{ccl}%
\ln\mathbb{P}\left(  \widetilde{\mathbf{y}}|y_{0},\mathbf{\theta}\right)  &
= &
%TCIMACRO{\tsum \limits_{d\leq d^{\ast}-1}}%
%BeginExpansion
{\textstyle\sum\limits_{d\leq d^{\ast}-1}}
%EndExpansion
H^{(1)}(d)\text{ }g_{\mathbf{\theta},1}(d)+\left[
%TCIMACRO{\tsum \limits_{d\geq d^{\ast}}}%
%BeginExpansion
{\textstyle\sum\limits_{d\geq d^{\ast}}}
%EndExpansion
H^{(1)}(d)\right]  g_{\mathbf{\theta},1}(d^{\ast})\\
&  & \\
& + & \sum\limits_{d\leq d^{\ast}-1}\Delta^{(1)}(d)\text{ }g_{\mathbf{\theta
},2}(d)+\left[  \sum\limits_{d\geq d^{\ast}}\Delta^{(1)}(d)\right]
g_{\mathbf{\theta},2}(d^{\ast})\\
&  & \\
& + & \Delta^{(1)}(d^{\ast})\text{ }\left[  \beta_{d}(1,d^{\ast}-1)-\beta
_{d}(1,d^{\ast})\right]
\end{array}
\label{Prop 5 log prob}%
\end{equation}
\textit{with }$g_{\mathbf{\theta},1}(d)\equiv\widetilde{\alpha}%
_{\mathbf{\theta}}+\sigma_{\mathbf{\theta}}(1,d)-\sigma_{\mathbf{\theta}%
}(0)+\gamma(d-1)+v_{\mathbf{\theta}}\left(  1,d\right)  $\textit{, and}
$g_{\mathbf{\theta},2}(d)\equiv\widetilde{\alpha}_{\mathbf{\theta}%
}+v_{\mathbf{\theta}}\left(  1,d\right)  +\gamma(d-1)$. \textit{We have that:
(i) }$U=\{H^{(1)}(d):d\leq d^{\ast}-1,$ $%
%TCIMACRO{\tsum \nolimits_{d\geq d^{\ast}}}%
%BeginExpansion
{\textstyle\sum\nolimits_{d\geq d^{\ast}}}
%EndExpansion
H^{(1)}(d),$ $\Delta^{(1)}(d):d\leq d^{\ast}-1$, $%
%TCIMACRO{\tsum \nolimits_{d\geq d^{\ast}}}%
%BeginExpansion
{\textstyle\sum\nolimits_{d\geq d^{\ast}}}
%EndExpansion
\Delta^{(1)}(d)\}\ $\textit{is a sufficient statistic for }$\mathbf{\theta}%
$\textit{. (ii) The elements in the vector }$U$ \textit{are linearly
independent such that }$U$\textit{ is a minimal sufficient statistic. (iii)
Conditional on }$U$\textit{, the statistic }$\Delta^{(1)}(d^{\ast})$
\textit{has variation and the structural parameter }$\Delta\beta_{d}(d^{\ast
})\equiv$ $\beta_{d}(1,d^{\ast})-\beta_{d}(1,d^{\ast}-1)$ \textit{is
identified, i.e., there is a pair of histories, }$A$ \textit{and }$B$\textit{,
such that }$U(A)=U(B)$ \textit{and}$\Delta\beta_{d}(d^{\ast})=$ $[\ln
\mathbb{P}\left(  A|U\right)  -\ln\mathbb{P}\left(  B|U\right)  ]/[\Delta
_{A}^{(1)}(d^{\ast})-\Delta_{B}^{(1)}(d^{\ast})]$.\qquad$\blacksquare$

\medskip

\noindent\textit{Proof. }The derivation of equation (\ref{Prop 5 log prob}) is
in the Appendix. Proof of (iii). Given a choice history $\{y_{0},d_{1}$ $|$
$\widetilde{\mathbf{y}}\}$ consider the sub-history $\{y_{0},d_{1}$ $|$
$y_{1},y_{2},...,y_{2d^{\ast}+1}\}$. Consider the choice histories $A=\left\{
0,0\text{ }|\text{ }\mathbf{1}_{d^{\ast}-1},0,\mathbf{1}_{d^{\ast}+1}\right\}
$ and $B=\left\{  0,0\text{ }|\text{ }\mathbf{1}_{d^{\ast}},0,\mathbf{1}%
_{d^{\ast}}\right\}  $. The corresponding histories of durations $\{d_{t}:$
$t=1,...,2d^{\ast}+1\}$ are: for $A$, $\{0,$ $1,$ $2,...,$ $d^{\ast}-1,$ $0,$
$1,$ $2,$ $...,d^{\ast}\}$; and for $B$, $\{0,$ $1,$ $2,$ $...,$ $d^{\ast},$
$0,$ $1,$ $2,$ $...,$ $d^{\ast}-1\}$. We verify that $U(A)=U(B)$: (a) for any
$d\leq d^{\ast}-1$, $H_{A}(d)=H_{B}(d)=2$; (b) $\sum_{d\geq d^{\ast}}%
H_{A}(d)=$ $\sum_{d\geq d^{\ast}}H_{B}(d)=1$; (c) for any $d\leq d^{\ast}-1$,
$\Delta_{A}^{(1)}(d)=\Delta_{B}^{(1)}(d)=0$; and (d) $%
%TCIMACRO{\tsum \nolimits_{d\geq d^{\ast}}}%
%BeginExpansion
{\textstyle\sum\nolimits_{d\geq d^{\ast}}}
%EndExpansion
\Delta_{A}^{(1)}(d)=$ $%
%TCIMACRO{\tsum \nolimits_{d\geq d^{\ast}}}%
%BeginExpansion
{\textstyle\sum\nolimits_{d\geq d^{\ast}}}
%EndExpansion
\Delta_{B}^{(1)}(d)=1$. The two histories have different values for the
statistic $\Delta^{(1)}(d^{\ast})$, i.e., $\Delta_{A}^{(1)}(d^{\ast})=0$ and
$\Delta_{B}^{(1)}(d^{\ast})=1$. Therefore, $\ln\mathbb{P}\left(  A|U\right)
-\ln\mathbb{P}\left(  B|U\right)  =$ $[\Delta_{A}^{(1)}(d^{\ast})-\Delta
_{B}^{(1)}(d^{\ast})]$ $[-\Delta\beta_{d}(d^{\ast})]=$ $\Delta\beta
_{d}(d^{\ast})$.\qquad$\blacksquare$

\medskip

In the forward-looking binary choice model with duration dependence, only
$\Delta\beta_{d}(d^{\ast})$ is identified. This result contrasts with the
myopic model where we can identify $\Delta\beta_{d}(d)$ for any duration
$2\leq d\leq T-1$ (Proposition 3).

Table 2 summarizes the identification results for the binary choice model.

\begin{center}%
\begin{tabular}
[c]{lccccc}\hline\hline
\multicolumn{6}{c}{\textbf{Table 2}}\\
\multicolumn{6}{c}{\textbf{Identification of Dynamic Binary Logit Models}%
}\\\hline
&  &  &  &  & \\
\multicolumn{6}{c}{\textbf{Panel 1: Models without duration dependence}%
}\\\hline
\multicolumn{1}{c}{} &  &  & \multicolumn{1}{|c}{} &  & \\
\multicolumn{3}{c}{\textit{Myopic Model}} &
\multicolumn{3}{|c}{\textit{Forward-Looking Model}}\\\hline
\multicolumn{1}{c}{{\small Minimal}} & \multicolumn{1}{|c}{{\small Identified}%
} & \multicolumn{1}{|c}{{\small Identifying}} &
\multicolumn{1}{|c}{{\small Minimal}} &
\multicolumn{1}{|c}{{\small Identified}} &
\multicolumn{1}{|c}{{\small Identifying}}\\
\multicolumn{1}{c}{{\small sufficient stat.}} &
\multicolumn{1}{|c}{{\small parameters}} &
\multicolumn{1}{|c}{{\small statistics}} &
\multicolumn{1}{|c}{{\small sufficient stat.}} &
\multicolumn{1}{|c}{{\small parameters}} &
\multicolumn{1}{|c}{{\small statistics}}\\\hline
\multicolumn{1}{c}{} & \multicolumn{1}{|c}{} & \multicolumn{1}{|c}{} &
\multicolumn{1}{|c}{} & \multicolumn{1}{|c}{} & \multicolumn{1}{|c}{}\\
\multicolumn{1}{c}{${\footnotesize T}^{{\footnotesize (1)}},$
${\footnotesize \Delta}^{{\footnotesize (1)}}$} &
\multicolumn{1}{|c}{$\widetilde{{\footnotesize \beta}}_{{\footnotesize y}}$} &
\multicolumn{1}{|c}{${\footnotesize D}^{{\footnotesize (1,1)}}$} &
\multicolumn{1}{|c}{${\footnotesize T}^{{\footnotesize (1)}},$
${\footnotesize \Delta}^{{\footnotesize (1)}}$} &
\multicolumn{1}{|c}{$\widetilde{{\footnotesize \beta}}_{{\footnotesize y}}$} &
\multicolumn{1}{|c}{${\footnotesize D}^{{\footnotesize (1,1)}}$}\\
\multicolumn{1}{c}{} & \multicolumn{1}{|c}{} & \multicolumn{1}{|c}{} &
\multicolumn{1}{|c}{} & \multicolumn{1}{|c}{} & \multicolumn{1}{|c}{}\\\hline
\multicolumn{1}{c}{} &  &  &  &  & \\
\multicolumn{6}{c}{\textbf{Panel 2: Models with duration dependence}}\\\hline
\multicolumn{1}{c}{} &  &  & \multicolumn{1}{|c}{} &  & \\
\multicolumn{3}{c}{\textit{Myopic Model}} &
\multicolumn{3}{|c}{\textit{Forward-Looking Model}}\\\hline
\multicolumn{1}{c}{{\small Minimal}} & \multicolumn{1}{|c}{{\small Identified}%
} & \multicolumn{1}{|c}{{\small Identifying}} &
\multicolumn{1}{|c}{{\small Minimal}} &
\multicolumn{1}{|c}{{\small Identified}} &
\multicolumn{1}{|c}{{\small Identifying}}\\
\multicolumn{1}{c}{{\small sufficient stat.}} &
\multicolumn{1}{|c}{{\small parameters}} &
\multicolumn{1}{|c}{{\small statistics}} &
\multicolumn{1}{|c}{{\small sufficient stat.}} &
\multicolumn{1}{|c}{{\small parameters}} &
\multicolumn{1}{|c}{{\small statistics}}\\\hline
\multicolumn{1}{c}{} & \multicolumn{1}{|c}{} & \multicolumn{1}{|c}{} &
\multicolumn{1}{|c}{} & \multicolumn{1}{|c}{} & \multicolumn{1}{|c}{}\\
\multicolumn{1}{c}{${\footnotesize \Delta^{(1)},}$} &
\multicolumn{1}{|c}{$\widetilde{{\footnotesize \beta}}_{{\footnotesize y}%
}+{\footnotesize \beta}_{{\footnotesize d}}{\footnotesize (1,d)}$} &
\multicolumn{1}{|c}{${\footnotesize \Delta}^{{\footnotesize (1)}%
}{\footnotesize (d)}$} & \multicolumn{1}{|c}{${\footnotesize H}%
^{{\footnotesize (1)}}{\footnotesize (d):d\leq d}^{{\footnotesize \ast}%
}{\footnotesize -1,}$} & \multicolumn{1}{|c}{$\Delta{\footnotesize \beta
}_{{\footnotesize d}}{\footnotesize (d}^{\ast}{\footnotesize )\equiv}$} &
\multicolumn{1}{|c}{${\footnotesize \Delta}^{{\footnotesize (1)}%
}{\footnotesize (d}^{\ast}{\footnotesize )}$}\\
\multicolumn{1}{c}{${\footnotesize H}^{{\footnotesize (1)}}%
{\footnotesize (d):d\geq1}$} & \multicolumn{1}{|c}{{\footnotesize for
}${\footnotesize d\leq T-2}$} & \multicolumn{1}{|c}{} & \multicolumn{1}{|c}{$%
%TCIMACRO{\tsum \nolimits_{{\footnotesize d\geq d}^{\ast}}}%
%BeginExpansion
{\textstyle\sum\nolimits_{{\footnotesize d\geq d}^{\ast}}}
%EndExpansion
{\footnotesize H}^{{\footnotesize (1)}}{\footnotesize (d),}$} &
\multicolumn{1}{|c}{${\footnotesize \beta}_{{\footnotesize d}}%
{\footnotesize (1,d}^{\ast}{\footnotesize )}$} & \multicolumn{1}{|c}{}\\
\multicolumn{1}{c}{} & \multicolumn{1}{|c}{} & \multicolumn{1}{|c}{} &
\multicolumn{1}{|c}{${\footnotesize \Delta}^{{\footnotesize (1)}%
}{\footnotesize (d):d\leq d}^{{\footnotesize \ast}}{\footnotesize -1,}$} &
\multicolumn{1}{|c}{${\footnotesize -\beta}_{{\footnotesize d}}%
{\footnotesize (1,d}^{{\footnotesize \ast}}{\footnotesize -1)}$} &
\multicolumn{1}{|c}{}\\
\multicolumn{1}{c}{} & \multicolumn{1}{|c}{} & \multicolumn{1}{|c}{} &
\multicolumn{1}{|c}{$%
%TCIMACRO{\tsum \nolimits_{{\footnotesize d\geq d}^{\ast}}}%
%BeginExpansion
{\textstyle\sum\nolimits_{{\footnotesize d\geq d}^{\ast}}}
%EndExpansion
{\footnotesize \Delta}^{{\footnotesize (1)}}{\footnotesize (d)}$} &
\multicolumn{1}{|c}{} & \multicolumn{1}{|c}{}\\
\multicolumn{1}{c}{} & \multicolumn{1}{|c}{} & \multicolumn{1}{|c}{} &
\multicolumn{1}{|c}{} & \multicolumn{1}{|c}{} & \multicolumn{1}{|c}{}%
\\\hline\hline
\end{tabular}

\end{center}

\medskip

\noindent\textit{Identification of }$d^{\ast}$\textit{ in the forward-looking
model. }We have assumed so far that the value of $d^{\ast}$ is known to the
researcher. We now establish the identification of $d^{\ast}$. Let $n$ be any
duration such that $2n+1\leq T$. Consider the pair of histories $A_{n}%
=\left\{  0,0\text{ }|\text{ }\mathbf{1}_{n-1},0,\mathbf{1}_{n+1}\right\}  $
and $B_{n}=\left\{  0,0\text{ }|\text{ }\mathbf{1}_{n},0,\mathbf{1}%
_{n}\right\}  $. We have that:%
\begin{equation}
\left\{
\begin{array}
[c]{rl}%
\text{For }n>d^{\ast}\text{,} & U(A_{n})=U(B_{n})\text{, and }\ln
\mathbb{P}\left(  A_{n}|U\right)  -\ln\mathbb{P}\left(  B_{n}|U\right)
=\Delta\beta_{d}(n)=0\\
& \\
\text{For }n=d^{\ast}\text{,} & U(A_{n})=U(B_{n})\text{, and }\ln
\mathbb{P}\left(  A_{n}|U\right)  -\ln\mathbb{P}\left(  B_{n}|U\right)
=\Delta\beta_{d}(d^{\ast})\neq0\\
& \\
\text{For }n<d^{\ast}\text{,} & U(A_{n})\neq U(B_{n})
\end{array}
\right.  \label{identification d* three cases}%
\end{equation}
Note that $\ln\mathbb{P}\left(  A_{n}|U_{n}\right)  -\ln\mathbb{P}\left(
B_{n}|U_{n}\right)  $ identifies the parameter $\Delta\beta_{d}(n)$ only if
$n\geq d^{\ast}$. Given a dataset with $T$ time periods, we can construct
histories $A_{n}$ and $B_{n}$ only if $2n+1\leq T$. Putting these two
conditions together, the identification of the value of $d^{\ast}$ requires
that $T\geq2d^{\ast}+1$ or equivalently, $d^{\ast}\leq(T-1)/2$. Under this
condition, we can describe the parameter $d^{\ast}$ as the maximum value of
$n$ such that $\ln\mathbb{P}\left(  A_{n}|U_{n}\right)  -\ln\mathbb{P}\left(
B_{n}|U_{n}\right)  \neq0$. This condition uniquely identifies $d^{\ast}$.

\medskip

\noindent\textit{PROPOSITION 6. Consider the forward-looking binary choice
model with duration dependence under Assumptions 1 and 2. For any duration
}$n$ \textit{with }$2n+1\leq T$\textit{, define the pair of histories}
$A_{n}=$ $\left\{  0,0\text{ }|\text{ }\mathbf{1}_{n-1},0,\mathbf{1}%
_{n+1}\right\}  $ \textit{and} $B_{n}=$ $\left\{  0,0\text{ }|\text{
}\mathbf{1}_{n},0,\mathbf{1}_{n}\right\}  $\textit{. Then, if }$d^{\ast}%
\leq(T-1)/2$\textit{, we have that the value of} $d^{\ast}$ \textit{is point
identified as:}%
\begin{equation}
d^{\ast}=\max\left\{  n:\text{\textit{ }}\ln\mathbb{P}\left(  A_{n}%
|U_{n}\right)  -\ln\mathbb{P}\left(  B_{n}|U_{n}\right)  \neq0\right\}
\qquad\blacksquare\label{Prop 4 identification d*}%
\end{equation}

\subsection{Multinomial choice models}

\subsubsection{Multinomial myopic model without duration dependence}

We can represent this model as $y_{t}=$ $\arg\max_{y\in\mathcal{Y}}%
\{\alpha_{\mathbf{\theta}}(y)+\beta_{y}(y,y_{t-1})+\varepsilon_{t}(y)\}$. The
log-probability of the choice history $\widetilde{\mathbf{y}}$ conditional on
$(y_{0},\mathbf{\theta})$ is:%
\begin{equation}%
\begin{array}
[c]{ccl}%
\ln\mathbb{P}\left(  \widetilde{\mathbf{y}}|y_{0},\mathbf{\theta}\right)  &
= & \dsum\limits_{t=1}^{T}\left[  \alpha_{\mathbf{\theta}}(y_{t})+\beta
_{y}(y_{t},y_{t-1})\right]  +\sigma_{\mathbf{\theta}}(y_{t-1})
\end{array}
\label{multinomial myopic no duration}%
\end{equation}
where $\sigma_{\mathbf{\theta}}(y_{t-1})\equiv-\ln\left[
%TCIMACRO{\tsum \nolimits_{y=0}^{J}}%
%BeginExpansion
{\textstyle\sum\nolimits_{y=0}^{J}}
%EndExpansion
\exp\{\alpha_{\mathbf{\theta}}(y)+\beta_{y}(y,y_{t-1})\}\right]  $.
Proposition 7 presents our identification result for this model.

\medskip

\noindent\textit{PROPOSITION 7. In the myopic multinomial model without
duration dependence under Assumption 1, the log-probability has the following
form}%
\begin{equation}%
\begin{array}
[c]{ccl}%
\ln\mathbb{P}\left(  \widetilde{\mathbf{y}}|y_{0},\mathbf{\theta}\right)  &
= &
%TCIMACRO{\tsum \limits_{y=1}^{J}}%
%BeginExpansion
{\textstyle\sum\limits_{y=1}^{J}}
%EndExpansion
T^{(y)}\text{ }g_{\mathbf{\theta},1}(y)+%
%TCIMACRO{\tsum \limits_{y=1}^{J}}%
%BeginExpansion
{\textstyle\sum\limits_{y=1}^{J}}
%EndExpansion
\Delta^{(y)}\text{ }g_{\mathbf{\theta},2}(y)+%
%TCIMACRO{\tsum \limits_{y_{-1}=1}^{J}}%
%BeginExpansion
{\textstyle\sum\limits_{y_{-1}=1}^{J}}
%EndExpansion%
%TCIMACRO{\tsum \limits_{y=1}^{J}}%
%BeginExpansion
{\textstyle\sum\limits_{y=1}^{J}}
%EndExpansion
D^{(y_{-1},y)}\text{ }\widetilde{\beta}_{y}(y,y_{-1})
\end{array}
\label{Pro 7 log prob multinomial myopic nodura}%
\end{equation}
\textit{where }$g_{\mathbf{\theta},1}(y)\equiv\alpha_{\mathbf{\theta}%
}(y)-\alpha_{\mathbf{\theta}}(0)+\sigma_{\mathbf{\theta}}(y)-\sigma
_{\mathbf{\theta}}(0)+\beta_{y}(0,y)+\beta_{y}(y,0)$\textit{,}
$g_{\mathbf{\theta},2}(y)\equiv-\sigma_{\mathbf{\theta}}(y)+\sigma
_{\mathbf{\theta}}(0)-\beta_{y}(0,y)$\textit{, and }$\widetilde{\beta}%
_{y}(y,y_{-1})\equiv\beta_{y}(y,y_{-1})-\beta_{y}(0,y_{-1})-\beta_{y}(y,0)$
\textit{for any} $y,y_{-1}\in\mathcal{Y}$. \textit{Then: (i) }$U=\{T^{(y)}%
:y\geq1,$ $\Delta^{(y)}:y\geq1\}\ $\textit{is a sufficient statistic for
}$\mathbf{\theta}$\textit{. (ii) The elements in the vector }$U$ \textit{are
linearly independent such that }$U$\textit{ is a minimal sufficient statistic.
(iii) Conditional on }$U$\textit{, the vector of statistics }$\{D^{(y_{-1}%
,y)}:y_{-1},y\in\mathcal{Y-\{}0\}\}$ \textit{are linearly independent such
that they can identify the vector of parameters }$\{\widetilde{\beta}%
_{y}(y,y_{-1}):y_{-1},y\in\mathcal{Y-\{}0\}\}$\textit{, i.e., for every pair
of choices }$y_{-1},y\in\mathcal{Y-\{}0\}$\textit{, there is a pair of
histories, }$A$ \textit{and }$B$\textit{, such that }$U(A)=U(B)$ \textit{and
}$\widetilde{\beta}_{y}(y,y_{-1})=$ $[\ln\mathbb{P}\left(  A|U\right)
-\ln\mathbb{P}\left(  B|U\right)  ]/$ $[D_{A}^{(y_{-1},y)}-D_{B}^{(y_{-1}%
,y)}]$.\qquad$\blacksquare$

\medskip

The following example illustrates a pair of histories that identifies
$\widetilde{\beta}_{y}(y,y_{-1})$.

\medskip

\noindent\textit{EXAMPLE 1.} Suppose that $T=3$ and consider the following two
realizations of the history $\left(  y_{0}|\widetilde{\mathbf{y}}\right)  $:
$A=\left\{  0\text{ }|\text{ }0,j,k\right\}  $ and $B=\left\{  0\text{
}|\text{ }j,0,k\right\}  $ with $j,k\neq0$. We first confirm that $U(A)=U(B)$:
$T_{A}^{(j)}=T_{B}^{(j)}=1$, $T_{A}^{(k)}=T_{B}^{(k)}=1$, and $T_{A}%
^{(y)}=T_{B}^{(y)}=0$ for any $y\neq0,j,k$. The identifying statistics
$D^{(y_{-1},y)}$ take the following values: $D_{A}^{(j,k)}-D_{B}^{(j,k)}=1$,
$D_{A}^{(j,0)}-D_{B}^{(j,0)}=-1$, $D_{A}^{(0,k)}-D_{B}^{(0,k)}=-1$, and
$D_{A}^{(y_{-1},y)}-D_{B}^{(y_{-1},y)}=0$ for any other pair $(y_{-1},y)$.
Therefore, we have that $\ln\mathbb{P}\left(  A|U\right)  -\ln\mathbb{P}%
\left(  B|U\right)  =$ $\widetilde{\beta}_{y}(k,j)-\widetilde{\beta}%
_{y}(0,j)-\widetilde{\beta}_{y}(k,0)=$ $\widetilde{\beta}_{y}(k,j)$. A
particular case of this example is when $j=k$, such that $A=\left\{  0\text{
}|\text{ }0,j,j\right\}  $ and $B=\left\{  0\text{ }|\text{ }j,0,j\right\}  $.
In this case, $\ln\mathbb{P}\left(  A|U\right)  -\ln\mathbb{P}\left(
B|U\right)  $ identifies $\widetilde{\beta}_{y}(j,j)$ that is equal to the
sunk cost $-\beta_{y}(0,j)-\beta_{y}(j,0)$.\qquad$\blacksquare$

\medskip

As in the binary choice model, we cannot identify the whole switching cost
function $\beta_{y}$. With $J+1$ choice alternatives, we can identify $J^{2}$
switching cost parameters. However, the structural parameter $\widetilde
{\beta}_{y}(y,y_{-1})\equiv$ $\beta_{y}(y,y_{-1})-\beta_{y}(0,y_{-1}%
)-\beta_{y}(y,0)$ has a clear interpretation: it is the difference in
switching cost between a \textit{direct switch} from $y_{-1}$ to $y$ and an
\textit{indirect switch} via alternative $0$. For this identification result,
there is nothing special with alternative $0$ and we could choose any other
alternative as the baseline. Note also that the set of identified structural
parameters $\widetilde{\beta}_{y}(y,y_{-1})$ includes the \textit{sunk cost of
entry} in market $y$, i.e., for any $y>0$, $\widetilde{\beta}_{y}(y,y)=$
$-\beta_{y}(0,y)-\beta_{y}(y,0)$, because $\beta_{y}(y,y)=0$.

\subsubsection{Multinomial forward-looking model without duration dependence}

The optimal decision rule for this model is $y_{t}=$ $\arg\max_{y\in
\mathcal{Y}}\{\alpha_{\mathbf{\theta}}(y)+v_{\mathbf{\theta}}(y)+\beta
_{y}(y,y_{t-1})+\varepsilon_{t}(y)\}$, where $v_{\mathbf{\theta}}(y)$ is the
continuation value of choosing alternative $y$. The log-probability of the
choice history $\widetilde{\mathbf{y}}$ conditional on $(y_{0},\mathbf{\theta
})$ has a similar form as in the myopic model, but now the incidental
parameter $\mathbf{\theta}$ enters through the function $\alpha
_{\mathbf{\theta}}(y)+v_{\mathbf{\theta}}(y)$.%
\begin{equation}%
\begin{array}
[c]{ccl}%
\ln\mathbb{P}\left(  \widetilde{\mathbf{y}}|y_{0},\mathbf{\theta}\right)  &
= & \dsum\limits_{t=1}^{T}\left[  \alpha_{\mathbf{\theta}}(y_{t}%
)+v_{\mathbf{\theta}}(y_{t})+\beta_{y}(y_{t},y_{t-1})\right]  +\sigma
_{\mathbf{\theta}}(y_{t-1})
\end{array}
\label{log prob multinomial forward no duration}%
\end{equation}
Therefore, the identification of the structural parameters is the same as in
the myopic model without duration dependence.

\medskip

\noindent\textit{PROPOSITION 8. In the multinomial forward-looking model
without duration dependence under Assumption 1, the log-probability of a
choice history has the following form}%
\begin{equation}%
\begin{array}
[c]{ccl}%
\ln\mathbb{P}\left(  \widetilde{\mathbf{y}}|y_{0},\mathbf{\theta}\right)  &
= &
%TCIMACRO{\tsum \limits_{y=1}^{J}}%
%BeginExpansion
{\textstyle\sum\limits_{y=1}^{J}}
%EndExpansion
T^{(y)}\text{ }g_{\mathbf{\theta},1}(y)+%
%TCIMACRO{\tsum \limits_{y=1}^{J}}%
%BeginExpansion
{\textstyle\sum\limits_{y=1}^{J}}
%EndExpansion
\Delta^{(y)}\text{ }g_{\mathbf{\theta},2}(y)+%
%TCIMACRO{\tsum \limits_{y_{-1}=1}^{J}}%
%BeginExpansion
{\textstyle\sum\limits_{y_{-1}=1}^{J}}
%EndExpansion%
%TCIMACRO{\tsum \limits_{y=1}^{J}}%
%BeginExpansion
{\textstyle\sum\limits_{y=1}^{J}}
%EndExpansion
D^{(y_{-1},y)}\text{ }\widetilde{\beta}_{y}(y,y_{-1})
\end{array}
\label{Pro 8 log prob multinomial forward nodura}%
\end{equation}
\textit{where }$g_{\mathbf{\theta},1}(y)\equiv\alpha_{\mathbf{\theta}%
}(y)-\alpha_{\mathbf{\theta}}(0)+v_{\mathbf{\theta}}(y)-v_{\mathbf{\theta}%
}(0)+\sigma_{\mathbf{\theta}}(y)-\sigma_{\mathbf{\theta}}(0)+\beta
_{y}(0,y)+\beta_{y}(y,0)$\textit{,} \textit{and} $g_{\mathbf{\theta}%
,2}(y)\equiv\sigma_{\mathbf{\theta}}(y)-\sigma_{\mathbf{\theta}}(0)-\beta
_{y}(0,y)$\textit{.} \textit{Then: (i) }$U=\{T^{(y)}:y\geq1,$ $\Delta
^{(y)}:y\geq1\}\ $\textit{is a sufficient statistic for }$\mathbf{\theta}%
$\textit{. (ii) The elements in the vector }$U$ \textit{are linearly
independent such that }$U$\textit{ is a minimal sufficient statistic. (iii)
Conditional on }$U$\textit{, the vector of statistics }$\{D^{(y_{-1}%
,y)}:y_{-1},y\in\mathcal{Y-\{}0\}\}$ \textit{are linearly independent such
that they can identify the vector of parameters }$\{\widetilde{\beta}%
_{y}(y,y_{-1}):y_{-1},y\in\mathcal{Y-\{}0\}\}$\textit{.}\qquad$\blacksquare$

\subsubsection{Multinomial myopic model with duration dependence}

The model is $y_{t}=$ $\arg\max_{y\in\mathcal{Y}}\{\alpha_{\mathbf{\theta}%
}(y)+1\{y\neq y_{t-1}\}$ $\beta_{y}(y,y_{t-1})+1\{y=y_{t-1}\}$ $\beta
_{d}(y,d_{t})+\varepsilon_{t}(y)\}$, and the log-probability of a choice
history $\widetilde{\mathbf{y}}$ conditional on $(y_{0},d_{1},\mathbf{\theta
})$ is:%
\begin{equation}%
\begin{array}
[c]{ccl}%
\ln\mathbb{P}\left(  \widetilde{\mathbf{y}}|y_{0},d_{1},\mathbf{\theta}\right)
& = &
%TCIMACRO{\tsum \limits_{t=1}^{T}}%
%BeginExpansion
{\textstyle\sum\limits_{t=1}^{T}}
%EndExpansion
\left[  \alpha_{\mathbf{\theta}}(y_{t})+1\{y_{t}\neq y_{t-1}\}\beta_{y}%
(y_{t},y_{t-1})+1\{y_{t}=y_{t-1}\}\beta_{d}(y_{t},d_{t})\right]
+\sigma_{\mathbf{\theta}}(y_{t-1})
\end{array}
\label{multinomial myopic duration: logprob sums}%
\end{equation}
Proposition 9 presents identification results for the structural parameters
$\beta_{y}$ and $\beta_{d}$.

\medskip

\noindent\textit{PROPOSITION 9. In the multinomial myopic model with duration
dependence under Assumption 1, the log-probability of a choice history}
\textit{has the form}%
\begin{equation}%
\begin{array}
[c]{ccl}%
\ln\mathbb{P}\left(  \widetilde{\mathbf{y}}|y_{0},d_{1}\right)  & = &
{\sum\limits_{y=1}^{J}}{\sum\limits_{d\geq1}}H^{(y)}(d)\text{ }%
g_{\mathbf{\theta},1}(y,d)+{\sum\limits_{y=1}^{J}}\Delta^{(y)}\text{
}g_{\mathbf{\theta},2}(y)\\
&  & \\
& + &
%TCIMACRO{\tsum \limits_{y_{-1}=1}^{J}}%
%BeginExpansion
{\textstyle\sum\limits_{y_{-1}=1}^{J}}
%EndExpansion%
%TCIMACRO{\tsum \limits_{y=1,y\neq y_{-1}}^{J}}%
%BeginExpansion
{\textstyle\sum\limits_{y=1,y\neq y_{-1}}^{J}}
%EndExpansion
D^{(y_{-1},y)}\text{ }\widetilde{\beta}_{y}(y,y_{-1})+{\sum\limits_{y=1}^{J}%
}{\sum\limits_{d\geq1}}\Delta^{(y)}(d)\text{ }\gamma(y,d-1)
\end{array}
\label{multinomial Prop 9 logprob myopic duration}%
\end{equation}
\textit{with} $g_{\mathbf{\theta},1}(y,d)\equiv$ $\alpha_{\mathbf{\theta}%
}(y)-\alpha_{\mathbf{\theta}}(0)+\sigma_{\theta}(y,d)-\sigma_{\mathbf{\theta}%
}(0)+\beta_{y}(0,y)+\beta_{y}(y,0)+\gamma(y,d-1)$\textit{, }$g_{\mathbf{\theta
},2}(y)\equiv$ $\alpha_{\mathbf{\theta}}(y)-\alpha_{\mathbf{\theta}}%
(0)+\beta_{y}(y,0)$\textit{, }$\widetilde{\beta}_{y}(y,y_{-1})\equiv\beta
_{y}(y,y_{-1})-\beta_{y}(0,y_{-1})-\beta_{y}(y,0)$\textit{, and }%
$\gamma(y,d)\equiv\beta_{d}(y,d)-\beta_{y}(y,0)-\beta_{y}(0,y)$. \textit{Then:
(i) }$U=$ $\{H^{(y)}(d):y\geq1,$ $d\geq1,$ $\Delta^{(y)}:y\geq1\}$ \textit{is
a sufficient statistic of }$\mathbf{\theta}$\textit{. (ii) The elements in the
vector }$U$ \textit{are linearly independent such that }$U$\textit{ is a
minimal sufficient statistic. (iii) Conditional on }$U$\textit{, the vector of
statistics }$\{D^{(y_{-1},y)}:y_{-1},$ $y\geq1$; $\Delta^{(y)}(d):y\geq1$,
$d\geq1\}$ \textit{are linearly independent and they identify the vectors of
structural parameters }$\{\widetilde{\beta}_{y}(y,y_{-1}):$ $y_{-1},$ $y\geq
1$, $y\neq y_{-1}$; $\gamma(y,d):$ $y\geq1$, $d\geq1\}$\textit{.}%
\qquad$\blacksquare$

\medskip

The following examples present choice histories that identify structural
parameters $\widetilde{\beta}_{y}(y,y_{-1})$ and $\gamma(y,d)$ according to
Proposition 9.

\medskip

\noindent\textit{EXAMPLE 2. }Suppose that $T=3$ and consider two realizations
of the history $\left(  y_{0},d_{1}|\widetilde{\mathbf{y}}\right)  $: for
$j\neq k$, $A=\left\{  0,0\text{ }|\text{ }0,j,k\right\}  $ and $B=\left\{
0,0\text{ }|\text{ }j,0,k\right\}  $. It is straightforward to verify that
$U(A)=U(B)$ and that $\ln\mathbb{P}\left(  A|U\right)  -\ln\mathbb{P}\left(
B|U\right)  =\beta_{y}(k,j)-\beta_{y}(k,0)-\beta_{y}(0,j)=\widetilde{\beta
}_{y}(k,j)$.\qquad$\blacksquare$

\medskip

\noindent\textit{EXAMPLE 3. }Given an arbitrary positive integer $n$, consider
the pair of choice histories $\left(  y_{0},d_{1}|\widetilde{\mathbf{y}%
}\right)  $ with $T=n+2$: $A=\left\{  0,0\text{ }|\text{ }0,\mathbf{y}%
_{n+1}\right\}  $ and $B=\left\{  0,0\text{ }|\text{ }\mathbf{y}%
_{n},0,y\right\}  $, where $\mathbf{y}_{n}$ represents a vector of dimension
$n$ with all its elements equal to $y$. It is simple to verify that
$U(A)=U(B)$ (i.e., same values for $H^{(y)}(d)$ and $\Delta^{(y)}$).
Furthermore, $\Delta_{A}^{(y)}(n+1)=1$ and $\Delta_{B}^{(y)}(n+1)=0$, such
that we have $\ln\mathbb{P}\left(  A|U\right)  -$ $\ln\mathbb{P}\left(
B|U\right)  =\gamma(y,n)$.\qquad$\blacksquare$

\subsubsection{Multinomial forward-looking model with duration dependence}

The model is $y_{t}=$ $\arg\max_{y\in\mathcal{Y}}\{\alpha_{\mathbf{\theta}%
}(y)+\beta_{y}(y,y_{t-1})+1\{y=y_{t-1}\}\beta_{d}(y,d_{t})+v_{\mathbf{\theta}%
}(y,d_{t+1}[y,y_{t-1},d_{t}])+\varepsilon_{t}(y)\}$, where $d_{t+1}%
[y,y_{t-1},d_{t}]=0$ if $y=0$, and $d_{t+1}[y,y_{t-1},d_{t}]=1\{y=y_{t-1}%
\}d_{t}+1$ if $y\neq0$. In contrast to the binary choice model, in the
multinomial choice model it is possible to identify switching cost parameters
without imposing Assumption 2. Proposition 10 establishes the identification
of switching costs parameters under Assumption 1.

\medskip

\noindent\textit{PROPOSITION 10. In the multinomial forward-looking model with
duration dependence under Assumption 1, the log-probability a choice history}
\textit{has the form}%
\begin{equation}%
\begin{array}
[c]{ccl}%
\ln\mathbb{P}\left(  \widetilde{\mathbf{y}}|y_{0},d_{1}\right)  & = &
{\sum\limits_{y=1}^{J}\sum\limits_{d\geq1}}H^{(y)}(d)\text{ }g_{\mathbf{\theta
},1}(y,d)+{\sum\limits_{y=1}^{J}}{\sum\limits_{d\geq1}}\Delta^{(y)}(d)\text{
}g_{\mathbf{\theta},2}(y,d)\\
&  & \\
& + &
%TCIMACRO{\tsum \limits_{y_{-1}=1}^{J}}%
%BeginExpansion
{\textstyle\sum\limits_{y_{-1}=1}^{J}}
%EndExpansion%
%TCIMACRO{\tsum \limits_{y\neq y_{-1}}}%
%BeginExpansion
{\textstyle\sum\limits_{y\neq y_{-1}}}
%EndExpansion
D^{(y_{-1},y)}\text{ }\widetilde{\beta}_{y}(y,y_{-1})
\end{array}
\label{Prop 10 identification beta_y but not beta_d}%
\end{equation}
\textit{with} $g_{\mathbf{\theta},1}(y,d)\equiv$ $\alpha_{\mathbf{\theta}%
}(y)-\alpha_{\mathbf{\theta}}(0)+\sigma_{\theta}(y,d)-\sigma_{\mathbf{\theta}%
}(0)+\beta_{y}(0,y)+\beta_{y}(y,0)+v_{\mathbf{\theta}}\left(  y,d\right)
-v_{\mathbf{\theta}}\left(  0\right)  +\gamma(y,d-1)$\textit{, and
}$g_{\mathbf{\theta},2}(y,d)\equiv$ $\alpha_{\mathbf{\theta}}(y)-\alpha
_{\mathbf{\theta}}(0)+\beta_{y}(0,y)+v_{\mathbf{\theta}}\left(  y,d\right)
-v_{\mathbf{\theta}}\left(  0\right)  +\gamma(y,d-1)$\textit{. Then: (i) }$U=$
$\{H^{(y)}(d):y\geq1,$ $d\geq1,$ $\Delta^{(y)}(d):y\geq1$, $d\geq1\}$
\textit{is a sufficient statistic of }$\mathbf{\theta}$\textit{. (ii) The
elements in the vector }$U$ \textit{are linearly independent such that }%
$U$\textit{ is a minimal sufficient statistic. (iii) Conditional on }%
$U$\textit{, the vector of statistics }$\{D^{(y_{-1},y)}:y_{-1},$ $y\geq1\}$
\textit{are linearly independent and they identify the vectors of structural
parameters }$\{\widetilde{\beta}_{y}(y,y_{-1}):$ $y_{-1},$ $y\geq1,y\neq
y_{-1}\}$\textit{. The duration dependence parameters }$\gamma(y,d)$
\textit{are not identified.}\qquad$\blacksquare$

\medskip

For instance, the pair of choice histories in Example 2, $A=\left\{  0,0\text{
}|\text{ }0,j,k\right\}  $ and $B=\left\{  0,0\text{ }|\text{ }j,0,k\right\}
$, have the same continuation values. In this forward-looking model, it is
simple to very that these histories satisfy the conditions in Proposition 10
such that $U(A)=U(B)$ and $\ln\mathbb{P}\left(  A|U\right)  -\ln
\mathbb{P}\left(  B|U\right)  =\widetilde{\beta}_{y}(k,j)$.

For the identification of duration dependence parameters, we impose the
restriction in Assumption 2. Proposition 11 presents this identification result.

\medskip

\noindent\textit{PROPOSITION 11. In the multinomial forward-looking model with
duration dependence under Assumptions 1 and 2, the log-probability a choice
history} \textit{has the form}%
\begin{equation}%
\begin{array}
[c]{ccl}%
\ln\mathbb{P}\left(  \widetilde{\mathbf{y}}|y_{0},d_{1}\right)  & = &
{\sum\limits_{y=1}^{J}\sum\limits_{d\leq d_{y}^{\ast}-1}}H^{(y)}(d)\text{
}g_{\mathbf{\theta},1}(y,d)+\left[  {\sum\limits_{y=1}^{J}}%
%TCIMACRO{\tsum \limits_{d\geq d_{y}^{\ast}}}%
%BeginExpansion
{\textstyle\sum\limits_{d\geq d_{y}^{\ast}}}
%EndExpansion
H^{(y)}(d)\right]  g_{\mathbf{\theta},1}(y,d_{y}^{\ast})\\
&  & \\
&  & {\sum\limits_{y=1}^{J}\sum\limits_{d\leq d_{y}^{\ast}-1}}\Delta
^{(y)}(d)\text{ }g_{\mathbf{\theta},2}(y,d)+\left[  {\sum\limits_{y=1}^{J}}%
%TCIMACRO{\tsum \limits_{d\geq d_{y}^{\ast}}}%
%BeginExpansion
{\textstyle\sum\limits_{d\geq d_{y}^{\ast}}}
%EndExpansion
\Delta^{(y)}(d)\right]  g_{\mathbf{\theta},2}(y,d_{y}^{\ast})\\
&  & \\
& + &
%TCIMACRO{\tsum \limits_{y_{-1}=1}^{J}}%
%BeginExpansion
{\textstyle\sum\limits_{y_{-1}=1}^{J}}
%EndExpansion%
%TCIMACRO{\tsum \limits_{y=1,y\neq y_{-1}}^{J}}%
%BeginExpansion
{\textstyle\sum\limits_{y=1,y\neq y_{-1}}^{J}}
%EndExpansion
D^{(y_{-1},y)}\text{ }\widetilde{\beta}_{y}(y,y_{-1})-%
%TCIMACRO{\tsum \limits_{y=1}^{J}}%
%BeginExpansion
{\textstyle\sum\limits_{y=1}^{J}}
%EndExpansion
\Delta^{(y)}(d_{y}^{\ast})\text{ }\Delta\beta_{d}(y,d_{y}^{\ast})
\end{array}
\label{Prop 11 identification beta_y and beta_d}%
\end{equation}
\textit{with} $g_{\mathbf{\theta},1}(y,d)\equiv$ $\alpha_{\mathbf{\theta}%
}(y)-\alpha_{\mathbf{\theta}}(0)+\sigma_{\theta}(y,d)-\sigma_{\mathbf{\theta}%
}(0)+\beta_{y}(0,y)+\beta_{y}(y,0)+v_{\mathbf{\theta}}\left(  y,d\right)
-v_{\mathbf{\theta}}\left(  0\right)  +\gamma(y,d-1)$\textit{, and
}$g_{\mathbf{\theta},2}(y,d)\equiv$ $\alpha_{\mathbf{\theta}}(y)-\alpha
_{\mathbf{\theta}}(0)+\beta_{y}(y,0)+v_{\mathbf{\theta}}\left(  y,d\right)
-v_{\mathbf{\theta}}\left(  0\right)  +\gamma(y,d-1)$\textit{, and }%
$\Delta\beta_{d}(y,d_{y}^{\ast})\equiv\beta_{d}(y,d_{y}^{\ast})-\beta
_{d}(y,d_{y}^{\ast}-1)$. \textit{(i) }$U=\{H^{(y)}(d):y\geq1,$ $d\leq
d_{y}^{\ast}-1,$ $%
%TCIMACRO{\tsum \nolimits_{d\geq d_{y}^{\ast}}}%
%BeginExpansion
{\textstyle\sum\nolimits_{d\geq d_{y}^{\ast}}}
%EndExpansion
H^{(y)}(d),$ $\Delta^{(y)}(d):y\geq1,$ $d\leq d_{y}^{\ast}-1$, $%
%TCIMACRO{\tsum \nolimits_{d\geq d_{y}^{\ast}}}%
%BeginExpansion
{\textstyle\sum\nolimits_{d\geq d_{y}^{\ast}}}
%EndExpansion
\Delta^{(y)}(d)\}\ $\textit{is a sufficient statistic of }$\mathbf{\theta}%
$\textit{. (ii) The elements in the vector }$U$ \textit{are linearly
independent such that }$U$\textit{ is a minimal sufficient statistic. (iii)
Conditional on }$U$\textit{, the vector of statistics }$\{D^{(y_{-1}%
,y)}:y_{-1},$ $y\geq1\}$ \textit{are linearly independent and they identify
the vector of structural parameters }$\{\widetilde{\beta}_{y}(y,y_{-1}):$
$y_{-1},$ $y\geq1,y\neq y_{-1}\}$\textit{. Furthermore, the vector of
statistics }$\{\Delta^{(y)}(d^{\ast}):y\geq1\}$ \textit{are also linearly
independent and they identify the vector of structural parameters }%
$\{\Delta\beta_{d}(y,d_{y}^{\ast}):$ $y\geq1\}$\textit{.}\qquad$\blacksquare$

\medskip

\noindent\textit{EXAMPLE 4. }Given $y\geq1$ with $d_{y}^{\ast}\geq2$, consider
the pair of choice histories $A=\left\{  0,0\text{ }|\text{ }\mathbf{y}%
_{d_{y}^{\ast}-1},0,\mathbf{y}_{d_{y}^{\ast}+1}\right\}  $ and $B=\left\{
0,0\text{ }|\text{ }\mathbf{y}_{d_{y}^{\ast}},0,\mathbf{y}_{d_{y}^{\ast}%
}\right\}  $. The two choice histories have the same statistics $H^{(y)}(d)$
for all $1\leq d\leq d_{y}^{\ast}-1$ and $\sum_{d\geq d_{y}^{\ast}}H^{(y)}%
(d)$, and $\min\{d_{1},d_{y}^{\ast}\}$ and $\min\{d_{T+1},d_{y}^{\ast}\}$
agrees between $A$ and $B$. Therefore, we have that $U(A)=U(B)$. It is
straightforward to show that $\Delta_{A}^{(y)}(d_{y}^{\ast})=0$ and
$\Delta_{B}^{(y)}(d_{y}^{\ast})=1$, and this implies that $\ln\mathbb{P}%
\left(  A|U\right)  -$ $\ln\mathbb{P}\left(  B|U\right)  =\Delta\beta
_{d}(y,d_{y}^{\ast})$.\qquad$\blacksquare$

\newpage

Table 3 summarizes the identification results for the multinomial model.

\begin{center}%
\begin{tabular}
[c]{lccccc}\hline\hline
\multicolumn{6}{c}{\textbf{Table 3}}\\
\multicolumn{6}{c}{\textbf{Identification of Dynamic Multinomial Logit
Models}}\\\hline
&  &  &  &  & \\
\multicolumn{6}{c}{\textbf{Panel 1: Models without duration dependence}%
}\\\hline
\multicolumn{1}{c}{} &  &  & \multicolumn{1}{|c}{} &  & \\
\multicolumn{3}{c}{\textit{Myopic Model}} &
\multicolumn{3}{|c}{\textit{Forward-Looking Model}}\\\hline
\multicolumn{1}{c}{{\small Minimal}} & \multicolumn{1}{|c}{{\small Identified}%
} & \multicolumn{1}{|c}{{\small Identifying}} &
\multicolumn{1}{|c}{{\small Minimal}} &
\multicolumn{1}{|c}{{\small Identified}} &
\multicolumn{1}{|c}{{\small Identifying}}\\
\multicolumn{1}{c}{{\small sufficient stat.}} &
\multicolumn{1}{|c}{{\small parameters}} &
\multicolumn{1}{|c}{{\small statistics}} &
\multicolumn{1}{|c}{{\small sufficient stat.}} &
\multicolumn{1}{|c}{{\small parameters}} &
\multicolumn{1}{|c}{{\small statistics}}\\\hline
\multicolumn{1}{c}{} & \multicolumn{1}{|c}{} & \multicolumn{1}{|c}{} &
\multicolumn{1}{|c}{} & \multicolumn{1}{|c}{} & \multicolumn{1}{|c}{}\\
\multicolumn{1}{c}{${\footnotesize T}^{{\footnotesize (y)}},$
${\footnotesize \Delta}^{{\footnotesize (y)}}{\footnotesize :y\geq1}$} &
\multicolumn{1}{|c}{$\widetilde{{\footnotesize \beta}}_{{\footnotesize y}%
}{\footnotesize (y,y}_{{\footnotesize -1}}{\footnotesize )}$} &
\multicolumn{1}{|c}{${\footnotesize D}^{(y_{{\footnotesize -1}},y)}%
{\footnotesize :}$} & \multicolumn{1}{|c}{${\footnotesize T}%
^{{\footnotesize (y)}},$ ${\footnotesize \Delta}^{{\footnotesize (y)}%
}{\footnotesize :y\geq1}$} & \multicolumn{1}{|c}{$\widetilde
{{\footnotesize \beta}}_{{\footnotesize y}}{\footnotesize (y,y}%
_{{\footnotesize -1}}{\footnotesize )}$} &
\multicolumn{1}{|c}{${\footnotesize D}^{(y_{{\footnotesize -1}},y)}$}\\
& \multicolumn{1}{|c}{${\footnotesize y}_{{\footnotesize -1}}%
{\footnotesize ,y\geq1}$} & \multicolumn{1}{|c}{${\footnotesize y}%
_{{\footnotesize -1}}{\footnotesize ,y\geq1}$} & \multicolumn{1}{|c}{} &
\multicolumn{1}{|c}{${\footnotesize y}_{{\footnotesize -1}}%
{\footnotesize ,y\geq1}$} & \multicolumn{1}{|c}{${\footnotesize y}%
_{{\footnotesize -1}}{\footnotesize ,y\geq1}$}\\
\multicolumn{1}{c}{} & \multicolumn{1}{|c}{} & \multicolumn{1}{|c}{} &
\multicolumn{1}{|c}{} & \multicolumn{1}{|c}{} & \multicolumn{1}{|c}{}\\\hline
\multicolumn{1}{c}{} &  &  &  &  & \\
\multicolumn{6}{c}{\textbf{Panel 2: Models with duration dependence}}\\\hline
\multicolumn{1}{c}{} &  &  & \multicolumn{1}{|c}{} &  & \\
\multicolumn{3}{c}{\textit{Myopic Model}} &
\multicolumn{3}{|c}{\textit{Forward-Looking Model}}\\\hline
\multicolumn{1}{c}{{\small Minimal}} & \multicolumn{1}{|c}{{\small Identified}%
} & \multicolumn{1}{|c}{{\small Identifying}} &
\multicolumn{1}{|c}{{\small Minimal}} &
\multicolumn{1}{|c}{{\small Identified}} &
\multicolumn{1}{|c}{{\small Identifying}}\\
\multicolumn{1}{c}{{\small sufficient stat.}} &
\multicolumn{1}{|c}{{\small parameters}} &
\multicolumn{1}{|c}{{\small statistics}} &
\multicolumn{1}{|c}{{\small sufficient stat.}} &
\multicolumn{1}{|c}{{\small parameters}} &
\multicolumn{1}{|c}{{\small statistics}}\\\hline
\multicolumn{1}{c}{} & \multicolumn{1}{|c}{} & \multicolumn{1}{|c}{} &
\multicolumn{1}{|c}{} & \multicolumn{1}{|c}{} & \multicolumn{1}{|c}{}\\
\multicolumn{1}{c}{${\footnotesize \Delta}^{{\footnotesize (y)}}%
{\footnotesize :y\geq1,}$} & \multicolumn{1}{|c}{$\widetilde
{{\footnotesize \beta}}_{{\footnotesize y}}{\footnotesize (y,y}%
_{{\footnotesize -1}}{\footnotesize ):}$} &
\multicolumn{1}{|c}{${\footnotesize D}^{(y_{{\footnotesize -1}},y)}%
{\footnotesize :}$} & \multicolumn{1}{|c}{${\footnotesize H}%
^{{\footnotesize (y)}}{\footnotesize (d):}$} & \multicolumn{1}{|c}{$\widetilde
{{\footnotesize \beta}}_{{\footnotesize y}}{\footnotesize (y,y}%
_{{\footnotesize -1}}{\footnotesize ):}$} &
\multicolumn{1}{|c}{${\footnotesize D}^{(y_{{\footnotesize -1}},y)}$}\\
\multicolumn{1}{c}{${\footnotesize H}^{{\footnotesize (y)}}%
{\footnotesize (d):}$} & \multicolumn{1}{|c}{${\footnotesize y}%
_{{\footnotesize -1}}{\footnotesize ,y\geq1}$} &
\multicolumn{1}{|c}{${\footnotesize y}_{{\footnotesize -1}}%
{\footnotesize ,y\geq1}$} & \multicolumn{1}{|c}{${\footnotesize y\geq1,d\leq
d_{y}^{\ast}-1;}$} & \multicolumn{1}{|c}{${\footnotesize y}%
_{{\footnotesize -1}}{\footnotesize \neq y\geq1}$} &
\multicolumn{1}{|c}{${\footnotesize y}_{{\footnotesize -1}}{\footnotesize \neq
y\geq1}$}\\
\multicolumn{1}{c}{${\footnotesize y\geq1,d\geq1}$} &
\multicolumn{1}{|c}{{\footnotesize and}} &
\multicolumn{1}{|c}{{\footnotesize and}} & \multicolumn{1}{|c}{$%
%TCIMACRO{\tsum \limits_{{\footnotesize d\geq d}^{\ast}}}%
%BeginExpansion
{\textstyle\sum\limits_{{\footnotesize d\geq d}^{\ast}}}
%EndExpansion
{\footnotesize H}^{{\footnotesize (y)}}{\footnotesize (d):y\geq1;}$} &
\multicolumn{1}{|c}{{\footnotesize and}} &
\multicolumn{1}{|c}{{\footnotesize and}}\\
\multicolumn{1}{c}{} & \multicolumn{1}{|c}{${\footnotesize \gamma(y,d):}$} &
\multicolumn{1}{|c}{${\footnotesize \Delta}^{{\footnotesize (y)}%
}{\footnotesize (d):}$} & \multicolumn{1}{|c}{${\footnotesize \Delta
}^{{\footnotesize (y)}}{\footnotesize (d):}$} &
\multicolumn{1}{|c}{${\footnotesize \Delta\beta}_{{\footnotesize d}%
}{\footnotesize (y,d}_{{\footnotesize y}}^{{\footnotesize \ast}}%
{\footnotesize ):}$} & \multicolumn{1}{|c}{${\footnotesize \Delta
}^{{\footnotesize (y)}}{\footnotesize (d}^{\ast}_{y}{\footnotesize ):y\geq1}$%
}\\
& \multicolumn{1}{|c}{${\footnotesize y\geq1,d\geq1}$} &
\multicolumn{1}{|c}{${\footnotesize y\geq1,d\geq1}$} &
\multicolumn{1}{|c}{${\footnotesize y\geq1,d\leq d_{y}^{\ast}-1;}$} &
\multicolumn{1}{|c}{${\footnotesize y\geq1}$} & \multicolumn{1}{|c}{}\\
& \multicolumn{1}{|c}{} & \multicolumn{1}{|c}{} & \multicolumn{1}{|c}{$%
%TCIMACRO{\tsum \limits_{{\footnotesize d\geq d}^{\ast}}}%
%BeginExpansion
{\textstyle\sum\limits_{{\footnotesize d\geq d}^{\ast}}}
%EndExpansion
{\footnotesize \Delta}^{{\footnotesize (y)}}{\footnotesize (d):y\geq1}$} &
\multicolumn{1}{|c}{} & \multicolumn{1}{|c}{}\\
\multicolumn{1}{c}{} & \multicolumn{1}{|c}{} & \multicolumn{1}{|c}{} &
\multicolumn{1}{|c}{} & \multicolumn{1}{|c}{} & \multicolumn{1}{|c}{}%
\\\hline\hline
\end{tabular}

\end{center}

\newpage

\subsection{Identification of the distribution of unobserved heterogeneity}

In empirical applications of dynamic structural models, the answer to some
important empirical questions requires the identification of the distribution
of the unobserved heterogeneity. For instance, the researcher can be
interested in the \textit{average marginal effects }$\int[\partial
P_{\mathbf{\theta}}\left(  y|\text{ }\mathbf{x},\mathbf{\beta}^{\ast}\right)
/\partial\mathbf{x}]$ $f(\mathbf{\theta})$ $d\mathbf{\theta}$ or
$\int[\partial P_{\mathbf{\theta}}\left(  y|\text{ }\mathbf{x},\mathbf{\beta
}^{\ast}\right)  /\partial\mathbf{\beta}^{\ast}]$ $f(\mathbf{\theta})$
$d\mathbf{\theta}$, where $f(\mathbf{\theta})$ is the density function of the
unobserved heterogeneity. Without further restrictions, the density function
$f(\mathbf{\theta})$ is not (nonparametrically) point identified, i.e.,
initial conditions problem. In this section, we briefly describe this
identification problem, and two possible approaches that the researcher can
take to deal with this problem: (a) nonparametric finite mixture; and (b) set identification.

Let $f(\mathbf{\theta}$ $|$ $\mathbf{x}_{1})$ be the density function of
$\mathbf{\theta}$ conditional on the initial value of the state variables
$\mathbf{x}_{1}\equiv(y_{0},d_{1})$. After the identification/estimation of
the structural parameters, $\mathbf{\beta}^{\ast}$, the model implies the
following restrictions for the identification of $f(\mathbf{\theta}$ $|$
$\mathbf{x}_{1})$. For any choice history $\widetilde{\mathbf{y}\text{,}}$ we
have that:%
\begin{equation}
\mathbb{P}\left(  \widetilde{\mathbf{y}}|\mathbf{x}_{1}\right)  =\int\left[
\dprod \limits_{t=1}^{T}P\left(  y_{t}\text{ }|\text{ }\mathbf{x}%
_{t},\mathbf{\beta}^{\ast},\mathbf{\theta}\right)  \right]  f(\mathbf{\theta
}|\mathbf{x}_{1})\text{ }d\mathbf{\theta}
\label{system of equation for density theta}%
\end{equation}
The probabilities of choice histories $\mathbb{P}\left(  \widetilde
{\mathbf{y}}|\mathbf{x}_{1}\right)  $ are identified from the data. Also, for
a fixed value of $\mathbf{\theta}$, the probabilities $P\left(  y_{t}\text{
}|\text{ }\mathbf{x}_{t},\mathbf{\beta}^{\ast},\mathbf{\theta}\right)  $ are
also known to the researcher after the identification of the structural
parameters $\mathbf{\beta}^{\ast}$. Therefore, the identification of the
density function $f(\mathbf{\theta}|\mathbf{x}_{1})$ can be seen as the
solution to a system of linear equations.

Let $|\Theta|$ be the dimension of the support of $\mathbf{\theta}$. This
dimension can be infinite. Equation
(\ref{system of equation for density theta}) can be written in vector form as,%
\begin{equation}
\mathbb{P}_{\mathbf{x}_{1}}=\mathbf{L}_{\mathbf{x}_{1}}\text{ }\mathbf{f}%
_{\mathbf{x}_{1}} \label{in vector form}%
\end{equation}
$\mathbb{P}_{\mathbf{x}_{1}}$ is a vector of dimension $(J+1)^{T}\times1$ with
the probabilities of all the possible choice histories with initial conditions
$\mathbf{x}_{1}$. $\mathbf{L}_{\mathbf{x}_{1}}$ is a matrix with dimension
$(J+1)^{T}\times|\Theta|$ such that each row contains the probabilities $%
%TCIMACRO{\tprod \nolimits_{t=1}^{T}}%
%BeginExpansion
{\textstyle\prod\nolimits_{t=1}^{T}}
%EndExpansion
P\left(  y_{t}\text{ }|\text{ }\mathbf{x}_{t},\mathbf{\beta}^{\ast
},\mathbf{\theta}\right)  $ for a given choice history and for every value of
$\mathbf{\theta}$. Finally, $\mathbf{f}_{\mathbf{x}_{1}}$ is a $|\Theta
|\times1$ vector with the probabilities $f(\mathbf{\theta}|\mathbf{x}_{1})$.
Given this representation, it is clear that $\mathbf{f}_{\mathbf{x}_{1}}$ is
point identified if and only if matrix $\mathbf{L}_{\mathbf{x}_{1}}$ is full
column rank.

If the distribution of $\mathbf{\theta}$ is continuous, then $|\Theta|=\infty$
and $\mathbf{L}_{\mathbf{x}_{1}}$ cannot be full-column rank. In fact, the
number of rows in matrix $\mathbf{L}_{\mathbf{x}_{1}}$ (i.e., the number of
possible choice histories, $(J+1)^{T}$) provides an upper bound to the
dimension of the support $|\Theta|$ for which the density is nonparametrically
(point) identified. The researcher may be willing to impose the restriction
that the support of $\mathbf{\theta}$ is discrete such that matrix
$\mathbf{L}_{\mathbf{x}_{1}}$ is full column rank. Under this condition,
$\mathbf{f}_{\mathbf{x}_{1}}$ can be identified as the linear projection:%
\begin{equation}
\mathbf{f}_{\mathbf{x}_{1}}=\left[  \mathbf{L}_{\mathbf{x}_{1}}^{\prime
}\mathbf{L}_{\mathbf{x}_{1}}\right]  ^{-1}\mathbf{L}_{\mathbf{x}_{1}}^{\prime
}\mathbb{P}_{\mathbf{x}_{1}} \label{linear projection}%
\end{equation}
Note that the estimator $\mathbf{\beta}^{\ast}$ is still a fixed-effect
estimator that is robust to this finite-mixture restriction on the
distribution of the unobservables. However, under this approach, the
estimation of marginal effects depends on this assumption. Alternatively, the
researcher may prefer not to impose this finite support restriction and
set-identify the distribution of the unobservables. This is the approach in
Chernozhukov, Fernandez-Val, Hahn, and Newey (2013).

Finally, we would like to comment on a practical issue in the implementation
of the finite-mixture estimation described above. For the evaluation of the
choice probabilities $P\left(  y_{t}\text{ }|\text{ }\mathbf{x}_{t}%
,\mathbf{\beta}^{\ast},\mathbf{\theta}\right)  $ in matrix $\mathbf{L}%
_{\mathbf{x}_{1}}$, the vector of unobserved heterogeneity $\mathbf{\theta}$
is multidimensional. That is, we need to choose a grid of points for the
parameters $\alpha_{\mathbf{\theta}}(y)$ but also for the continuation values
$v_{\mathbf{\theta}}(y,d)$. In the forward-looking model without duration
dependence, unobserved heterogeneity enters through the term $\tau
_{\mathbf{\theta}}(y)\equiv\alpha_{\mathbf{\theta}}(y)+v_{\mathbf{\theta}}%
(y)$. Therefore, for this model we need to fix a grid of points for the $J$
incidental parameters $\{\tau_{\mathbf{\theta}}(y):y>1\}$. Using a grid of
$\kappa$ points for each parameter $\tau_{\mathbf{\theta}}(y)$ we have that
the dimension of the density vector $\mathbf{f}_{\mathbf{x}_{1}}$ is
$|\Theta|=\kappa^{J}$ that should be smaller that $(J+1)^{T}$ in order to have
identification. In the forward-looking model with duration dependence,
unobserved heterogeneity enters through the term $\tau_{\mathbf{\theta}%
}(y,d)\equiv\alpha_{\mathbf{\theta}}(y)+v_{\mathbf{\theta}}(y,d)$. Therefore,
we need to fix a grid of points for the $JT$ incidental parameters
$\{\tau_{\mathbf{\theta}}(y,d):y>1$; $1\leq d\leq T\}$. Using a grid of
$\kappa$ points for each parameter $\tau_{\mathbf{\theta}}(y,d)$ we have that
the dimension of $\mathbf{f}_{\mathbf{x}_{1}}$ is $|\Theta|=\kappa^{JT}$ that
should be smaller that $(J+1)^{T}$. This is a strong restriction on the
dimension of unobserved heterogeneity, $\kappa$. However, this approach is not
taking into account that the continuation values $v_{\mathbf{\theta}}(y,d)$
are endogenous objects that can be obtained given $\alpha_{\mathbf{\theta}%
}^{\prime}s$ and $\mathbf{\beta}^{\ast}$ by solving the Bellman equation of
the model. Taking into account this structure of the model, we can reduce
substantially the dimensionality of $\mathbf{\theta}$. Given a value of the
$J$ incidental parameters $\{\alpha_{\mathbf{\theta}}(y):y>1\}$, we can solve
the Bellman equation to obtain all the continuation values $v_{\mathbf{\theta
}}(y,d)$. Therefore, the dimension of $\mathbf{\theta}$ in the structural
model with duration dependence is also equal to the dimension of
$\{\alpha_{\mathbf{\theta}}(y):y>1\}$, as in the model without duration dependence.

\section{Estimation and Inference}

Since the identification is based on the conditional MLE approach, the
estimator for the structural parameters of interest $(\beta_{y},\beta_{d})$
will be an Andersen (1970) type of estimator. We illustrate the estimator for
the forward-looking multinomial choice model with duration dependence under
Assumption 1 and 2, since estimators for the structural parameters in the
other models can be defined in a similar fashion.

\subsection{Estimation of $\mathbf{\beta}^{\ast}$ (given $d^{\ast}$)}

Let $\mathbf{\beta}^{\ast}=\{\widetilde{\mathbf{\beta}}_{y}^{\prime
},\mathbf{\gamma}^{\prime}\}^{\prime}$ be the vector of identified structural
parameters. Let $U_{i}$ be the vector of sufficient statistics (associated to
$\mathbf{\theta}$), and let $S_{i}$ be the vector of identifying statistics
associated to $\mathbf{\beta}^{\ast}$. Then, the conditional MLE for
$\mathbf{\beta}^{\ast}$is defined as the maximizer of the conditional
log-likelihood function:%
\begin{equation}
\mathcal{L}_{N}(\mathbf{\beta}^{\ast})=\sum_{i=1}^{N}\mathcal{L}%
_{i}(\mathbf{\beta}^{\ast})=\sum_{i=1}^{N}S_{i}^{\prime}\mathbf{\beta}^{\ast
}-\left(
%TCIMACRO{\tsum \limits_{j:U(j)=U_{i}}}%
%BeginExpansion
{\textstyle\sum\limits_{j:U(j)=U_{i}}}
%EndExpansion
\exp\left\{  S(j)^{\prime}\mathbf{\beta}^{\ast}\right\}  \right)
\label{conditional log-likelihood function}%
\end{equation}
where the condition $\{j:U(j)=U_{i}\}$ represents all the choice histories
$(y_{0},d_{1},\widetilde{\mathbf{y}})$ with the same value of $U$\thinspace as
observation $i$. This log-likelihood function is globally concave in
$\mathbf{\beta}^{\ast}$, and therefore the computation of the CMLE is
straightforward using Newton-Raphson or BHHH algorithm. Using standard
arguments (Newey and McFadden, 1994), we have
\begin{equation}
\sqrt{N}(\widehat{\mathbf{\beta}}^{\ast}-\mathbf{\beta}^{\ast})\Rightarrow
\mathcal{N}(0,J(\mathbf{\beta}^{\ast})^{-1})
\label{Asymptotic distribution of beta*}%
\end{equation}
The consistent estimator for the Fisher information is $J_{N}(\widehat
{\mathbf{\beta}}^{\ast})=-N^{-1}\sum_{i=1}^{N}\nabla_{\mathbf{\beta\beta}%
}\mathcal{L}_{i}(\widehat{\mathbf{\beta}}^{\ast})$.

\subsection{Estimation of $d^{\ast}$}

We describe here a CML estimator for the joint estimation of $(d^{\ast
},\mathbf{\beta}^{\ast})$. Let $d_{0}^{\ast}$ represent the true value of the
parameter $d^{\ast}$. And let $\beta_{0}(n)$ be the true value of the
parameter $\beta(n)\equiv\beta_{d}(y,n)-\beta_{d}(y,n-1)$. By definition, we
have that $\beta_{0}(d_{0}^{\ast})\neq0$ and $\beta_{0}(n)=0$ for any
$n>d_{0}^{\ast}$. For notational simplicity, we use $\beta^{\ast}$ and
$\beta_{0}^{\ast}$ to represent $\beta(d^{\ast})$ and and $\beta_{0}%
(d_{0}^{\ast})$, respectively. We are interested in the joint identification
of $(d_{0}^{\ast},\beta_{0}^{\ast})$ from the maximization of the conditional
likelihood function.

Based on Proposition 6, we consider the following representation of the
sufficient statistic $U$. For any $2\leq n\leq L$ with $L\leq(T-1)/2$, define
the pair of histories $A_{n}=\{0,0|\mathbf{y}_{n-1},0,\mathbf{y}_{T-n}\}$ and
$B_{n}=\{0,0|\mathbf{y}_{n},0,\mathbf{y}_{T-n-1}\}$. Then, $U_{i}=\left\{
\mathbf{y}_{i}\in A_{n}\cup B_{n}\text{ \ for some }2\leq n\leq L\right\}  $.
Given this statistic, the conditional likelihood function is:%
\begin{equation}%
\begin{array}
[c]{ccl}%
\mathcal{L}_{N}(\mathbf{\nu}) & = & \sum\limits_{n=2}^{L}\sum\limits_{i=1}%
^{N}1\{\mathbf{y}_{i}=A_{n}\}\ln\left[  \dfrac{\exp\left\{  \nu(n)\right\}
}{1+\exp\left\{  \nu(n)\right\}  }\right]  +1\{\mathbf{y}_{i}=B_{n}%
\}\ln\left[  \dfrac{1}{1+\exp\left\{  \nu(n)\right\}  }\right]
\end{array}
\label{enhanced conditional likelihood}%
\end{equation}
where $\nu(n)$ is a parameter that represents the value $\beta_{d}%
(y,n)-\beta_{d}(y,n-1)+$ $\int[v_{\mathbf{\theta}}(y,n+1)-v_{\mathbf{\theta}%
}(y,n)]$ $dF(\mathbf{\theta}|\mathbf{x}_{1})$, and $\mathbf{\nu}$ is the
vector of parameters $\{\nu(n):n=2,3,...,L\}$. The model implies the following
relationship between the parameters $\nu(n)$ and the structural parameters
$(d^{\ast},\beta^{\ast})$.%
\begin{equation}
\nu(n)=\left\{
\begin{array}
[c]{ccc}%
\text{unrestricted} & \text{if} & n<d^{\ast}\\
&  & \\
\beta^{\ast} & \text{if} & n=d^{\ast}\\
&  & \\
0 & \text{if} & n>d^{\ast}%
\end{array}
\right.  \label{relationship between v's and structural}%
\end{equation}

The unconstrained likelihood function $\mathcal{L}_{N}(\mathbf{\nu})$ is
globally concave in each of the parameters $\nu(n)$. It is straightforward to
show that the unconstrained CML\ estimator of $\nu(n)$ is $\widehat{v}(n)=$
$\ln\widehat{\mathbb{P}}\left(  A_{n}\right)  -\ln\widehat{\mathbb{P}}\left(
B_{n}\right)  $, where $\widehat{\mathbb{P}}\left(  A_{n}\right)  $ and
$\widehat{\mathbb{P}}\left(  B_{n}\right)  $ are the sample frequencies
$N^{-1}%
%TCIMACRO{\tsum \nolimits_{i=1}^{N}}%
%BeginExpansion
{\textstyle\sum\nolimits_{i=1}^{N}}
%EndExpansion
1\{\mathbf{y}_{i}=A_{n}\}$ and $N^{-1}%
%TCIMACRO{\tsum \nolimits_{i=1}^{N}}%
%BeginExpansion
{\textstyle\sum\nolimits_{i=1}^{N}}
%EndExpansion
1\{\mathbf{y}_{i}=B_{n}\}$, respectively. For a given value of $d^{\ast}$, let
$\widehat{\mathbf{\nu}}_{d^{\ast}}^{c}$ be the constrained estimator of
$\mathbf{\nu}$ that imposes the restriction in equation
(\ref{relationship between v's and structural}) such that: $\widehat{\nu
}_{d^{\ast}}^{c}(n)=\widehat{\nu}(n)$ (unconstrained) for $n\leq d^{\ast}$;
and $\widehat{\nu}_{d^{\ast}}^{c}(n)=0$ (constrained) for $n>d^{\ast}$.
Furthermore, the estimator of the structural parameter $\beta^{\ast}$ is
$\widehat{\beta^{\ast}}=\widehat{\nu}(d^{\ast})$.

Let $\ell_{N}(d^{\ast})$ be the concentrated likelihood function $\ell
_{N}(d^{\ast})\equiv\mathcal{L}_{N}(\widehat{\mathbf{\nu}}_{d^{\ast}}^{c})$,
i.e., the value of the likelihood given a value of $d^{\ast}$ and where the
parameters $\mathbf{\nu}$ have been estimated under this restriction. By
definition, we have that:%
\begin{equation}%
\begin{array}
[c]{ccl}%
\ell_{N}(d^{\ast}) & = & N\sum\limits_{n=2}^{d^{\ast}}\widehat{\mathbb{P}%
}\left(  A_{n}\right)  \text{ }\ln\left[  \dfrac{\widehat{\mathbb{P}}\left(
A_{n}\right)  }{\widehat{\mathbb{P}}\left(  A_{n}\right)  +\widehat
{\mathbb{P}}\left(  B_{n}\right)  }\right]  +\widehat{\mathbb{P}}\left(
B_{n}\right)  \text{ }\ln\left[  \dfrac{\widehat{\mathbb{P}}\left(
B_{n}\right)  }{\widehat{\mathbb{P}}\left(  A_{n}\right)  +\widehat
{\mathbb{P}}\left(  B_{n}\right)  }\right] \\
&  & \\
& + & N\sum\limits_{n=d^{\ast}+1}^{L}\widehat{\mathbb{P}}\left(  A_{n}\right)
\text{ }\ln\left[  \dfrac{1}{2}\right]  +\widehat{\mathbb{P}}\left(
B_{n}\right)  \text{ }\ln\left[  \dfrac{1}{2}\right]
\end{array}
\label{concentrated likelihood function}%
\end{equation}
The following Proposition 12 establishes some properties of this concentrated
likelihood function.

\medskip

\noindent\textit{PROPOSITION 12. (A) As }$N\rightarrow\infty$\textit{, the
concentrated likelihood function }$N^{-1}\ell_{N}(d^{\ast})$ \textit{converges
uniformly in }$d^{\ast}$ \textit{to its population counterpart }$\ell
_{0}(d^{\ast})$. \textit{(B) }$\ell_{0}(d_{0}^{\ast})>\ell_{0}(d^{\ast})$
\textit{for any }$d^{\ast}<d_{0}^{\ast}$\textit{, and }$\ell_{0}(d_{0}^{\ast
})=\ell_{0}(d^{\ast})$ \textit{for any }$d^{\ast}>d_{0}^{\ast}$\textit{.
Therefore, }$d_{0}^{\ast}$ \textit{is point identified as the minimum value of
}$d^{\ast}$\textit{ that maximizes the concentrated likelihood function}:
$d_{0}^{\ast}=\min\{n:n\in\arg\max_{2\leq d^{\ast}\leq L}$ $\ell_{0}(d^{\ast
})\}$.\qquad$\blacksquare$

\medskip

Given this result, a possible estimator for $d_{0}^{\ast}$ would be the sample
analog $\widehat{d^{\ast}}=\min\{n:$ $n\in\arg\max_{2\leq d^{\ast}\leq L}$
$\ell_{N}(d^{\ast})\}$. However, this estimator has a an important limitation
in finite samples. Though the population likelihood function $\ell_{0}%
(d^{\ast})$ is flat for values of $d^{\ast}$ greater than the true
$d_{0}^{\ast}$, in a finite sample this likelihood increases with $d^{\ast}$
and reaches its maximum at the largest possible value of $d^{\ast}$, i.e.,
$d^{\ast}=L$. This is because any value of $d^{\ast}$ smaller than $L$ implies
restrictions on the parameters $\nu(n)$ of the model, i.e., $\nu(n)=0$ for
$n>d^{\ast}$. The larger the value of $d^{\ast}$, the smaller the number of
these restrictions and the largest the value of the likelihood function in a
finite sample.

To deal with this problem we consider an estimator of $d_{0}^{\ast}$ that
maximizes the \textit{Bayesian Information Criterion} (BIC). This criterion
function introduces a penalty that increases with the number of free
parameters $\{v(n)\}$ in the model. In this model, the number of free
parameters is equal to $d^{\ast}$. The BIC\ function is defined as:%
\begin{equation}
BIC_{N}(d^{\ast})=\ell_{N}(d^{\ast})-\frac{d^{\ast}}{2}\ln(N)
\label{BIC function}%
\end{equation}
Our estimator of $d_{0}^{\ast}$ is defined as the value of $d^{\ast}$ that
maximizes $BIC_{N}(d^{\ast})$.

\medskip

\noindent\textit{PROPOSITION 13. Consider the estimator }$\widehat{d_{N}%
^{\ast}}=\arg\max_{2\leq d^{\ast}\leq L}$ $BIC_{N}(d^{\ast})$. \textit{As}
$N\rightarrow\infty$, $\mathbb{P}(\widehat{d_{N}^{\ast}}=d_{0}^{\ast
})\rightarrow1$.\qquad$\blacksquare$

The joint estimation of $(d^{\ast},\beta^{\ast})$ has the analogy of model
selection where $d^{\ast}$ determines the model dimension and $\beta^{\ast}$
is the parameter of interest. We can use standard inference for the CML
estimator for $\beta^{\ast}$ in this joint estimation method since Proposition
13 shows that $\widehat{d_{N}^{\ast}}$ is a consistent estimator for
$d_{0}^{\ast}$. This is in the same spirit that under consistent model
selection: the asymptotic property of the estimator for parameters in the
selected model is unaffected (see P\"{o}tscher, 1991). However, P\"{o}tscher
(1991) also pointed out that inference for parameters post model selection can
be problematic in finite samples if the parameter is too close to zero and the
true model is not selected with probability close to one. In our Monte Carlo
experiments, we found that the probability of selecting the true $d_{0}^{\ast
}$ is very close to 1 throughout different data generating
processes.\footnote{For example, for DGP 1 with Sample B, described in Table
5, out of 1000 repetition, 99\% of the times $\widehat{d_{N}^{\ast}}$ agrees
with the true $d_{0}^{\ast}$.}

\bigskip

\bigskip

\bigskip

\newpage

\section{Empirical Application}

Here we revisit the model and data in the seminal article by Rust (1987). The
model belongs to the class of \textit{machine replacement models }that we have
briefly described in section 2. The superintendent of maintenance at the
Madison (Wisconsin) Metropolitan Bus Company has a fleet of $N$ buses indexed
by $i$. For every bus $i$ and at every period $t$, the superintendent decides
whether to keep the bus engine ($y_{it}=1$) or to replace it ($y_{it}=0$). In
Rust's model, if the engine is replaced, the payoff is equal to
$-RC+\varepsilon_{it}(0)$, where $RC$ is a parameter that represents the
replacement cost. If the manager decides to keep the engine, the payoff is
equal to $-c_{0}-c_{1}(m_{it})+\varepsilon_{it}(1)$, where $m_{it}$ is a state
variable that represents the engine cumulative mileage, and $c_{0}%
+c_{1}(m_{it})$ is the maintenance cost. We incorporate two modifications in
this model. First, we replace cumulative mileage $m_{it}$ with duration since
last replacement, $d_{it}$. The transition rule for this state variable is
$d_{it+1}=y_{it}[d_{it}+1]$, such that $d_{it}\in\{0,1,2,...\}$. Using Rust's
actual data, the correlation between the variables $m_{it}$ and $d_{it}$ is
0.9552. Second, we allow for time-invariant unobserved heterogeneity in the
replacement cost, $RC_{i}$, and in the constant term in the maintenance cost
function, $c_{0i}$. Using our notation, the payoff function is $\alpha
_{i}(0)+\varepsilon_{it}(0)$ if $y_{it}=0$ (replacing the engine), and
$\alpha_{i}(1)+\beta_{d}(d_{it})+\varepsilon_{it}(1)$ if $y_{it}=1$ (keeping
the engine), where $\alpha_{i}(0)=-RC_{i}$, $\alpha_{i}(1)=-c_{0i}$, and
$\beta_{d}(d_{it})=-c_{1}(d_{it})$.

In section 5.1, we present evidence from several Monte Carlo experiments using
this model. The purpose of these experiments is threefold. First, showing that
the FE-CMLE provides precise and robust estimates of structural parameters,
even when the sample size is not large. Second, showing that the bias of
misspecifying the distribution of the unobserved heterogeneity. And third,
showing that a Hausman test, based on the comparison of the FE-CMLE and a
CRE-MLE, has enough power to reject specifications that wrongly ignore
unobserved heterogeneity, or that misspecified its probability distribution or
its joint distribution with the initial conditions of the state variables. In
section 5.2, we apply the FE-CMLE method, our procedure to estimate $d^{\ast}%
$, and the Hausman test to the actual dataset in Rust (1987).

\subsection{Monte Carlo experiments}

We present experiments using simulated data from four different Data
Generating Processes (DGPs). Table 4 describes these DGPs. The difference
between the four DGPs is in the specification of the distribution of the
unobserved heterogeneity for the replacement cost $RC_{i}$. In DGP 1, the
distribution of the replacement cost is normal with mean $8$ and standard
deviation $2$. In DGPs 2 and 3, this distribution has only two types. Finally,
DGP 4 is a model without unobserved heterogeneity.

For each of these DGPs, we do not estimate the model using the whole sample of
$T=25$ periods. Instead, we construct three samples: sample A, from period 1
to 7; sample B, from period 1 to 14; and Sample C, from period 8 to 21.
Therefore, we present results from $12$ Monte Carlo experiments, i.e., four
DGPs times 3 samples. We analyze the effect of increasing the number of time
periods $T$, by comparing the experiments with sample A (with $T=7$) and
sample B (with $T=14$). We study the effect of the initial conditions problem
by comparing the experiments for sample B (where at $t=1$ all the buses have
the same initial condition, $(y_{i0},d_{i1})=(0,0)$) and sample C, that is
subject to the initial conditions problem.

\medskip

\begin{center}%
\begin{tabular}
[c]{r|c|c|c|c}\hline\hline
\multicolumn{5}{c}{\textbf{Table 4}}\\
\multicolumn{5}{c}{\textbf{Description of DGPs in the Monte Carlo
experiments}}\\\hline
\textit{Parameter / Constant} & \textit{DGP 1} & \textit{DGP 2} & \textit{DGP
3} & \textit{DGP 4}\\\hline
\multicolumn{1}{c|}{} &  &  &  & \\
$\alpha_{i}(0)=-RC_{i}$ & $N(\mu,\sigma^{2})$ & Two types & Two types & 1
type\\
Random draws from: & $\mu=8,\sigma=2$ & $RC_{1}=4.5,RC_{2}=9$ & $RC_{1}%
=8,RC_{2}=9$ & $RC=8$\\
&  & $\lambda_{1}=\lambda_{2}=0.5$ & $\lambda_{1}=\lambda_{2}=0.5$ & \\
$\alpha_{i}(1)=-c_{0i}$ & 0 & 0 & 0 & 0\\
$\beta_{d}(d)=\beta$ $d$ if $d\leq d^{\ast}$ & $\beta=1$ & $\beta=1$ &
$\beta=1$ & $\beta=1$\\
$d^{\ast}$ & $3$ & $3$ & $3$ & $3$\\
\textit{Discount factor (}$\delta$\textit{)} & 0.95 & 0.95 & 0.95 & 0.95\\
&  &  &  & \\
\textit{Initial }$y_{0},d_{1}$ & $0,0$ & $0,0$ & $0,0$ & $0,0$\\
\textit{Maximum} $T$ & 25 & 25 & 25 & 25\\
$N$ \textit{(number of buses)} & 1000 & 1000 & 1000 & 1000\\
\textit{\# simulated samples} & 1000 & 1000 & 1000 & 1000\\
& \multicolumn{1}{|l|}{} &  &  & \\\hline\hline
\end{tabular}

\end{center}

\medskip

The structural parameter of interest is parameter $\beta$ in the maintenance
cost function, $\beta_{d}(d)=\beta$ $d$. We apply four estimators to each of
the samples: the \textit{FE-CMLE }using the true value of $d^{\ast}$ (that we
denote as \textit{CMLE-true-d*}); \textit{FE-CMLE }using the BIC\ estimator of
$d^{\ast}$ (that we denote as \textit{CMLE-BIC-d*}); a MLE that imposes the
restriction of no unobserved heterogeneity (that we denote as
\textit{MLE-noUH}), and a MLE that assumes that there are two types of
replacement costs and ignores the potential initial conditions problem (that
we denote as \textit{MLE-2types}). We compare the bias and variance of these
estimators.\footnote{The code for this experiment is in Matlab. For the two ML
estimators, we use the Nested Fixed Point Algorithm. The maximization of the
log-likelihood function applies a quasi-newton method (procedure
\texttt{fminunc}) using the true value of the vector of parameters as the
starting value. For the MLE with 2-types, during the search algorithm we often
get a singular Hessian matrix. When this happens, we switch to the BHHH
method.} We also implement two Hausman tests: a test of the null hypothesis of
no unobserved heterogeneity, that compares estimators \textit{CMLE-BIC-d* }and
\textit{MLE-noUH}; and a test of the null hypothesis of two-types, that
compares estimators \textit{CMLE-BIC-d* }and \textit{MLE-2types}. We present
the results of these experiments in tables 5 to 8, one table for each DGP.

\begin{center}%
\begin{tabular}
[c]{r|ccc|ccc|ccc}\hline\hline
\multicolumn{10}{c}{\textbf{Table 5}}\\
\multicolumn{10}{c}{\textbf{Monte Carlo Experiments with DGP 1 (Normal RCs)}%
}\\\hline
&  &  &  &  &  &  &  &  & \\
& \multicolumn{3}{|c|}{\textbf{Sample A} {\small (}${\small t=1}${\small to
}${\small 7}${\small )}} & \multicolumn{3}{|c|}{\textbf{Sample B}
{\small (}${\small t=1}${\small to }${\small 14}${\small )}} &
\multicolumn{3}{|c}{\textbf{Sample C} {\small (}${\small t=8}${\small to
}${\small 21}${\small )}}\\
{\small Estimator} & \multicolumn{3}{|c}{{\small Estimate}$^{(1)}$} &
\multicolumn{3}{|c}{{\small Estimate}$^{(1)}$} &
\multicolumn{3}{|c}{{\small Estimate}$^{(1)}$}\\
{\small of }${\small \beta}$ & {\small Mean} & {\small Median} & {\small St.
dev.} & {\small Mean} & {\small Median} & {\small St. dev.} & {\small Mean} &
{\small Median} & {\small St. dev.}\\\hline
\multicolumn{1}{c|}{} &  &  &  &  &  &  &  &  & \\
{\small CMLE-true-d*} & {\small 1.0073} & {\small 1.0086} & {\small 0.1436} &
{\small 0.9990} & {\small 1.0003} & {\small 0.0801} & {\small 0.9954} &
{\small 0.9978} & {\small 0.0731}\\
&  &  &  &  &  &  &  &  & \\
{\small CMLE-BIC-d*} & {\small 1.0073} & {\small 1.0086} & {\small 0.1436} &
{\small 0.9935} & {\small 1.0001} & {\small 0.1054} & {\small 0.9873} &
{\small 0.9971} & {\small 0.1146}\\
&  &  &  &  &  &  &  &  & \\
{\small MLE-2types} & {\small 0.9778} & {\small 0.9765} & {\small 0.0528} &
{\small 0.8956} & {\small 0.8962} & {\small 0.0325} & {\small 0.8565} &
{\small 0.8554} & {\small 0.0308}\\
&  &  &  &  &  &  &  &  & \\
{\small MLE-noUH} & {\small 0.6204} & {\small 0.6191} & {\small 0.0295} &
{\small 0.5842} & {\small 0.5835} & {\small 0.0232} & {\small 0.5444} &
{\small 0.5439} & {\small 0.0229}\\
&  &  &  &  &  &  &  &  & \\\hline
& \multicolumn{3}{|c}{{\small Frequency of Ho rejection}} &
\multicolumn{3}{|c}{{\small Frequency of Ho rejection}} &
\multicolumn{3}{|c}{{\small Frequency of Ho rejection}}\\
{\small Testing} & \multicolumn{3}{|c}{{\small with significance level}} &
\multicolumn{3}{|c}{{\small with significance level}} &
\multicolumn{3}{|c}{{\small with significance level}}\\
{\small null hypothesis} & ${\small 1\%}$ & ${\small 5\%}$ & ${\small 10\%}$ &
${\small 1\%}$ & ${\small 5\%}$ & ${\small 10\%}$ & ${\small 1\%}$ &
${\small 5\%}$ & ${\small 10\%}$\\\hline
&  &  &  &  &  &  &  &  & \\
{\small No Unob. Het.} & {\small 0.541} & {\small 0.777} & {\small 0.874} &
{\small 0.999} & {\small 1.000} & {\small 1.000} & {\small 1.000} &
{\small 1.000} & {\small 1.000}\\
&  &  &  &  &  &  &  &  & \\
{\small Two types} & {\small 0.008} & {\small 0.042} & {\small 0.096} &
{\small 0.125} & {\small 0.308} & {\small 0.429} & {\small 0.281} &
{\small 0.515} & {\small 0.658}\\
&  &  &  &  &  &  &  &  & \\\hline\hline
\multicolumn{10}{l}{{\footnotesize Note (1): Mean, median, and standard
deviation of estimated parameter over the 1,000 replications.}}%
\end{tabular}

\end{center}

Table 5 deals with DGP 1, with normally distributed replacement costs. The
MLEs are substantially biased, especially in sample $C$ (with the initial
conditions problem) and sample $B$ (with large $T$). When $T$ increases there
are multiple spells per bus and this implies stronger correlation between
observed durations and unobserved heterogeneity. This generates a larger bias
of the MLE of a misspecified model. In contrast, the biases of the
\textit{CMLEs }(either with true or estimated $d^{\ast}$) are negligible. The
BIC method provides precise estimates of $d^{\ast}$: in all our DGPs, the
estimated value of $d^{\ast}$ is equal to its true value for more than $95\%$
of the Monte Carlo replications. As a result, the bias of the CMLE estimator
of $\beta$ with estimated \textit{d}* is very similar to the bias of the CMLE
with \textit{true d}*. As expected, the CMLEs have larger variance than the
MLEs, and the CMLE with estimated $d^{\ast}$ has larger variance than the CMLE
with true $d^{\ast}$. However, the \textit{CMLE-BIC-d*} has a Mean Square
Error (MSE, variance plus square bias) that is substantially smaller than the
one of the \textit{MLE-noUH }in the three samples, and of the
\textit{MLE-2types }in samples B\ and C. In sample A, the \textit{MLE-2types
}has a MSE comparable to the one of the \textit{CMLE. }That is, in a DGP
without initial conditions problem and with one duration spell for most of the
buses, a misspecified random effects model with only two types has good
properties. However, this is not longer the case in samples B and C. Hausman
test has very strong power to reject the model without unobserved
heterogeneity.\footnote{Though the distribution of types in DGP 1 is
continuous, the level of unobserved heterogeneity is modest. In the
distribution of $RC_{i}$, the ratio between the standard deviation and the
mean is only $25\%$. Continuous distributions with higher variance imply
higher rejection rates of the model with only two types, even in sample
A.}\ It has also substantial power to reject the model with two types in
samples B\ and C. However, the rejection rates for the model with two types in
sample A are practically equal to the nominal size or significance level of
the test.

\begin{center}%
\begin{tabular}
[c]{r|ccc|ccc|ccc}\hline\hline
\multicolumn{10}{c}{\textbf{Table 6}}\\
\multicolumn{10}{c}{\textbf{Monte Carlo Experiments with DGP 2 (Two types: RC
= 4.5, 9)}}\\\hline
&  &  &  &  &  &  &  &  & \\
& \multicolumn{3}{|c|}{\textbf{Sample A} {\small (}${\small t=1}${\small to
}${\small 7}${\small )}} & \multicolumn{3}{|c|}{\textbf{Sample B}
{\small (}${\small t=1}${\small to }${\small 14}${\small )}} &
\multicolumn{3}{|c}{\textbf{Sample C} {\small (}${\small t=8}${\small to
}${\small 21}${\small )}}\\
{\small Estimator} & \multicolumn{3}{|c}{{\small Estimate}$^{(1)}$} &
\multicolumn{3}{|c|}{{\small Estimate}$^{(1)}$} &
\multicolumn{3}{|c}{{\small Estimate}$^{(1)}$}\\
{\small of }${\small \beta}$ & {\small Mean} & {\small Median} & {\small St.
dev.} & {\small Mean} & {\small Median} & {\small St. dev.} & {\small Mean} &
{\small Median} & {\small St. dev.}\\\hline
\multicolumn{1}{c|}{} &  &  &  &  &  &  &  &  & \\
{\small CMLE-true-d*} & \multicolumn{1}{|l}{{\small 1.0094}} &
\multicolumn{1}{l}{{\small 1.0060}} & \multicolumn{1}{l|}{{\small 0.1598}} &
\multicolumn{1}{|l}{{\small 1.0027}} & \multicolumn{1}{l}{{\small 1.0033}} &
\multicolumn{1}{l|}{{\small 0.0813}} & \multicolumn{1}{|l}{{\small 0.9992}} &
\multicolumn{1}{l}{{\small 0.9948}} & \multicolumn{1}{l}{{\small 0.0813}}\\
&  &  &  &  &  &  &  &  & \\
{\small CMLE-BIC-d*} & {\small 1.0094} & {\small 1.0060} & {\small 0.1598} &
{\small 0.9952} & {\small 1.0025} & {\small 0.1216} & {\small 0.9886} &
{\small 0.9941} & {\small 0.1384}\\
&  &  &  &  &  &  &  &  & \\
{\small MLE-2types} & \multicolumn{1}{|l}{{\small 1.0018}} &
\multicolumn{1}{l}{{\small 0.9990}} & \multicolumn{1}{l|}{{\small 0.0513}} &
\multicolumn{1}{|l}{{\small 1.0007}} & \multicolumn{1}{l}{{\small 1.0001}} &
\multicolumn{1}{l|}{{\small 0.0289}} & \multicolumn{1}{|l}{{\small 0.9954}} &
\multicolumn{1}{l}{{\small 0.9941}} & \multicolumn{1}{l}{{\small 0.0288}}\\
&  &  &  &  &  &  &  &  & \\
{\small MLE-noUH} & \multicolumn{1}{|l}{{\small 0.5556}} &
\multicolumn{1}{l}{{\small 0.5557}} & \multicolumn{1}{l|}{{\small 0.0229}} &
\multicolumn{1}{|l}{{\small 0.5283}} & \multicolumn{1}{l}{{\small 0.5284}} &
\multicolumn{1}{l|}{{\small 0.0156}} & \multicolumn{1}{|l}{{\small 0.5009}} &
\multicolumn{1}{l}{{\small 0.5004}} & \multicolumn{1}{l}{{\small 0.0146}}\\
&  &  &  &  &  &  &  &  & \\\hline
& \multicolumn{3}{|c}{{\small Frequency of Ho rejection}} &
\multicolumn{3}{|c|}{{\small Frequency of Ho rejection}} &
\multicolumn{3}{|c}{{\small Frequency of Ho rejection}}\\
{\small Testing} & \multicolumn{3}{|c}{{\small with significance level}} &
\multicolumn{3}{|c|}{{\small with significance level}} &
\multicolumn{3}{|c}{{\small with significance level}}\\
{\small null hypothesis} & ${\small 1\%}$ & ${\small 5\%}$ & ${\small 10\%}$ &
${\small 1\%}$ & ${\small 5\%}$ & ${\small 10\%}$ & ${\small 1\%}$ &
${\small 5\%}$ & ${\small 10\%}$\\\hline
&  &  &  &  &  &  &  &  & \\
{\small No Unob. Het.} & \multicolumn{1}{|l}{{\small 0.590}} &
\multicolumn{1}{l}{{\small 0.820}} & \multicolumn{1}{l|}{{\small 0.902}} &
\multicolumn{1}{|l}{{\small 1.000}} & \multicolumn{1}{l}{{\small 1.000}} &
\multicolumn{1}{l|}{{\small 1.000}} & \multicolumn{1}{|l}{{\small 1.000}} &
\multicolumn{1}{l}{{\small 1.000}} & \multicolumn{1}{l}{{\small 1.000}}\\
&  &  &  &  &  &  &  &  & \\
{\small Two types} & \multicolumn{1}{|l}{{\small 0.005}} &
\multicolumn{1}{l}{{\small 0.044}} & \multicolumn{1}{l|}{{\small 0.094}} &
\multicolumn{1}{|l}{{\small 0.005}} & \multicolumn{1}{l}{{\small 0.054}} &
\multicolumn{1}{l|}{{\small 0.096}} & \multicolumn{1}{|l}{{\small 0.005}} &
\multicolumn{1}{l}{{\small 0.047}} & \multicolumn{1}{l}{{\small 0.107}}\\
&  &  &  &  &  &  &  &  & \\\hline\hline
\multicolumn{10}{l}{{\footnotesize Note (1): Mean, median, and standard
deviation of estimated parameter over the 1,000 replications.}}%
\end{tabular}

\end{center}

Table 6 presents results under DGP 2, with two types of replacement costs,
$RC_{1}=4.5$ and $RC_{2}=9$, with equal probabilities. In this case, the
\textit{MLE-2types} and our \textit{CMLEs} are consistent estimators. Both
estimators have negligible finite-sample biases in the three samples. As
expected, the \textit{MLE-2types }has smaller variance, especially in sample
A. In the three samples, the \textit{MLE-noUH }is still extremely biased and
the Hausman test that compares this estimator with \textit{CMLE-BIC-d* }has
strong power to reject the model without unobserved heterogeneity. For the
rejection of the true model with two types, Hausman test exhibits a rejection
rate that is practically identical to the nominal size or significance level.

\begin{center}%
\begin{tabular}
[c]{r|ccc|ccc|ccc}\hline\hline
\multicolumn{10}{c}{\textbf{Table 7}}\\
\multicolumn{10}{c}{\textbf{Monte Carlo Experiments with DGP 3 (Two types: RC
= 8, 9)}}\\\hline
&  &  &  &  &  &  &  &  & \\
& \multicolumn{3}{|c|}{\textbf{Sample A} {\small (}${\small t=1}${\small to
}${\small 7}${\small )}} & \multicolumn{3}{|c|}{\textbf{Sample B}
{\small (}${\small t=1}${\small to }${\small 14}${\small )}} &
\multicolumn{3}{|c}{\textbf{Sample C} {\small (}${\small t=8}${\small to
}${\small 21}${\small )}}\\
{\small Estimator} & \multicolumn{3}{|c}{{\small Estimate}$^{(1)}$} &
\multicolumn{3}{|c}{{\small Estimate}$^{(1)}$} &
\multicolumn{3}{|c}{{\small Estimate}$^{(1)}$}\\
{\small of }${\small \beta}$ & {\small Mean} & {\small Median} & {\small St.
dev.} & {\small Mean} & {\small Median} & \multicolumn{1}{c|}{{\small St.
dev.}} & {\small Mean} & {\small Median} & {\small St. dev.}\\\hline
\multicolumn{1}{c|}{} &  &  &  &  &  &  &  &  & \\
{\small CMLE-true-d*} & \multicolumn{1}{|l}{{\small 1.0088}} &
\multicolumn{1}{l}{{\small 1.0058}} & \multicolumn{1}{l|}{{\small 0.1371}} &
\multicolumn{1}{|l}{{\small 1.0014}} & \multicolumn{1}{l}{{\small 1.0035}} &
\multicolumn{1}{l|}{{\small 0.0744}} & \multicolumn{1}{|l}{{\small 0.9978}} &
\multicolumn{1}{l}{{\small 0.9957}} & \multicolumn{1}{l}{{\small 0.0726}}\\
&  &  &  &  &  &  &  &  & \\
{\small CMLE-BIC-d*} & {\small 1.0088} & {\small 1.0058} & {\small 0.1371} &
{\small 0.9905} & {\small 1.0026} & {\small 0.1313} & {\small 0.9923} &
{\small 0.9941} & {\small 0.1040}\\
&  &  &  &  &  &  &  &  & \\
{\small MLE-2types} & \multicolumn{1}{|l}{{\small 1.0111}} &
\multicolumn{1}{l}{{\small 1.0064}} & \multicolumn{1}{l|}{{\small 0.0626}} &
\multicolumn{1}{|l}{{\small 1.0026}} & \multicolumn{1}{l}{{\small 1.0012}} &
\multicolumn{1}{l|}{{\small 0.0374}} & \multicolumn{1}{|l}{{\small 0.9990}} &
\multicolumn{1}{l}{{\small 0.9982}} & \multicolumn{1}{l}{{\small 0.0389}}\\
&  &  &  &  &  &  &  &  & \\
{\small MLE-noUH} & \multicolumn{1}{|l}{{\small 0.9628}} &
\multicolumn{1}{l}{{\small 0.9609}} & \multicolumn{1}{l|}{{\small 0.0451}} &
\multicolumn{1}{|l}{{\small 0.9576}} & \multicolumn{1}{l}{{\small 0.9564}} &
\multicolumn{1}{l|}{{\small 0.0317}} & \multicolumn{1}{|l}{{\small 0.9501}} &
\multicolumn{1}{l}{{\small 0.9492}} & \multicolumn{1}{l}{{\small 0.0334}}\\
&  &  &  &  &  &  &  &  & \\\hline
& \multicolumn{3}{|c}{{\small Frequency of Ho rejection}} &
\multicolumn{3}{|c}{{\small Frequency of Ho rejection}} &
\multicolumn{3}{|c}{{\small Frequency of Ho rejection}}\\
\multicolumn{1}{r|}{{\small Testing}} & \multicolumn{3}{|c}{{\small with
significance level}} & \multicolumn{3}{|c}{{\small with significance level}} &
\multicolumn{3}{|c}{{\small with significance level}}\\
\multicolumn{1}{r|}{{\small null hypothesis}} & ${\small 1\%}$ &
${\small 5\%}$ & ${\small 10\%}$ & ${\small 1\%}$ & ${\small 5\%}$ &
${\small 10\%}$ & ${\small 1\%}$ & ${\small 5\%}$ & ${\small 10\%}$\\\hline
&  &  &  &  &  &  &  &  & \\
{\small No Unob. Het.} & \multicolumn{1}{|l}{{\small 0.014}} &
\multicolumn{1}{l}{{\small 0.057}} & \multicolumn{1}{l|}{{\small 0.117}} &
\multicolumn{1}{|l}{{\small 0.031}} & \multicolumn{1}{l}{{\small 0.088}} &
\multicolumn{1}{l|}{{\small 0.163}} & \multicolumn{1}{|l}{{\small 0.032}} &
\multicolumn{1}{l}{{\small 0.121}} & \multicolumn{1}{l}{{\small 0.187}}\\
&  &  &  &  &  &  &  &  & \\
{\small Two types} & \multicolumn{1}{|l}{{\small 0.014}} &
\multicolumn{1}{l}{{\small 0.051}} & \multicolumn{1}{l|}{{\small 0.104}} &
\multicolumn{1}{|l}{{\small 0.008}} & \multicolumn{1}{l}{{\small 0.053}} &
\multicolumn{1}{l|}{{\small 0.100}} & \multicolumn{1}{|l}{{\small 0.009}} &
\multicolumn{1}{l}{{\small 0.065}} & \multicolumn{1}{l}{{\small 0.115}}\\
&  &  &  &  &  &  &  &  & \\\hline\hline
\multicolumn{10}{l}{{\footnotesize Note (1): Mean, median, and standard
deviation of estimated parameter over the 1,000 replications.}}%
\end{tabular}

\end{center}

Table 7 deals with DGP 3, that has also two types of replacement costs, but
now these types are very similar: $RC_{1}=8$ and $RC_{2}=9$, with equal
probabilities. The main purpose of the experiments with this DGP is to
investigate the bias of the \textit{MLE-noUH }and the power of this Hausman
test in an scenario with a very modest amount of unobserved heterogeneity.
Even in this scenario, for samples B and C, the bias of the \textit{MLE-noUH
}is approximately $5\%$ of the true value of the parameter, and the Hausman
test rejects the null hypothesis of no unobserved heterogeneity with
probability that is more than twice the nominal size of the test.

Finally, Table 8 presents results of experiments under DGP 4 where there is
not unobserved heterogeneity and $RC=8$. The purpose of these experiments is
to study possible biases in the size of Hausman test for the null hypothesis
of no unobserved heterogeneity. We can see that, for the three samples, the
size of this test is very close to the nominal size.

\begin{center}%
\begin{tabular}
[c]{r|ccc|ccc|ccc}\hline\hline
\multicolumn{10}{c}{\textbf{Table 8}}\\
\multicolumn{10}{c}{\textbf{Monte Carlo Experiments with DGP 4 (No UH, RC =
8)}}\\\hline
&  &  &  &  &  &  &  &  & \\
& \multicolumn{3}{|c|}{\textbf{Sample A} {\small (}${\small t=1}${\small to
}${\small 7}${\small )}} & \multicolumn{3}{|c|}{\textbf{Sample B}
{\small (}${\small t=1}${\small to }${\small 14}${\small )}} &
\multicolumn{3}{|c}{\textbf{Sample C} {\small (}${\small t=8}${\small to
}${\small 21}${\small )}}\\
{\small Estimator} & \multicolumn{3}{|c}{{\small Estimate}$^{(1)}$} &
\multicolumn{3}{|c|}{{\small Estimate}$^{(1)}$} &
\multicolumn{3}{|c}{{\small Estimate}$^{(1)}$}\\
{\small of }${\small \beta}$ & {\small Mean} & {\small Median} & {\small St.
dev.} & {\small Mean} & {\small Median} & {\small St. dev.} & {\small Mean} &
{\small Median} & {\small St. dev.}\\\hline
\multicolumn{1}{c|}{} &  &  &  &  &  &  &  &  & \\
{\small CMLE-true-d*} & \multicolumn{1}{|l}{{\small 1.0030}} &
\multicolumn{1}{l}{{\small 1.0029}} & \multicolumn{1}{l|}{{\small 0.1237}} &
\multicolumn{1}{|l}{{\small 0.9979}} & \multicolumn{1}{l}{{\small 0.9942}} &
\multicolumn{1}{l|}{{\small 0.0660}} & \multicolumn{1}{|l}{{\small 0.9994}} &
\multicolumn{1}{l}{{\small 0.9994}} & \multicolumn{1}{l}{{\small 0.0660}}\\
&  &  &  &  &  &  &  &  & \\
{\small CMLE-BIC-d*} & {\small 1.0030} & {\small 1.0029} & {\small 0.1237} &
{\small 0.9900} & {\small 0.9937} & {\small 0.1140} & {\small 0.9889} &
{\small 0.9986} & {\small 0.1201}\\
&  &  &  &  &  &  &  &  & \\
{\small MLE-2types} & \multicolumn{1}{|l}{{\small 1.0203}} &
\multicolumn{1}{l}{{\small 1.0156}} & \multicolumn{1}{l|}{{\small 0.0513}} &
\multicolumn{1}{|l}{{\small 1.0070}} & \multicolumn{1}{l}{{\small 1.0063}} &
\multicolumn{1}{l|}{{\small 0.0312}} & \multicolumn{1}{|l}{{\small 1.0079}} &
\multicolumn{1}{l}{{\small 1.0061}} & \multicolumn{1}{l}{{\small 0.0318}}\\
&  &  &  &  &  &  &  &  & \\
{\small MLE-noUH} & \multicolumn{1}{|l}{{\small 1.0011}} &
\multicolumn{1}{l}{{\small 1.0004}} & \multicolumn{1}{l|}{{\small 0.0414}} &
\multicolumn{1}{|l}{{\small 1.0001}} & \multicolumn{1}{l}{{\small 0.9990}} &
\multicolumn{1}{l|}{{\small 0.0293}} & \multicolumn{1}{|l}{{\small 1.0017}} &
\multicolumn{1}{l}{{\small 1.0005}} & \multicolumn{1}{l}{{\small 0.0302}}\\
&  &  &  &  &  &  &  &  & \\\hline
& \multicolumn{3}{|c}{{\small Frequency of Ho rejection}} &
\multicolumn{3}{|c|}{{\small Frequency of Ho rejection}} &
\multicolumn{3}{|c}{{\small Frequency of Ho rejection}}\\
\multicolumn{1}{r|}{{\small Testing}} & \multicolumn{3}{|c}{{\small with
significance level}} & \multicolumn{3}{|c|}{{\small with significance level}}
& \multicolumn{3}{|c}{{\small with significance level}}\\
\multicolumn{1}{r|}{{\small null hypothesis}} & ${\small 1\%}$ &
${\small 5\%}$ & ${\small 10\%}$ & ${\small 1\%}$ & ${\small 5\%}$ &
${\small 10\%}$ & ${\small 1\%}$ & ${\small 5\%}$ & ${\small 10\%}$\\\hline
&  &  &  &  &  &  &  &  & \\
{\small No Unob. Het.} & \multicolumn{1}{|l}{{\small 0.007}} &
\multicolumn{1}{l}{{\small 0.045}} & \multicolumn{1}{l|}{{\small 0.094}} &
\multicolumn{1}{|l}{{\small 0.009}} & \multicolumn{1}{l}{{\small 0.05}} &
\multicolumn{1}{l|}{{\small 0.097}} & \multicolumn{1}{|l}{{\small 0.014}} &
\multicolumn{1}{l}{{\small 0.052}} & \multicolumn{1}{l}{{\small 0.108}}\\
&  &  &  &  &  &  &  &  & \\
{\small Two types} & \multicolumn{1}{|l}{{\small 0.008}} &
\multicolumn{1}{l}{{\small 0.056}} & \multicolumn{1}{l|}{{\small 0.104}} &
\multicolumn{1}{|l}{{\small 0.012}} & \multicolumn{1}{l}{{\small 0.063}} &
\multicolumn{1}{l|}{{\small 0.109}} & \multicolumn{1}{|l}{{\small 0.019}} &
\multicolumn{1}{l}{{\small 0.053}} & \multicolumn{1}{l}{{\small 0.107}}\\
&  &  &  &  &  &  &  &  & \\\hline\hline
\multicolumn{10}{l}{{\footnotesize Note (1): Mean, median, and standard
deviation of estimated parameter over the 1,000 replications.}}%
\end{tabular}

\end{center}

\subsection{Estimation using Rust's dataset}

Rust's full sample contains a total of 124 buses that are classified in eight
groups according to the bus size and the engine manufacturer. For the
estimation of the structural model, Rust focuses on groups 1 to 4 that account
for $104$ buses. For every bus, the choice history in the data starts with the
actual initial condition of the engine, i.e., the first month where the engine
was installed. For these $104$ buses, the distribution of the number of engine
replacements is the following: $0$ engine replacements for $45$ buses; $1$
replacement for $58$ buses; and $2$ replacements for $1$ bus. For the
implementation of our FE-CMLE, choice histories with zero replacements do not
contain any useful information. Therefore, for the CMLE we use only $59$
buses. For our analysis, we consider that the frequency of the
superintendent's decisions is at the annual level. Table 9 presents the
empirical distribution of choice histories for the $59$ buses with at least
one engine replacement, of which $27$ are observed during 6 years, and $32$
over 10 years. 

\medskip

\begin{center}%
\begin{tabular}
[c]{c|c|c|c}\hline\hline
\multicolumn{4}{c}{\textbf{Table 9}}\\
\multicolumn{4}{c}{\textbf{Bus Engine Replacement (Rust, 1987)}}\\\hline
\multicolumn{4}{c}{\textbf{Empirical distribution of choice histories}}\\
\multicolumn{4}{c}{\textbf{with replacement}}\\\hline
& \multicolumn{3}{|c}{\textit{Frequency}}\\
\textit{Choice history} & \textit{Absolute} & \textit{\%} & \textit{\%
cumulative}\\\hline
&  &  & \\
\multicolumn{1}{r|}{110111} & 2 & 3.39 & 3.39\\
\multicolumn{1}{r|}{111011} & 7 & 11.86 & 15.25\\
\multicolumn{1}{r|}{111101} & 7 & 11.86 & 27.12\\
\multicolumn{1}{r|}{111110} & 11 & 1864 & 45.76\\
\multicolumn{1}{r|}{} &  &  & \\
\multicolumn{1}{r|}{1101111111} & 1 & 1.69 & 47.46\\
\multicolumn{1}{r|}{1110111111} & 4 & 6.78 & 54.24\\
\multicolumn{1}{r|}{1111011111} & 2 & 3.39 & 57.63\\
\multicolumn{1}{r|}{1111101111} & 7 & 11.86 & 69.49\\
\multicolumn{1}{r|}{1111110111} & 7 & 11.86 & 81.35\\
\multicolumn{1}{r|}{1111111011} & 5 & 8.47 & 89.83\\
\multicolumn{1}{r|}{1111111101} & 3 & 5.08 & 94.91\\
\multicolumn{1}{r|}{1111111110} & 2 & 3.39 & 98.30\\
&  &  & \\
\multicolumn{1}{r|}{1101110111} & 1 & 1.69 & 100.00\\
\multicolumn{1}{r|}{} &  &  & \\\hline
\multicolumn{1}{r|}{\textit{Total}} & 59 & 100.00 & \\
&  &  & \\\hline\hline
\end{tabular}

\end{center}

\medskip

Table 10 presents ML estimates of the model with three different
specifications of the maintenance cost function $\beta_{d}(d)$ according to:
the value of the parameter $d^{\ast}$ (at which function $\beta_{d}(d)$
becomes flat); and the functional for durations smaller than $d^{\ast}$, i.e.,
linear, quadratic, and square-root. We report estimates of the replacement
cost parameter and of the parameter $\beta_{d}^{\ast}\equiv\beta_{d}(d^{\ast
})-\beta_{d}(d^{\ast}-1)$. We consider a model with two unobserved types.
However, for all the specifications, we always converge to a model with a
single type. We have tried thousands of initial values for the vector of
parameters (i.e., $RC_{1}$, $RC_{2}$, $\lambda$, and $\beta_{d}$), and we have
also estimated the model using grid search. Regardless the computational
strategy, we always converge to the same estimate with only one type. The
specification of the function $\beta_{d}(d)$ that provides the maximum value
of the likelihood function is the the square-root function with a value
$d^{\ast}$ equal to six. For this specification, the estimate of the
replacement cost parameter is $\widehat{RC}=10.8566$ ($s.e.=1.5247$), and the
estimate of the parameter of $\beta_{d}^{\ast}$ is $\widehat{\beta_{d}^{\ast}%
}=0.3054$ ($s.e.=0.0496$).

\medskip

\begin{center}%
\begin{tabular}
[c]{ccccccc}\hline\hline
\multicolumn{7}{c}{\textbf{Table 10}}\\
\multicolumn{7}{c}{\textbf{Bus Engine Replacement (Rust, 1987)}}\\\hline
&  &  &  &  &  & \\
\multicolumn{7}{c}{\textbf{Maximum Likelihood Estimates}}\\\hline
\multicolumn{2}{c}{\textit{Model}} & \multicolumn{2}{|c}{$RC$} &
\multicolumn{2}{|c}{$\beta_{d}^{\ast}\equiv-\Delta\beta_{d}(d^{\ast})$} & \\
$\beta_{d}(d)$ & $d^{\ast}$ & \multicolumn{1}{|c}{$\widehat{RC}$} & $se\left(
\widehat{RC}\right)  $ & \multicolumn{1}{|c}{$\widehat{\beta_{d}^{\ast}}$} &
$se\left(  \widehat{\beta_{d}^{\ast}}\right)  $ & $\log$\textit{-likelihood}%
\\\hline
&  & \multicolumn{1}{|c}{} &  & \multicolumn{1}{|c}{} &  & \\
\textit{Square root} & 3 & \multicolumn{1}{|l}{28.2218} &
\multicolumn{1}{l}{6.9110} & \multicolumn{1}{|l}{2.0110} &
\multicolumn{1}{l}{0.5149} & \multicolumn{1}{l}{-162.7081}\\
$\beta_{d}(d)=\beta\sqrt{d}$ & 4 & \multicolumn{1}{|l}{16.5364} &
\multicolumn{1}{l}{3.0438} & \multicolumn{1}{|l}{0.7777} &
\multicolumn{1}{l}{0.1546} & \multicolumn{1}{l}{{\small -}160.7515}\\
& 5 & \multicolumn{1}{|l}{12.8403} & \multicolumn{1}{l}{1.9959} &
\multicolumn{1}{|l}{0.4486} & \multicolumn{1}{l}{0.0774} &
\multicolumn{1}{l}{-158.5760}\\
& \textbf{6} & \multicolumn{1}{|l}{\textbf{10.8566}} &
\multicolumn{1}{l}{\textbf{1.5247}} & \multicolumn{1}{|l}{\textbf{0.3054}} &
\multicolumn{1}{l}{\textbf{0.0496}} & \multicolumn{1}{l}{\textbf{-158.2108}%
$^{\ast\ast}$}\\
& 7 & \multicolumn{1}{|l}{9.6817} & \multicolumn{1}{l}{1.2821} &
\multicolumn{1}{|l}{0.2317} & \multicolumn{1}{l}{0.0372} &
\multicolumn{1}{l}{-158.7021}\\
& 8 & \multicolumn{1}{|l}{8.9953} & \multicolumn{1}{l}{1.1623} &
\multicolumn{1}{|l}{0.1909} & \multicolumn{1}{l}{0.0313} &
\multicolumn{1}{l}{-159.4693}\\
& 9 & \multicolumn{1}{|l}{8.6517} & \multicolumn{1}{l}{1.1183} &
\multicolumn{1}{|l}{0.1682} & \multicolumn{1}{l}{0.0285} &
\multicolumn{1}{l}{-160.0868}\\
&  & \multicolumn{1}{|c}{} &  & \multicolumn{1}{|c}{} &  & \\\hline
&  & \multicolumn{1}{|c}{} &  & \multicolumn{1}{|c}{} &  & \\
\textit{Linear} & 3 & \multicolumn{1}{|l}{18.2995} &
\multicolumn{1}{l}{4.1695} & \multicolumn{1}{|l}{2.0388} &
\multicolumn{1}{l}{0.4977} & \multicolumn{1}{l}{-162.7529}\\
$\beta_{d}(d)=\beta$ $d$ & 4 & \multicolumn{1}{|l}{11.4552} &
\multicolumn{1}{l}{1.9053} & \multicolumn{1}{|l}{0.8418} &
\multicolumn{1}{l}{0.1566} & \multicolumn{1}{l}{-160.9650}\\
& 5 & \multicolumn{1}{|l}{9.2473} & \multicolumn{1}{l}{1.2769} &
\multicolumn{1}{|l}{0.5103} & \multicolumn{1}{l}{0.0817} &
\multicolumn{1}{l}{-158.8536}\\
& \textbf{6} & \multicolumn{1}{|l}{\textbf{7.9817}} &
\multicolumn{1}{l}{\textbf{0.9809}} & \multicolumn{1}{|l}{\textbf{0.3623}} &
\multicolumn{1}{l}{\textbf{0.0548}} & \multicolumn{1}{l}{\textbf{-158.8132}}\\
& 7 & \multicolumn{1}{|l}{7.1859} & \multicolumn{1}{l}{0.8219} &
\multicolumn{1}{|l}{0.2856} & \multicolumn{1}{l}{0.0434} &
\multicolumn{1}{l}{-159.7641}\\
& 8 & \multicolumn{1}{|l}{6.7030} & \multicolumn{1}{l}{0.7411} &
\multicolumn{1}{|l}{0.2448} & \multicolumn{1}{l}{0.0388} &
\multicolumn{1}{l}{-160.9912}\\
& 9 & \multicolumn{1}{|l}{6.4612} & \multicolumn{1}{l}{0.7114} &
\multicolumn{1}{|l}{0.2259} & \multicolumn{1}{l}{0.0379} &
\multicolumn{1}{l}{-161.9368}\\
&  & \multicolumn{1}{|c}{} &  & \multicolumn{1}{|c}{} &  & \\\hline
&  & \multicolumn{1}{|c}{} &  & \multicolumn{1}{|c}{} &  & \\
\textit{Square} & 3 & \multicolumn{1}{|l}{13.1481} &
\multicolumn{1}{l}{2.7300} & \multicolumn{1}{|l}{2.1006} &
\multicolumn{1}{l}{0.4804} & \multicolumn{1}{l}{-162.8699}\\
$\beta_{d}(d)=\beta$ $d^{2}$ & 4 & \multicolumn{1}{|l}{8.7707} &
\multicolumn{1}{l}{1.2806} & \multicolumn{1}{|l}{0.9603} &
\multicolumn{1}{l}{0.1628} & \multicolumn{1}{l}{-161.4943}\\
& \textbf{5} & \multicolumn{1}{|l}{\textbf{7.3081}} &
\multicolumn{1}{l}{\textbf{0.8850}} & \multicolumn{1}{|l}{\textbf{0.6257}} &
\multicolumn{1}{l}{\textbf{0.0921}} & \multicolumn{1}{l}{\textbf{-159.4992}}\\
& 6 & \multicolumn{1}{|l}{6.3777} & \multicolumn{1}{l}{0.6844} &
\multicolumn{1}{|l}{0.4709} & \multicolumn{1}{l}{0.0673} &
\multicolumn{1}{l}{-160.0882}\\
& 7 & \multicolumn{1}{|l}{5.7404} & \multicolumn{1}{l}{0.5689} &
\multicolumn{1}{|l}{0.3905} & \multicolumn{1}{l}{0.0583} &
\multicolumn{1}{l}{-161.9366}\\
& 8 & \multicolumn{1}{|l}{5.3323} & \multicolumn{1}{l}{0.5072} &
\multicolumn{1}{|l}{0.3535} & \multicolumn{1}{l}{0.0578} &
\multicolumn{1}{l}{-164.0680}\\
& 9 & \multicolumn{1}{|l}{5.1227} & \multicolumn{1}{l}{0.4837} &
\multicolumn{1}{|l}{0.3515} & \multicolumn{1}{l}{0.0636} &
\multicolumn{1}{l}{-165.6751}\\
&  & \multicolumn{1}{|c}{} &  & \multicolumn{1}{|c}{} &  & \\\hline\hline
\end{tabular}

\end{center}

\medskip

Table 11 presents estimates of the parameter $\beta_{d}^{\ast}\equiv\beta
_{d}(d^{\ast})-\beta_{d}(d^{\ast}-1)$ using the CMLE\ and under different
values of $d^{\ast}$. Given the observed histories in this dataset (as shown
in Table 9), the parameter $\beta_{d}^{\ast}$ is identified only under two
possible values of $d^{\ast}\,$: $d^{\ast}=3$ and $d^{\ast}=4$.\footnote{To
identify $\beta_{d}^{\ast}$ for $d^{\ast}=2$, we need histories with a
replacement when duration is equal to $1$ ($d^{\ast}-1$). To identify
$\beta_{d}^{\ast}$ when $d^{\ast}\geq5$, we need histories with at least 5
years without replacement both before and after an observed replacement. In
this small sample, we do not observe these types of histories.} We report the
value of the concentrated log-likelihood function and of the BIC function.
According to the BIC\ function, the estimate of $d^{\ast}$ is $\widehat
{d^{\ast}}=3$, and the corresponding estimator of $\beta_{d}^{\ast}$ is
$\widehat{\beta_{d}^{\ast}}=1.7009$ (s.e. = 1.0244). Note also that for
$d^{\ast}=3$, the estimate of $\beta_{d}^{\ast}$ is significantly different to
zero for a significance level of 10\% parameter (p-value = 0.0968). In
contrast, for $d^{\ast}=4$, this parameter is not significantly different to
zero for any standard significance level (p-value = 0.8446). Therefore, the
estimate $\widehat{d^{\ast}}=3$ and $\widehat{\beta_{d}^{\ast}}=1.7009$ is
consistent with the definition of $d^{\ast}$ as the maximum duration with
$\beta_{d}(d)-\beta_{d}(d-1)$ different to zero.

\begin{center}%
\begin{tabular}
[c]{c|cc|c|cc}\hline\hline
\multicolumn{6}{c}{\textbf{Table 11}}\\
\multicolumn{6}{c}{\textbf{Bus Engine Replacement (Rust, 1987)}}\\\hline
\multicolumn{6}{c}{}\\
\multicolumn{6}{c}{\textbf{Fixed-Effects-Conditional Maximum Likelihood}%
}\\\hline
& \multicolumn{2}{|c|}{$\beta_{d}^{\ast}$} & \textit{p-value} &
\textit{concentrated} & \\
$d^{\ast}$ & $\widehat{\beta_{d}^{\ast}}$ & $se\left(  \widehat{\beta
_{d}^{\ast}}\right)  $ & $H_{0}:\beta_{d}^{\ast}=0$ & \textit{log-likelihood}
& \textit{BIC(d}$^{\ast}$\textit{)}\\\hline
&  &  &  &  & \\
\textbf{3} & \textbf{1.7009} & \textbf{1.0244} & \textbf{0.0968} & \textbf{
-102.1215} & \textbf{ -108.2378}\\
&  &  &  &  & \\
4 & 0.1178 & 0.6009 & 0.8446 & -102.1020 & -110.2571\\
&  &  &  &  & \\\hline\hline
\end{tabular}

\end{center}

\medskip

Table 12 compares the CMLE estimate of the parameter $\beta_{d}^{\ast}$ with
the corresponding MLE using the estimates in Table 10. Given the very small
sample size and the corresponding large standard error of the CMLE estimates,
we cannot reject the null hypothesis of no unobserved heterogeneity, despite
the magnitude of the difference between MLE and CMLE estimates is important
and it generates important differences in distribution of durations.

\medskip

\begin{center}%
\begin{tabular}
[c]{ccccc}\hline\hline
\multicolumn{5}{c}{\textbf{Table 12}}\\
\multicolumn{5}{c}{\textbf{Bus Engine Replacement (Rust, 1987)}}\\\hline
&  &  &  & \\
\multicolumn{5}{c}{\textbf{Hausman Test of Unobserved Heterogeneity}}\\\hline
& \multicolumn{1}{|c}{$\widehat{\beta_{d}^{\ast}}$ $\left(  se\right)  $} &
\multicolumn{1}{|c}{$\widehat{\beta_{d}^{\ast}}$ $\left(  se\right)  $} &
\multicolumn{1}{|c}{} & \multicolumn{1}{|c}{}\\
\textit{Model} & \multicolumn{1}{|c}{\textit{MLE}} &
\multicolumn{1}{|c}{\textit{CMLE}} & \multicolumn{1}{|c}{\textit{Hausman}} &
\multicolumn{1}{|c}{\textit{p-value}}\\\hline
& \multicolumn{1}{|c}{} & \multicolumn{1}{|c}{} & \multicolumn{1}{|c}{} &
\multicolumn{1}{|c}{}\\
Square root & \multicolumn{1}{|c}{0.4548 (0.0739)} &
\multicolumn{1}{|c}{1.7009 (1.0244)} & \multicolumn{1}{|c}{1.4873} &
\multicolumn{1}{|c}{0.2226}\\
& \multicolumn{1}{|c}{} & \multicolumn{1}{|c}{} & \multicolumn{1}{|c}{} &
\multicolumn{1}{|c}{}\\
Linear & \multicolumn{1}{|c}{0.3623 (0.0549)} & \multicolumn{1}{|c}{1.7009
(1.0244)} & \multicolumn{1}{|c}{1.7123} & \multicolumn{1}{|c}{0.1907}\\
& \multicolumn{1}{|c}{} & \multicolumn{1}{|c}{} & \multicolumn{1}{|c}{} &
\multicolumn{1}{|c}{}\\
Square & \multicolumn{1}{|c}{0.3476 (0.0512)} & \multicolumn{1}{|c}{1.7009
(1.0244)} & \multicolumn{1}{|c}{1.7494} & \multicolumn{1}{|c}{0.186}\\
& \multicolumn{1}{|c}{} & \multicolumn{1}{|c}{} & \multicolumn{1}{|c}{} &
\multicolumn{1}{|c}{}\\\hline\hline
\end{tabular}

\end{center}

\newpage

\section{Conclusions}

This paper presents the first identification results of structural parameters
in forward-looking dynamic discrete choice models where the joint distribution
of time-invariant unobserved heterogeneity and endogenous state variables is
nonparametrically specified. This unobserved heterogeneity can have multiple
components and can have continuous support. The dependence between the
unobserved heterogeneity and the initial values of the state variables is also
unrestricted. We consider models with two endogenous state variables: the
lagged decision variable, and the time duration in the last choice. We show
that structural parameters that capture switching costs are identified under
mild conditions. The identification of structural parameters that capture
duration dependence require additional restrictions. In particular, to obtain
identification of these parameters we assume that the marginal return of an
additional period of experience (duration) becomes equal to zero after a
finite number of periods.

Based on our identification results, we propose tests for the validity of
restricted models without unobserved heterogeneity or with a parametric
specification of the correlated random effects. Our Monte Carlo experiments
show that the Conditional MLE provides precise estimates of structural
parameters and the specification test has strong power to reject misspecified
correlated random effects models.

\newpage

\noindent\textbf{Appendix 1. Proofs}

\medskip

\noindent\textbf{Proof of Lemma 2. }

\textbf{(i) }For any $y>0$, we have that $1\{y_{t-1}=y$, $d_{t}=0\}=0$ because
$y_{t-1}>0$ implies $d_{t}>0$. Therefore, $H^{(y)}(0)=$ $\sum_{t=1}%
^{T}1\{y_{t-1}=y$, $d_{t}=0\}=0$.

\textbf{(ii) }For any $y>0$, we have that $1\{y_{t-1}=y_{t}=y$, $d_{t}=0\}=0$
because $y_{t-1}>0$ implies $d_{t}>0$. Therefore, $X^{(y)}(0)=$ $\sum
_{t=1}^{T}1\{y_{t-1}=y_{t}=y,d_{t}=0\}=0$.

\textbf{(iii)}\textit{ }For any $y>0$, $\sum_{d\geq1}H^{(y)}(d)=$\textit{
}$\sum_{d\geq1}\sum_{t=1}^{T}1\{y_{t-1}=y$, $d_{t}=d\}=$ $\sum_{t=1}%
^{T}1\{y_{t-1}=y\}=$ $T^{(y)}+$\textit{ }$1\{y_{0}=y\}-$\textit{ }%
$1\{y_{T}=y\}$\textit{.}

\textbf{(iv)}\textit{ }For any $y>0$, $\sum_{d\geq1}X^{(y)}(d)=$\textit{
}$\sum_{d\geq1}\sum_{t=1}^{T}1\{y_{t-1}=y_{t}=y,d_{t}=d\}=$ $\sum_{t=1}%
^{T}1\{y_{t-1}=y_{t}=y\}=$ $T^{(y)}+$\textit{ }$1\{y_{0}=y\}-$\textit{
}$1\{y_{T}=y\}$\textit{.}

\textbf{(v)}\textit{ }First, note that $y_{t-1}=y>0$ implies $d_{t}\geq1$.
Therefore, for any $y>0$ and $d\geq1$, the event $\{y_{t-1}=y_{t}=y,d_{t}=d\}$
is equivalent to the event $\{y_{t}=y,$ $d_{t+1}=d+1\}$ for any $1\leq t\leq
T$. Therefore, $X^{(y)}(d)=$ $\sum_{t=1}^{T}\{y_{t}=y,d_{t+1}=d+1\}=$
$\sum_{t=2}^{T+1}1\{y_{t-1}=y,d_{t}=d+1\}=$ $H^{(y)}(d+1)-$ $1\{y_{0}%
=y,d_{1}=d+1\}+$ $1\{y_{T}=y,d_{T+1}=d+1\}$.\qquad$\blacksquare$

\bigskip

\noindent\textbf{Proof of Propositions 1 and 2. }Remember that $T^{(y)}$
represents the number of times that choice alternative $y$ is visited in the
choice history $\widetilde{\mathbf{y}}$, and $D^{(y)}$ is the number of times
that choice alternative $y$ is observed at two consecutive periods over the
history $(y_{0},\widetilde{\mathbf{y}})$. For the binary choice model, we have
that $%
%TCIMACRO{\tsum \nolimits_{t=1}^{T}}%
%BeginExpansion
{\textstyle\sum\nolimits_{t=1}^{T}}
%EndExpansion
y_{t}=T^{(1)}$, $%
%TCIMACRO{\tsum \nolimits_{t=1}^{T}}%
%BeginExpansion
{\textstyle\sum\nolimits_{t=1}^{T}}
%EndExpansion
y_{t-1}y_{t}=D^{(1,1)}$, and $y_{T}-y_{0}=\Delta^{(1)}$.%
\begin{equation}%
\begin{array}
[c]{ccl}%
\ln\mathbb{P}\left(  \widetilde{\mathbf{y}}\text{ }|\text{ }y_{0}%
,\mathbf{\theta}\right)  & = & T^{(1)}\left[  \widetilde{\alpha}%
_{\mathbf{\theta}}+\sigma_{\mathbf{\theta}}(1)-\sigma_{\mathbf{\theta}%
}(0)\right]  +\Delta^{(1)}\text{ }\left[  \sigma_{\mathbf{\theta}}%
(0)-\sigma_{\mathbf{\theta}}(1)\right]  +\widetilde{\beta}_{y}\text{
}D^{(1,1)}%
\end{array}
\tag{A.1}%
\end{equation}
where we have omitted the term $T$ $\sigma_{\mathbf{\theta}}(0)$ because $T$
is constant over all the histories. Consider choice histories $A=\{0|0,1,1\}$
and $B=\{0|1,0,1\}$. It is clear that $T_{A}^{(1)}=T_{B}^{(1)}=2$, and
$\Delta_{A}^{(1)}=$ $\Delta_{B}^{(1)}=$ $1$, such that $U_{A}=U_{B}$. Also,
$D_{A}^{(1,1)}=1$ and $D_{B}^{(1,1)}=0$. Therefore, $\ln\mathbb{P}%
(A|U)-\ln\mathbb{P}(B|U)=\widetilde{\beta}_{y}$.\qquad$\blacksquare$

\bigskip

\noindent\textbf{Proof of Proposition 3. }The log-probability of this model
is:%
\begin{equation}%
\begin{array}
[c]{ccl}%
\ln\mathbb{P}\left(  \widetilde{\mathbf{y}}\text{ }|\text{ }y_{0}%
,d_{1},\mathbf{\theta}\right)  & = & \dsum\limits_{t=1}^{T}y_{t}\left[
\widetilde{\alpha}_{\mathbf{\theta}}+\widetilde{\beta}_{y}\text{ }%
y_{t-1}+\beta_{d}(1,d_{t})\text{ }y_{t-1}\right]  +\sigma_{\mathbf{\theta}%
}(y_{t-1},d_{t})
\end{array}
\tag{A.2}%
\end{equation}
We have that $\ln\mathbb{P}\left(  \widetilde{\mathbf{y}}\text{ }|\text{
}y_{0},d_{1},\mathbf{\theta}\right)  =$ $T^{(1)}\widetilde{\alpha
}_{\mathbf{\theta}}+$ $[T^{(0)}-\Delta^{(0)}]\sigma_{\mathbf{\theta}}(0)+$
$D^{(1,1)}\widetilde{\beta}_{y}+$ $\sum\nolimits_{d\geq1}X^{(1)}(d)$
$\beta_{d}(1,d)+$ $\sum\nolimits_{d\geq1}H^{(1)}(d)$ $\sigma_{\mathbf{\theta}%
}(1,d)$. Taking into account that $\sum_{d\geq1}H^{(1)}(d)=T^{(1)}%
-\Delta^{(1)}$ and $D^{(1,1)}=\sum_{d\geq1}X^{(1)}(d)$, we obtain:%
\begin{equation}%
\begin{array}
[c]{ccl}%
\ln\mathbb{P}\left(  \widetilde{\mathbf{y}}|y_{0},\mathbf{\theta}\right)  &
= & \sum\limits_{d\geq1}H^{(1)}(d)\text{ }\left[  \widetilde{\alpha
}_{\mathbf{\theta}}+\sigma_{\mathbf{\theta}}(1,d)-\sigma_{\mathbf{\theta}%
}(0)\right]  +\Delta^{(1)}\widetilde{\alpha}_{\mathbf{\theta}}\\
&  & \\
& + & \sum\limits_{d\geq1}X^{(1)}(d)\text{ }\gamma(d)
\end{array}
\tag{A.3}%
\end{equation}
where we have omitted the term $T$ $\sigma_{\mathbf{\theta}}(0)$ because $T$
is constant over all the histories, and we define $\gamma(d)\equiv
\widetilde{\beta}_{y}+\beta_{d}(1,d)$. Now, Lemma 2(v) establishes that
$X^{(1)}(d)=$ $H^{(1)}(d+1)+\Delta^{(1)}(d+1)$. Note that $%
%TCIMACRO{\tsum \nolimits_{d\geq1}}%
%BeginExpansion
{\textstyle\sum\nolimits_{d\geq1}}
%EndExpansion
\left[  H^{(1)}(d+1)+\Delta^{(1)}(d+1)\right]  $ $\gamma(d)$ is equal to $%
%TCIMACRO{\tsum \nolimits_{d\geq1}}%
%BeginExpansion
{\textstyle\sum\nolimits_{d\geq1}}
%EndExpansion
\left[  H^{(1)}(d)+\Delta^{(1)}(d)\right]  $ $\gamma(d-1)$, if we define
$\gamma(0)=0$. Then, we have that,%
\begin{equation}%
\begin{array}
[c]{ccl}%
\ln\mathbb{P}\left(  \widetilde{\mathbf{y}}|y_{0},\mathbf{\theta}\right)  &
= & \sum\limits_{d\geq1}H^{(1)}(d)\text{ }\left[  \widetilde{\alpha
}_{\mathbf{\theta}}+\sigma_{\mathbf{\theta}}(1,d)-\sigma_{\mathbf{\theta}%
}(0)\right]  +\Delta^{(1)}\widetilde{\alpha}_{\mathbf{\theta}}\\
&  & \\
& + & \sum\limits_{d\geq1}\left[  H^{(1)}(d)+\Delta^{(1)}(d)\right]  \text{
}\gamma(d-1)\\
&  & \\
& = & \sum\limits_{d\geq1}H^{(1)}(d)\text{ }\left[  \widetilde{\alpha
}_{\mathbf{\theta}}+\sigma_{\mathbf{\theta}}(1,d)-\sigma_{\mathbf{\theta}%
}(0)+\gamma(d-1)\right]  +\Delta^{(1)}\widetilde{\alpha}_{\mathbf{\theta}}\\
&  & \\
& + & \sum\limits_{d\geq1}\Delta^{(1)}(d)\text{ }\gamma(d-1)\qquad\blacksquare
\end{array}
\tag{A.4}%
\end{equation}

\bigskip

\noindent\textbf{Proof of Proposition 4. }The log-probability of this model
is:%
\begin{equation}%
\begin{array}
[c]{ccl}%
\ln\mathbb{P}\left(  \widetilde{\mathbf{y}}\text{ }|\text{ }y_{0}%
,d_{1},\mathbf{\theta}\right)  & = & \dsum\limits_{t=1}^{T}y_{t}\left[
\widetilde{\alpha}_{\mathbf{\theta}}+\widetilde{\beta}_{y}y_{t-1}+\beta
_{d}(1,d_{t})y_{t-1}+v_{\mathbf{\theta}}\left(  1,d_{t}+1\right)  \right]
+\sigma_{\mathbf{\theta}}(y_{t-1},d_{t})
\end{array}
\tag{A.5}%
\end{equation}
Comparing this log-probability with the one for the myopic model with
duration, we can see that the only difference is in the term $\sum_{t=1}%
^{T}y_{t}$ $v_{\mathbf{\theta}}\left(  1,d_{t}+1\right)  $, that can be
written as $\sum_{_{d\geq0}}\sum_{t=1}^{T}y_{t}$ $1\{d_{t}=d\}$
$v_{\mathbf{\theta}}\left(  1,d+1\right)  $. For the statistic $\sum_{t=1}%
^{T}y_{t}$ $1\{d_{t}=d\}$ we can distinguish two cases: (a) if $d=0$, then
$\sum_{t=1}^{T}y_{t}$ $1\{d_{t}=0\}=$ $\sum_{t=1}^{T}y_{t}$ $(1-y_{t-1})=$
$T^{(1)}-D^{(1,1)}$; and (b) if $d\geq1$, then $\sum_{t=1}^{T}y_{t}$
$1\{d_{t}=d\}=$ $\sum_{t=1}^{T}y_{t}$ $y_{t-1}$ $1\{d_{t}=d\}=$ $X^{(1)}(d)$.
Therefore,%
\begin{equation}%
\begin{array}
[c]{ccl}%
%TCIMACRO{\tsum \limits_{_{d\geq0}}}%
%BeginExpansion
{\textstyle\sum\limits_{_{d\geq0}}}
%EndExpansion%
%TCIMACRO{\tsum \limits_{t=1}^{T}}%
%BeginExpansion
{\textstyle\sum\limits_{t=1}^{T}}
%EndExpansion
y_{t}1\{d_{t}=d\}v_{\mathbf{\theta}}\left(  1,d+1\right)  & = & \left[
T^{(1)}-D^{(1,1)}\right]  \text{ }v_{\mathbf{\theta}}\left(  1,1\right)  +%
%TCIMACRO{\tsum \limits_{_{d\geq1}}}%
%BeginExpansion
{\textstyle\sum\limits_{_{d\geq1}}}
%EndExpansion
X^{(1)}(d)\text{ }v_{\mathbf{\theta}}\left(  1,d+1\right) \\
&  & \\
& = & T^{(1)}\text{ }v_{\mathbf{\theta}}\left(  1,1\right)  +%
%TCIMACRO{\tsum \limits_{_{d\geq1}}}%
%BeginExpansion
{\textstyle\sum\limits_{_{d\geq1}}}
%EndExpansion
X^{(1)}(d)\text{ }\left[  v_{\mathbf{\theta}}\left(  1,d+1\right)
-v_{\mathbf{\theta}}\left(  1,1\right)  \right]
\end{array}
\tag{A.6}%
\end{equation}
where for the second equality we have applied Lemma 2(iv), $D^{(1,1)}%
=\sum_{d\geq1}X^{(1)}(d)$. Then, the log-probability is equal to%
\begin{equation}%
\begin{array}
[c]{ccl}%
\ln\mathbb{P}\left(  \widetilde{\mathbf{y}}|y_{0},\mathbf{\theta}\right)  &
= & \sum\limits_{d\geq1}H^{(1)}(d)\text{ }\left[  \widetilde{\alpha
}_{\mathbf{\theta}}+\sigma_{\mathbf{\theta}}(1,d)-\sigma_{\mathbf{\theta}%
}(0)+\gamma(d-1)\right]  +\Delta^{(1)}\widetilde{\alpha}_{\mathbf{\theta}}\\
&  & \\
& + & \sum\limits_{d\geq1}\Delta^{(1)}(d)\text{ }\gamma(d-1)\\
&  & \\
& + & T^{(1)}\text{ }v_{\mathbf{\theta}}\left(  1,1\right)  +\sum
\limits_{d\geq1}X^{(1)}(d)\text{ }\left[  v_{\mathbf{\theta}}\left(
1,d+1\right)  -v_{\mathbf{\theta}}\left(  1,1\right)  \right]
\end{array}
\tag{A.7}%
\end{equation}
From Lemma 2, we have that: (iii) $T^{(1)}=\sum_{d\geq1}H^{(1)}(d)+\Delta
^{(1)}$; and (v) $X^{(1)}(d)=$ $H^{(1)}(d+1)+$ $\Delta^{(1)}(d+1)$, and
solving these expressions in (A.13), we have that:%
\begin{equation}%
\begin{array}
[c]{ccl}%
\ln\mathbb{P}\left(  \widetilde{\mathbf{y}}|y_{0},\mathbf{\theta}\right)  &
= & \sum\limits_{d\geq1}H^{(1)}(d)\text{ }\left[  \widetilde{\alpha
}_{\mathbf{\theta}}+\sigma_{\mathbf{\theta}}(1,d)-\sigma_{\mathbf{\theta}%
}(0)+\gamma(d-1)+v_{\mathbf{\theta}}\left(  1,d\right)  \right] \\
&  & \\
& + & \Delta^{(1)}\text{ }\left[  \widetilde{\alpha}_{\mathbf{\theta}%
}+v_{\mathbf{\theta}}\left(  1,1\right)  \right] \\
&  & \\
& + & \sum\limits_{d\geq1}\Delta^{(1)}(d)\text{ }\left[  v_{\mathbf{\theta}%
}\left(  1,d\right)  -v_{\mathbf{\theta}}\left(  1,1\right)  +\gamma
(d-1)\right]
\end{array}
\tag{A.8}%
\end{equation}
Taking into account that $\Delta^{(1)}=%
%TCIMACRO{\tsum \nolimits_{d\geq1}}%
%BeginExpansion
{\textstyle\sum\nolimits_{d\geq1}}
%EndExpansion
\Delta^{(1)}(d)$, we have:%
\begin{equation}%
\begin{array}
[c]{ccl}%
\ln\mathbb{P}\left(  \widetilde{\mathbf{y}}|y_{0},\mathbf{\theta}\right)  &
= & \sum\limits_{d\geq1}H^{(1)}(d)\text{ }g_{\mathbf{\theta},1}(d)+\sum
\limits_{d\geq1}\Delta^{(1)}(d)\text{ }\left[  \widetilde{\alpha
}_{\mathbf{\theta}}+v_{\mathbf{\theta}}\left(  1,d\right)  +\gamma
(d-1)\right]
\end{array}
\tag{A.9}%
\end{equation}
with $g_{\mathbf{\theta},1}(d)\equiv\widetilde{\alpha}_{\mathbf{\theta}%
}+\sigma_{\mathbf{\theta}}(1,d)-\sigma_{\mathbf{\theta}}(0)+\gamma
(d-1)+v_{\mathbf{\theta}}\left(  1,d\right)  $\textit{.}$\qquad\blacksquare$

\bigskip

\noindent\textbf{Proof of Proposition 5. }Define $Z\equiv%
%TCIMACRO{\tsum \nolimits_{d\geq1}}%
%BeginExpansion
{\textstyle\sum\nolimits_{d\geq1}}
%EndExpansion
\Delta^{(1)}(d)$ $\left[  v_{\mathbf{\theta}}\left(  1,d\right)
+\gamma(d-1)\right]  $. Under Assumption 2, we have that $v_{\mathbf{\theta}%
}\left(  1,d\right)  =v_{\mathbf{\theta}}\left(  1,d^{\ast}\right)  $ for any
$d\geq d^{\ast}$, and $\gamma\left(  d-1\right)  =\gamma\left(  d^{\ast
}\right)  $ for any $d\geq d^{\ast}+1$. Therefore, we have:%
\begin{equation}%
\begin{array}
[c]{ccl}%
Z & = & \sum\limits_{d\leq d^{\ast}-1}\Delta^{(1)}(d)\text{ }v_{\mathbf{\theta
}}\left(  1,d\right)  +\left[  \sum\limits_{d\geq d^{\ast}}\Delta
^{(1)}(d)\right]  v_{\mathbf{\theta}}\left(  1,d^{\ast}\right) \\
&  & \\
& + & \sum\limits_{d\leq d^{\ast}}\Delta^{(1)}(d)\text{ }\gamma(d-1)+\left[
\sum\limits_{d\geq d^{\ast}+1}\Delta^{(1)}(d)\right]  \gamma(d^{\ast})\\
&  & \\
& = & \sum\limits_{d\leq d^{\ast}-1}\Delta^{(1)}(d)\text{ }\left[
v_{\mathbf{\theta}}\left(  1,d\right)  +\gamma(d-1)\right]  +\left[
\sum\limits_{d\geq d^{\ast}}\Delta^{(1)}(d)\right]  \left[  v_{\mathbf{\theta
}}\left(  1,d^{\ast}\right)  +\gamma(d^{\ast})\right] \\
&  & \\
& + & \Delta^{(1)}(d^{\ast})\text{ }\left[  \gamma(d^{\ast}-1)-\gamma(d^{\ast
})\right]
\end{array}
\tag{A.10}%
\end{equation}
Then, the log-probability becomes:%
\begin{equation}%
\begin{array}
[c]{ccl}%
\ln\mathbb{P}\left(  \widetilde{\mathbf{y}}|y_{0},\mathbf{\theta}\right)  &
= & \sum\limits_{d\geq1}H^{(1)}(d)\text{ }g_{\mathbf{\theta},1}(d)\\
&  & \\
& + & \sum\limits_{d\leq d^{\ast}-1}\Delta^{(1)}(d)\text{ }g_{\mathbf{\theta
},2}(d)+\left[  \sum\limits_{d\geq d^{\ast}}\Delta^{(1)}(d)\right]
g_{\mathbf{\theta},2}(d^{\ast})\\
&  & \\
& + & \Delta^{(1)}(d^{\ast})\text{ }\left[  \gamma(d^{\ast}-1)-\gamma(d^{\ast
})\right]
\end{array}
\tag{A.11}%
\end{equation}
with $g_{\mathbf{\theta},1}(d)\equiv\widetilde{\alpha}_{\mathbf{\theta}%
}+\sigma_{\mathbf{\theta}}(1,d)-\sigma_{\mathbf{\theta}}(0)+\gamma
(d-1)+v_{\mathbf{\theta}}\left(  1,d\right)  $, and $g_{\mathbf{\theta}%
,2}(d)\equiv\widetilde{\alpha}_{\mathbf{\theta}}+v_{\mathbf{\theta}}\left(
1,d\right)  +\gamma(d-1)$. Note that $g_{\mathbf{\theta},1}%
(d)=g_{\mathbf{\theta},1}(d^{\ast})$ for any $d\geq d^{\ast}$. Therefore, we
have $%
%TCIMACRO{\tsum \nolimits_{d\geq1}}%
%BeginExpansion
{\textstyle\sum\nolimits_{d\geq1}}
%EndExpansion
H^{(1)}(d)$ $g_{\mathbf{\theta},1}(d)=%
%TCIMACRO{\tsum \nolimits_{d\leq d^{\ast}-1}}%
%BeginExpansion
{\textstyle\sum\nolimits_{d\leq d^{\ast}-1}}
%EndExpansion
H^{(1)}(d)$ $g_{\mathbf{\theta},1}(d)+$ $\left[
%TCIMACRO{\tsum \nolimits_{d\geq d^{\ast}}}%
%BeginExpansion
{\textstyle\sum\nolimits_{d\geq d^{\ast}}}
%EndExpansion
H^{(1)}(d)\right]  g_{\mathbf{\theta},1}(d^{\ast})$, such that%
\begin{equation}%
\begin{array}
[c]{ccl}%
\ln\mathbb{P}\left(  \widetilde{\mathbf{y}}|y_{0},\mathbf{\theta}\right)  &
= &
%TCIMACRO{\tsum \limits_{d\leq d^{\ast}-1}}%
%BeginExpansion
{\textstyle\sum\limits_{d\leq d^{\ast}-1}}
%EndExpansion
H^{(1)}(d)\text{ }g_{\mathbf{\theta},1}(d)+\left[
%TCIMACRO{\tsum \limits_{d\geq d^{\ast}}}%
%BeginExpansion
{\textstyle\sum\limits_{d\geq d^{\ast}}}
%EndExpansion
H^{(1)}(d)\right]  g_{\mathbf{\theta},1}(d^{\ast})\\
&  & \\
& + & \sum\limits_{d\leq d^{\ast}-1}\Delta^{(1)}(d)\text{ }g_{\mathbf{\theta
},2}(d)+\left[  \sum\limits_{d\geq d^{\ast}}\Delta^{(1)}(d)\right]
g_{\mathbf{\theta},2}(d^{\ast})\\
&  & \\
& + & \Delta^{(1)}(d^{\ast})\text{ }\left[  \gamma(d^{\ast}-1)-\gamma(d^{\ast
})\right]  \qquad\blacksquare
\end{array}
\tag{A.12}%
\end{equation}

\bigskip

\noindent\textbf{Proof of Propositions 7 and 8. }For this model, the log
probability is $%
%TCIMACRO{\tsum \nolimits_{j=0}^{J}}%
%BeginExpansion
{\textstyle\sum\nolimits_{j=0}^{J}}
%EndExpansion%
%TCIMACRO{\tsum \nolimits_{t=1}^{T}}%
%BeginExpansion
{\textstyle\sum\nolimits_{t=1}^{T}}
%EndExpansion
1\{y_{t}=j\}$ $\alpha_{\mathbf{\theta}}(j)+$ $%
%TCIMACRO{\tsum \nolimits_{j=0}^{J}}%
%BeginExpansion
{\textstyle\sum\nolimits_{j=0}^{J}}
%EndExpansion%
%TCIMACRO{\tsum \nolimits_{k=0}^{J}}%
%BeginExpansion
{\textstyle\sum\nolimits_{k=0}^{J}}
%EndExpansion%
%TCIMACRO{\tsum \nolimits_{t=1}^{T}}%
%BeginExpansion
{\textstyle\sum\nolimits_{t=1}^{T}}
%EndExpansion
1\{y_{t-1}=j$, $y_{t}=k\}$ $\beta_{y}(k,j)+$ $%
%TCIMACRO{\tsum \nolimits_{j=0}^{J}}%
%BeginExpansion
{\textstyle\sum\nolimits_{j=0}^{J}}
%EndExpansion%
%TCIMACRO{\tsum \nolimits_{t=1}^{T}}%
%BeginExpansion
{\textstyle\sum\nolimits_{t=1}^{T}}
%EndExpansion
1\{y_{t-1}=j\}$ $\sigma_{\mathbf{\theta}}(j)$. Using the definitions of our
statistics, we have that:%
\begin{equation}%
\begin{array}
[c]{ccl}%
\ln\mathbb{P}\left(  \widetilde{\mathbf{y}}|y_{0},\mathbf{\theta}\right)  &
= &
%TCIMACRO{\tsum \limits_{j=0}^{J}}%
%BeginExpansion
{\textstyle\sum\limits_{j=0}^{J}}
%EndExpansion
T^{(j)}\alpha_{\mathbf{\theta}}(j)+%
%TCIMACRO{\tsum \limits_{j=0}^{J}}%
%BeginExpansion
{\textstyle\sum\limits_{j=0}^{J}}
%EndExpansion%
%TCIMACRO{\tsum \limits_{k=0}^{J}}%
%BeginExpansion
{\textstyle\sum\limits_{k=0}^{J}}
%EndExpansion
D^{(j,k)}\beta_{y}(k,j)+%
%TCIMACRO{\tsum \limits_{j=0}^{J}}%
%BeginExpansion
{\textstyle\sum\limits_{j=0}^{J}}
%EndExpansion
\left[  T^{(j)}-\Delta^{(j)}\right]  \sigma_{\mathbf{\theta}}(j)
\end{array}
\tag{A.13}%
\end{equation}
where $\Delta^{(j)}\equiv1\{y_{T}=j\}-1\{y_{0}=j\}$. Note that $T^{(0)}=T-%
%TCIMACRO{\tsum \nolimits_{j=1}^{J}}%
%BeginExpansion
{\textstyle\sum\nolimits_{j=1}^{J}}
%EndExpansion
T^{(j)}$, and $\Delta^{(0)}=$ $1-%
%TCIMACRO{\tsum \nolimits_{j=1}^{J}}%
%BeginExpansion
{\textstyle\sum\nolimits_{j=1}^{J}}
%EndExpansion
\Delta^{(j)}$, such that:%
\begin{equation}%
\begin{array}
[c]{ccl}%
\ln\mathbb{P}\left(  \widetilde{\mathbf{y}}|y_{0},\mathbf{\theta}\right)  &
= &
%TCIMACRO{\tsum \limits_{j=1}^{J}}%
%BeginExpansion
{\textstyle\sum\limits_{j=1}^{J}}
%EndExpansion
T^{(j)}\left[  \alpha_{\mathbf{\theta}}(j)-\alpha_{\mathbf{\theta}}%
(0)+\sigma_{\mathbf{\theta}}(j)-\sigma_{\mathbf{\theta}}(0)\right]  +%
%TCIMACRO{\tsum \limits_{j=1}^{J}}%
%BeginExpansion
{\textstyle\sum\limits_{j=1}^{J}}
%EndExpansion
\Delta^{(j)}\text{ }\left[  -\sigma_{\mathbf{\theta}}(j)+\sigma
_{\mathbf{\theta}}(0)\right] \\
&  & \multicolumn{1}{c}{}\\
& + &
%TCIMACRO{\tsum \limits_{j=0}^{J}}%
%BeginExpansion
{\textstyle\sum\limits_{j=0}^{J}}
%EndExpansion%
%TCIMACRO{\tsum \limits_{k=0}^{J}}%
%BeginExpansion
{\textstyle\sum\limits_{k=0}^{J}}
%EndExpansion
D^{(j,k)}\beta_{y}(k,j)
\end{array}
\tag{A.14}%
\end{equation}
where we have omitted the term $T\alpha_{\mathbf{\theta}}(0)+\sigma
_{\mathbf{\theta}}(0)$ because it is constant over all the choice histories.
For the term, $%
%TCIMACRO{\tsum \nolimits_{j=0}^{J}}%
%BeginExpansion
{\textstyle\sum\nolimits_{j=0}^{J}}
%EndExpansion%
%TCIMACRO{\tsum \nolimits_{k=0}^{J}}%
%BeginExpansion
{\textstyle\sum\nolimits_{k=0}^{J}}
%EndExpansion
D^{(j,k)}\beta_{y}(k,j)$, note that: $%
%TCIMACRO{\tsum \nolimits_{j=0}^{J}}%
%BeginExpansion
{\textstyle\sum\nolimits_{j=0}^{J}}
%EndExpansion
D^{(j,k)}=T^{(k)}$ such that $D^{(0,k)}=T^{(k)}-%
%TCIMACRO{\tsum \nolimits_{j=1}^{J}}%
%BeginExpansion
{\textstyle\sum\nolimits_{j=1}^{J}}
%EndExpansion
D^{(j,k)}$; and $%
%TCIMACRO{\tsum \nolimits_{k=0}^{J}}%
%BeginExpansion
{\textstyle\sum\nolimits_{k=0}^{J}}
%EndExpansion
D^{(j,k)}=$ $T^{(j)}-\Delta^{(j)}$ such that $D^{(j,0)}=T^{(j)}-\Delta^{(j)}-%
%TCIMACRO{\tsum \nolimits_{k=1}^{J}}%
%BeginExpansion
{\textstyle\sum\nolimits_{k=1}^{J}}
%EndExpansion
D^{(j,k)}$. Therefore,%
\begin{equation}%
\begin{array}
[c]{ccl}%
%TCIMACRO{\tsum \limits_{j=0}^{J}}%
%BeginExpansion
{\textstyle\sum\limits_{j=0}^{J}}
%EndExpansion%
%TCIMACRO{\tsum \limits_{k=0}^{J}}%
%BeginExpansion
{\textstyle\sum\limits_{k=0}^{J}}
%EndExpansion
D^{(j,k)}\beta_{y}(k,j) & = &
%TCIMACRO{\tsum \limits_{j=0}^{J}}%
%BeginExpansion
{\textstyle\sum\limits_{j=0}^{J}}
%EndExpansion
\left[
%TCIMACRO{\tsum \limits_{k=1}^{J}}%
%BeginExpansion
{\textstyle\sum\limits_{k=1}^{J}}
%EndExpansion
D^{(j,k)}\beta_{y}(k,j)+\left[  T^{(j)}-\Delta^{(j)}-%
%TCIMACRO{\tsum \limits_{k=1}^{J}}%
%BeginExpansion
{\textstyle\sum\limits_{k=1}^{J}}
%EndExpansion
D^{(j,k)}\right]  \beta_{y}(0,j)\right] \\
&  & \\
& = &
%TCIMACRO{\tsum \limits_{j=0}^{J}}%
%BeginExpansion
{\textstyle\sum\limits_{j=0}^{J}}
%EndExpansion%
%TCIMACRO{\tsum \limits_{k=1}^{J}}%
%BeginExpansion
{\textstyle\sum\limits_{k=1}^{J}}
%EndExpansion
D^{(j,k)}\left[  \beta_{y}(k,j)-\beta_{y}(0,j)\right]  +%
%TCIMACRO{\tsum \limits_{j=1}^{J}}%
%BeginExpansion
{\textstyle\sum\limits_{j=1}^{J}}
%EndExpansion
\left[  T^{(j)}-\Delta^{(j)}\right]  \beta_{y}(0,j)
\end{array}
\tag{A.15}%
\end{equation}
where we have omitted the term $(T-1)\beta_{y}(0,0)$ because it is constant
over every choice history (also we have normalized $\beta_{y}(y,y)=0$ for
every $y$). Now, applying a similar property to the term $%
%TCIMACRO{\tsum \nolimits_{j=0}^{J}}%
%BeginExpansion
{\textstyle\sum\nolimits_{j=0}^{J}}
%EndExpansion%
%TCIMACRO{\tsum \nolimits_{k=1}^{J}}%
%BeginExpansion
{\textstyle\sum\nolimits_{k=1}^{J}}
%EndExpansion
D^{(j,k)}$ $\left[  \beta_{y}(k,j)-\beta_{y}(0,j)\right]  $, we have:%
\begin{equation}%
\begin{array}
[c]{ccl}%
%TCIMACRO{\tsum \limits_{j=0}^{J}}%
%BeginExpansion
{\textstyle\sum\limits_{j=0}^{J}}
%EndExpansion%
%TCIMACRO{\tsum \limits_{k=1}^{J}}%
%BeginExpansion
{\textstyle\sum\limits_{k=1}^{J}}
%EndExpansion
D^{(j,k)}\left[  \beta_{y}(k,j)-\beta_{y}(0,j)\right]  & = &
%TCIMACRO{\tsum \limits_{k=1}^{J}}%
%BeginExpansion
{\textstyle\sum\limits_{k=1}^{J}}
%EndExpansion
\left[
%TCIMACRO{\tsum \limits_{j=1}^{J}}%
%BeginExpansion
{\textstyle\sum\limits_{j=1}^{J}}
%EndExpansion
D^{(j,k)}\left[  \beta_{y}(k,j)-\beta_{y}(0,j)\right]  +\left[  T^{(k)}-%
%TCIMACRO{\tsum \limits_{j=1}^{J}}%
%BeginExpansion
{\textstyle\sum\limits_{j=1}^{J}}
%EndExpansion
D^{(j,k)}\right]  \left[  \beta_{y}(k,0)-\beta_{y}(0,0)\right]  \right] \\
&  & \\
& = &
%TCIMACRO{\tsum \limits_{k=1}^{J}}%
%BeginExpansion
{\textstyle\sum\limits_{k=1}^{J}}
%EndExpansion%
%TCIMACRO{\tsum \limits_{j=1}^{J}}%
%BeginExpansion
{\textstyle\sum\limits_{j=1}^{J}}
%EndExpansion
D^{(j,k)}\left[  \beta_{y}(k,j)-\beta_{y}(0,j)-\beta_{y}(k,0)\right]  +%
%TCIMACRO{\tsum \limits_{k=1}^{J}}%
%BeginExpansion
{\textstyle\sum\limits_{k=1}^{J}}
%EndExpansion
T^{(k)}\beta_{y}(k,0)
\end{array}
\tag{A.16}%
\end{equation}
Putting together (A.15) and (A.16), we have that:%
\begin{equation}%
\begin{array}
[c]{ccl}%
%TCIMACRO{\tsum \limits_{j=0}^{J}}%
%BeginExpansion
{\textstyle\sum\limits_{j=0}^{J}}
%EndExpansion%
%TCIMACRO{\tsum \limits_{k=0}^{J}}%
%BeginExpansion
{\textstyle\sum\limits_{k=0}^{J}}
%EndExpansion
D^{(j,k)}\beta_{y}(k,j) & = &
%TCIMACRO{\tsum \limits_{k=1}^{J}}%
%BeginExpansion
{\textstyle\sum\limits_{k=1}^{J}}
%EndExpansion%
%TCIMACRO{\tsum \limits_{j=1}^{J}}%
%BeginExpansion
{\textstyle\sum\limits_{j=1}^{J}}
%EndExpansion
D^{(j,k)}\widetilde{\beta}_{y}(k,j)+%
%TCIMACRO{\tsum \limits_{j=1}^{J}}%
%BeginExpansion
{\textstyle\sum\limits_{j=1}^{J}}
%EndExpansion
T^{(j)}\left[  \beta_{y}(0,j)+\beta_{y}(j,0)\right]  -%
%TCIMACRO{\tsum \limits_{j=1}^{J}}%
%BeginExpansion
{\textstyle\sum\limits_{j=1}^{J}}
%EndExpansion
\Delta^{(j)}\beta_{y}(0,j)
\end{array}
\tag{A.17}%
\end{equation}
where $\widetilde{\beta}_{y}(k,j)\equiv\beta_{y}(k,j)-\beta_{y}(0,j)-\beta
_{y}(k,0)$. And plugging this expression into equation (A.14) for the
log-probability, we obtain:%
\begin{equation}%
\begin{array}
[c]{ccl}%
\ln\mathbb{P}\left(  \widetilde{\mathbf{y}}|y_{0},\mathbf{\theta}\right)  &
= &
%TCIMACRO{\tsum \limits_{j=1}^{J}}%
%BeginExpansion
{\textstyle\sum\limits_{j=1}^{J}}
%EndExpansion
T^{(j)}\text{ }g_{\mathbf{\theta},1}(j)+%
%TCIMACRO{\tsum \limits_{j=1}^{J}}%
%BeginExpansion
{\textstyle\sum\limits_{j=1}^{J}}
%EndExpansion
\Delta^{(j)}\text{ }g_{\mathbf{\theta},2}(j)+%
%TCIMACRO{\tsum \limits_{j=1}^{J}}%
%BeginExpansion
{\textstyle\sum\limits_{j=1}^{J}}
%EndExpansion%
%TCIMACRO{\tsum \limits_{k=1}^{J}}%
%BeginExpansion
{\textstyle\sum\limits_{k=1}^{J}}
%EndExpansion
D^{(j,k)}\widetilde{\beta}_{y}(k,j)
\end{array}
\tag{A.18}%
\end{equation}
where $g_{\mathbf{\theta},1}(j)\equiv\alpha_{\mathbf{\theta}}(j)-\alpha
_{\mathbf{\theta}}(0)+\sigma_{\mathbf{\theta}}(j)-\sigma_{\mathbf{\theta}%
}(0)+\beta_{y}(0,j)+\beta_{y}(j,0)$, and $g_{\mathbf{\theta},2}(j)\equiv
-\sigma_{\mathbf{\theta}}(j)+\sigma_{\mathbf{\theta}}(0)-\beta_{y}%
(0,j)$.\qquad$\blacksquare$

\bigskip

\noindent\textbf{Proof of Proposition 9. }For this model, the log probability
is $%
%TCIMACRO{\tsum \nolimits_{j=0}^{J}}%
%BeginExpansion
{\textstyle\sum\nolimits_{j=0}^{J}}
%EndExpansion%
%TCIMACRO{\tsum \nolimits_{t=1}^{T}}%
%BeginExpansion
{\textstyle\sum\nolimits_{t=1}^{T}}
%EndExpansion
1\{y_{t}=j\}$ $\alpha_{\mathbf{\theta}}(j)+$ $%
%TCIMACRO{\tsum \nolimits_{j=0}^{J}}%
%BeginExpansion
{\textstyle\sum\nolimits_{j=0}^{J}}
%EndExpansion%
%TCIMACRO{\tsum \nolimits_{k=0}^{J}}%
%BeginExpansion
{\textstyle\sum\nolimits_{k=0}^{J}}
%EndExpansion%
%TCIMACRO{\tsum \nolimits_{t=1}^{T}}%
%BeginExpansion
{\textstyle\sum\nolimits_{t=1}^{T}}
%EndExpansion
1\{y_{t-1}=j$, $y_{t}=k\}$ $\beta_{y}(k,j)+%
%TCIMACRO{\tsum \nolimits_{j=1}^{J}}%
%BeginExpansion
{\textstyle\sum\nolimits_{j=1}^{J}}
%EndExpansion%
%TCIMACRO{\tsum \nolimits_{d=1}^{J}}%
%BeginExpansion
{\textstyle\sum\nolimits_{d=1}^{J}}
%EndExpansion%
%TCIMACRO{\tsum \nolimits_{t=1}^{T}}%
%BeginExpansion
{\textstyle\sum\nolimits_{t=1}^{T}}
%EndExpansion
1\{y_{t-1}=y_{t}=j$, $d_{t}=d\}$ $\beta_{d}(j,d)+$ $%
%TCIMACRO{\tsum \nolimits_{j=0}^{J}}%
%BeginExpansion
{\textstyle\sum\nolimits_{j=0}^{J}}
%EndExpansion%
%TCIMACRO{\tsum \nolimits_{d=0}^{J}}%
%BeginExpansion
{\textstyle\sum\nolimits_{d=0}^{J}}
%EndExpansion%
%TCIMACRO{\tsum \nolimits_{t=1}^{T}}%
%BeginExpansion
{\textstyle\sum\nolimits_{t=1}^{T}}
%EndExpansion
1\{y_{t-1}=j$, $d_{t}=d\}$ $\sigma_{\mathbf{\theta}}(j,d)$. Using the
definition of the statistics in Table 1, we can write this log-probability as
follows:%
\begin{equation}%
\begin{array}
[c]{ccl}%
\ln\mathbb{P}\left(  \widetilde{\mathbf{y}}|y_{0},d_{1},\mathbf{\theta}\right)
& = &
%TCIMACRO{\tsum \limits_{j=0}^{J}}%
%BeginExpansion
{\textstyle\sum\limits_{j=0}^{J}}
%EndExpansion
T^{(j)}\alpha_{\mathbf{\theta}}(j)+\left[  T^{(0)}-\Delta^{(0)}\right]
\sigma_{\mathbf{\theta}}(0)+{\sum\limits_{j=1}^{J}}{\sum\limits_{d\geq1}%
}H^{(j)}(d)\sigma_{\theta}(j,d)\\
&  & \\
& + &
%TCIMACRO{\tsum \limits_{j=0}^{J}}%
%BeginExpansion
{\textstyle\sum\limits_{j=0}^{J}}
%EndExpansion%
%TCIMACRO{\tsum \limits_{k=0}^{J}}%
%BeginExpansion
{\textstyle\sum\limits_{k=0}^{J}}
%EndExpansion
D^{(j,k)}\beta_{y}(k,j)+{\sum\limits_{j=1}^{J}}{\sum\limits_{d\geq1}}%
X^{(j)}(d)\beta_{d}(j,d)
\end{array}
\tag{A.19}%
\end{equation}
Taking into account that: $T^{(0)}=T-%
%TCIMACRO{\tsum \nolimits_{j=1}^{J}}%
%BeginExpansion
{\textstyle\sum\nolimits_{j=1}^{J}}
%EndExpansion
T^{(j)}$, we have that $%
%TCIMACRO{\tsum \nolimits_{j=0}^{J}}%
%BeginExpansion
{\textstyle\sum\nolimits_{j=0}^{J}}
%EndExpansion
T^{(j)}\alpha_{\mathbf{\theta}}(j)+$ $T^{(0)}\sigma_{\mathbf{\theta}%
}(0)=T[\alpha_{\mathbf{\theta}}(0)+\sigma_{\mathbf{\theta}}(0)]+$ $%
%TCIMACRO{\tsum \nolimits_{j=1}^{J}}%
%BeginExpansion
{\textstyle\sum\nolimits_{j=1}^{J}}
%EndExpansion
T^{(j)}[\alpha_{\mathbf{\theta}}(j)-\alpha_{\mathbf{\theta}}(0)-\sigma
_{\mathbf{\theta}}(0)]$. And using equation (A.17) from the proof of
Propositions 7-8, we have:%
\begin{equation}%
\begin{array}
[c]{ccl}%
\ln\mathbb{P}\left(  \widetilde{\mathbf{y}}|y_{0},d_{1},\mathbf{\theta}\right)
& = &
%TCIMACRO{\tsum \limits_{j=1}^{J}}%
%BeginExpansion
{\textstyle\sum\limits_{j=1}^{J}}
%EndExpansion
T^{(j)}\left[  \alpha_{\mathbf{\theta}}(j)-\alpha_{\mathbf{\theta}}%
(0)-\sigma_{\mathbf{\theta}}(0)+\beta_{y}(0,j)+\beta_{y}(j,0)\right]  +%
%TCIMACRO{\tsum \limits_{j=1}^{J}}%
%BeginExpansion
{\textstyle\sum\limits_{j=1}^{J}}
%EndExpansion
\Delta^{(j)}\left[  \sigma_{\mathbf{\theta}}(0)-\beta_{y}(0,j)\right] \\
&  & \\
& + & {\sum\limits_{j=1}^{J}}{\sum\limits_{d\geq1}}H^{(j)}(d)\sigma_{\theta
}(j,d)\\
&  & \\
& + &
%TCIMACRO{\tsum \limits_{k=1}^{J}}%
%BeginExpansion
{\textstyle\sum\limits_{k=1}^{J}}
%EndExpansion%
%TCIMACRO{\tsum \limits_{j=1}^{J}}%
%BeginExpansion
{\textstyle\sum\limits_{j=1}^{J}}
%EndExpansion
D^{(j,k)}\widetilde{\beta}_{y}(k,j)+{\sum\limits_{j=1}^{J}}{\sum
\limits_{d\geq1}}X^{(j)}(d)\beta_{d}(j,d)
\end{array}
\tag{A.20}%
\end{equation}
where we have omitted the term $T\alpha_{\mathbf{\theta}}(0)+(T-1)\sigma
_{\mathbf{\theta}}(0)$ because the are constant across all the histories.
Given that $T^{(j)}=$ $\Delta^{(j)}+\sum_{d\geq1}H^{(j)}(d)$ and
$D^{(j,j)}=\sum_{d\geq1}X^{(j)}(d)$ and $\tilde{\beta}_{y}(j,j)=-\beta
_{y}(j,0)-\beta_{y}(0,j)$ by construction, we get:%
\begin{equation}%
\begin{array}
[c]{ccl}%
\ln\mathbb{P}\left(  \widetilde{\mathbf{y}}|y_{0},d_{1},\mathbf{\theta}\right)
& = & {\sum\limits_{j=1}^{J}}{\sum\limits_{d\geq1}}H^{(j)}(d)\text{ }%
[\alpha_{\mathbf{\theta}}(j)-\alpha_{\mathbf{\theta}}(0)+\sigma_{\theta
}(j,d)-\sigma_{\mathbf{\theta}}(0)+\beta_{y}(0,j)+\beta_{y}(j,0)]\\
&  & \\
& + & {\sum\limits_{j=1}^{J}}\Delta^{(j)}\left[  \alpha_{\mathbf{\theta}%
}(j)-\alpha_{\mathbf{\theta}}(0)+\beta_{y}(j,0)\right] \\
&  & \\
& + &
%TCIMACRO{\tsum \limits_{k=1}^{J}}%
%BeginExpansion
{\textstyle\sum\limits_{k=1}^{J}}
%EndExpansion%
%TCIMACRO{\tsum \limits_{j\neq k}}%
%BeginExpansion
{\textstyle\sum\limits_{j\neq k}}
%EndExpansion
D^{(j,k)}\widetilde{\beta}_{y}(k,j)+{\sum\limits_{j=1}^{J}}{\sum
\limits_{d\geq1}}X^{(j)}(d)\text{ }\gamma(j,d)
\end{array}
\tag{A.21}%
\end{equation}
Now, consider the term ${\sum\nolimits_{j=1}^{J}}{%
%TCIMACRO{\tsum \nolimits_{d\geq1}}%
%BeginExpansion
{\textstyle\sum\nolimits_{d\geq1}}
%EndExpansion
}X^{(j)}(d)$ $\beta_{d}(j,d)$. By Lemma 2, for $d\geq1$, $X^{(j)}%
(d)=H^{(j)}(d+1)-\Delta^{(j)}(d+1)$. Therefore,
\begin{equation}%
\begin{array}
[c]{ccl}%
{\sum\limits_{j=1}^{J}}{\sum\limits_{d\geq1}}X^{(j)}(d)\text{ }\gamma(j,d) &
= & {\sum\limits_{j=1}^{J}}{\sum\limits_{d\geq1}}\left[  H^{(j)}%
(d+1)+\Delta^{(j)}(d+1)\right]  \text{ }\gamma(j,d)\\
&  & \\
& = & {\sum\limits_{j=1}^{J}}{\sum\limits_{d\geq1}}\left[  H^{(j)}%
(d)+\Delta^{(j)}(d)\right]  \text{ }\gamma(j,d-1)
\end{array}
\tag{A.22}%
\end{equation}
where, for the second equality, we take into account the normalization
$\beta_{d}(j,0)=0$ for any $j\geq1$. Solving equation (A.22) into (A.21), and
taking into account that ${\sum\nolimits_{d\geq1}}\Delta^{(j)}(d)=\Delta
^{(j)}$, we obtain:%
\begin{equation}%
\begin{array}
[c]{ccl}%
\ln\mathbb{P}\left(  \widetilde{\mathbf{y}}|y_{0},d_{1},\mathbf{\theta}\right)
& = & {\sum\limits_{j=1}^{J}}{\sum\limits_{d\geq1}}H^{(j)}(d)\text{
}g_{\mathbf{\theta},1}(j,d)+{\sum\limits_{j=1}^{J}}\Delta^{(j)}%
g_{\mathbf{\theta},2}(j)\\
&  & \\
& + &
%TCIMACRO{\tsum \limits_{k=1}^{J}}%
%BeginExpansion
{\textstyle\sum\limits_{k=1}^{J}}
%EndExpansion%
%TCIMACRO{\tsum \limits_{j=1,j\neq k}^{J}}%
%BeginExpansion
{\textstyle\sum\limits_{j=1,j\neq k}^{J}}
%EndExpansion
D^{(j,k)}\widetilde{\beta}_{y}(k,j)+{\sum\limits_{j=1}^{J}}{\sum
\limits_{d\geq1}}\Delta^{(j)}(d)\text{ }\gamma(j,d-1)
\end{array}
\tag{A.23}\label{A.23}%
\end{equation}
with $g_{\mathbf{\theta},1}(j,d)\equiv$ $\alpha_{\mathbf{\theta}}%
(j)-\alpha_{\mathbf{\theta}}(0)+\sigma_{\theta}(j,d)-\sigma_{\mathbf{\theta}%
}(0)+\beta_{y}(0,j)+\beta_{y}(j,0)+\gamma(j,d-1)$, $g_{\mathbf{\theta}%
,2}(j)\equiv$ $\alpha_{\mathbf{\theta}}(j)-\alpha_{\mathbf{\theta}}%
(0)+\beta_{y}(j,0)$,\textit{ }$\widetilde{\beta}_{y}(y,y_{-1})\equiv\beta
_{y}(y,y_{-1})-\beta_{y}(0,y_{-1})-\beta_{y}(y,0)$, and $\gamma(j,d)\equiv
\beta_{d}(j,d)-\beta_{y}(j,0)-\beta_{y}(0,j)$.\qquad$\blacksquare$

\bigskip

\noindent\textbf{Proof of Proposition 10. }The expression of the
log-probability is similar as in Proposition 9 but now we have the additional
term ${%
%TCIMACRO{\tsum \nolimits_{t=1}^{T}}%
%BeginExpansion
{\textstyle\sum\nolimits_{t=1}^{T}}
%EndExpansion
}v_{\mathbf{\theta}}(y_{t},d_{t+1}[y,y_{t-1},d_{t}])$. This term is equal to
$T^{(0)}v_{\mathbf{\theta}}(0)+$ ${%
%TCIMACRO{\tsum \nolimits_{j=1}^{J}}%
%BeginExpansion
{\textstyle\sum\nolimits_{j=1}^{J}}
%EndExpansion%
%TCIMACRO{\tsum \nolimits_{d\geq1}}%
%BeginExpansion
{\textstyle\sum\nolimits_{d\geq1}}
%EndExpansion%
%TCIMACRO{\tsum \nolimits_{t=1}^{T}}%
%BeginExpansion
{\textstyle\sum\nolimits_{t=1}^{T}}
%EndExpansion
}1\{y_{t}=j$, $d_{t+1}=d\}$ $v_{\mathbf{\theta}}(j,d)=T^{(0)}v_{\theta
}(0)+\sum_{j=1}^{J}\sum_{d\geq1}v_{\theta}(j,d)\Big[H^{(j)}(d)+\Delta
^{(j)}(d)\Big]$. Given $T^{(0)}=T-\sum_{j=1}^{J}T^{(j)}=T-\sum_{j=1}^{J}%
\sum_{d\geq1}H^{(j)}(d)-\sum_{j=1}^{J}\sum_{d\geq1}\Delta^{(j)}(d)$ and using
equation (\ref{A.23}) from the proof of Proposition 9, we have
\begin{equation}%
\begin{array}
[c]{ccl}%
\ln\mathbb{P}\left(  \widetilde{\mathbf{y}}|y_{0},d_{1},\mathbf{\theta}\right)
& = & {\sum\limits_{j=1}^{J}}{\sum\limits_{d\geq1}}H^{(j)}(d)\text{
}g_{\mathbf{\theta},1}(j,d)+{\sum\limits_{j=1}^{J}}\Delta^{(j)}%
g_{\mathbf{\theta},2}(j)\\
&  & \\
& + &
%TCIMACRO{\tsum \limits_{k=1}^{J}}%
%BeginExpansion
{\textstyle\sum\limits_{k=1}^{J}}
%EndExpansion%
%TCIMACRO{\tsum \limits_{j\neq k}}%
%BeginExpansion
{\textstyle\sum\limits_{j\neq k}}
%EndExpansion
D^{(j,k)}\widetilde{\beta}_{y}(k,j)+{\sum\limits_{j=1}^{J}}{\sum
\limits_{d\geq1}}\Delta^{(j)}(d)\text{ }(\gamma(j,d-1)+v_{\theta}(j,d))
\end{array}
\tag{A.24}%
\end{equation}
with $g_{\theta,1}(j,d)\equiv\alpha_{\theta}(j)-\alpha_{\theta}(0)+\sigma
_{\theta}(j,d)-\sigma_{\theta}(0)+\beta_{y}(0,y)+\beta_{y}(y,0)+\gamma
(y,d-1)+v_{\theta}(j,d)-v_{\theta}(0)$, $g_{\theta,2}(j)\equiv\alpha_{\theta
}(j)-\alpha_{\theta}(0)+\beta_{y}(y,0)-v_{\theta}(0)$ \bigskip

Taking into account that $\sum_{d\geq1}\Delta^{(j)}(d)=\Delta^{(j)}$ for any
$j\geq1$, we have
\begin{equation}%
\begin{array}
[c]{ccl}%
\ln\mathbb{P}\left(  \widetilde{\mathbf{y}}|y_{0},d_{1},\mathbf{\theta}\right)
& = & {\sum\limits_{j=1}^{J}}{\sum\limits_{d\geq1}}H^{(j)}(d)\text{
}g_{\mathbf{\theta},1}(j,d)+{\sum\limits_{j=1}^{J}}{\sum\limits_{d\geq1}%
}\Delta^{(j)}(d)\text{ }g_{\mathbf{\theta},2}(j,d)+%
%TCIMACRO{\tsum \limits_{k=1}^{J}}%
%BeginExpansion
{\textstyle\sum\limits_{k=1}^{J}}
%EndExpansion%
%TCIMACRO{\tsum \limits_{j\neq k}}%
%BeginExpansion
{\textstyle\sum\limits_{j\neq k}}
%EndExpansion
D^{(j,k)}\widetilde{\beta}_{y}(k,j)
\end{array}
\tag{A.25}%
\end{equation}
where $g_{\mathbf{\theta},2}(j,d)\equiv\gamma(j,d-1)+v_{\theta}(j,d)+\alpha
_{\theta}(j)-\alpha_{\theta}(0)+\beta_{y}(y,0)-v_{\theta}(0)$.

\bigskip

\noindent\textbf{Proof of Proposition 11. }Define $Z^{(j)}\equiv\sum_{d\geq
1}\Delta^{(j)}(d)[v_{\theta}(j,d)+\gamma(j,d-1)]$. Under Assumption 2, we have
$v_{\theta}(j,d)=v_{\theta}(j,d^{\ast})$ for any $d\geq d_{j}^{\ast}$, and
$\gamma(j,d-1)=\gamma(j,d^{\ast})$ for any $d\geq d_{j}^{\ast}+1$. Therefore,
we have for all $j\geq1$,%

\begin{equation}%
\begin{array}
[c]{ccl}%
Z^{(j)} & = & \sum_{d\leq d_{j}^{\ast}-1}\Delta^{(j)}(d)v_{\theta
}(j,d)+\Big[\sum_{d\geq d_{j}^{\ast}}\Delta^{(j)}(d)\Big]v_{\theta}%
(j,d_{j}^{\ast})\\
&  & \\
& + & \ \sum_{d\leq d_{j}^{\ast}}\Delta^{(j)}(d)\gamma(j,d-1)+\Big[\sum_{d\geq
d_{j}^{\ast}+1}\Delta^{(j)}(d)\Big]\gamma(j,d_{j}^{\ast})\\
&  & \\
& = & \sum_{d\leq d_{j}^{\ast}-1}\Delta^{(j)}(d)[v_{\theta}(j,d)+\gamma
(j,d-1)]+\Big[\sum_{d\geq d_{j}^{\ast}}\Delta^{(j)}(d)\Big][v_{\theta}%
(j,d_{j}^{\ast})+\gamma(j,d_{j}^{\ast})]\\
&  & \\
& + & \Delta^{(j)}(d_{j}^{\ast})[\gamma(j,d_{j}^{\ast}-1)-\gamma(j,d_{j}%
^{\ast})]
\end{array}
\tag{A.26}%
\end{equation}
Then the log-probability becomes:
\begin{equation}%
\begin{array}
[c]{ccl}%
\ln\mathbb{P}\left(  \widetilde{\mathbf{y}}|y_{0},d_{1},\mathbf{\theta}\right)
& = & {\sum\limits_{j=1}^{J}}{\sum\limits_{d\geq1}}H^{(j)}(d)\text{
}g_{\mathbf{\theta},1}(j,d)+%
%TCIMACRO{\tsum \limits_{k=1}^{J}}%
%BeginExpansion
{\textstyle\sum\limits_{k=1}^{J}}
%EndExpansion%
%TCIMACRO{\tsum \limits_{j\neq k}}%
%BeginExpansion
{\textstyle\sum\limits_{j\neq k}}
%EndExpansion
D^{(j,k)}\widetilde{\beta}_{y}(k,j)\\
&  & \\
& + & {\sum\limits_{j=1}^{J}}%
%TCIMACRO{\tsum \limits_{d\geq d_{j}^{\ast}-1}}%
%BeginExpansion
{\textstyle\sum\limits_{d\geq d_{j}^{\ast}-1}}
%EndExpansion
\Delta^{(j)}(d)g_{\theta,2}(j,d)+{\sum\limits_{j=1}^{J}}\Big[\sum_{d\geq
d_{j}^{\ast}}\Delta^{(j)}(d)\Big]g_{\theta,2}(j,d_{j}^{\ast})\\
&  & \\
& + & {\sum\limits_{j=1}^{J}}\Delta^{(j)}(d_{j}^{\ast})\text{ }(\gamma
(j,d_{j}^{\ast}-1)-\gamma(j,d_{j}^{\ast}))
\end{array}
\tag{A.27}%
\end{equation}
with $g_{\theta,1}(j,d)\equiv\alpha_{\theta}(j)-\alpha_{\theta}(0)+\sigma
_{\theta}(j,d)-\sigma_{\theta}(0)+\beta_{y}(0,y)+\beta_{y}(y,0)+\gamma
(y,d-1)+v_{\theta}(j,d)-v_{\theta}(0)$ and $g_{\theta,2}(j,d)\equiv
\alpha_{\theta}(j)-\alpha_{\theta}(0)+\beta_{y}(y,0)-v_{\theta}(0)+v_{\theta
}(j,d)+\gamma(j,d-1)$. Note that $g_{\theta,1}(j,d)=g_{\theta,1}(j,d_{j}%
^{\ast})$ for $d\geq d_{j}^{\ast}$. Therefore, we have $\sum_{d\geq1}%
H^{(j)}(d)g_{\theta,1}(d)=\sum_{d\leq d_{j}^{\ast}-1}H^{(j)}(d)g_{\theta
,1}(d)+\Big[\sum_{d\geq d_{j}^{\ast}}H^{(j)}(d)\Big]g_{\theta,1}(d_{j}^{\ast
})$, such that
\begin{equation}%
\begin{array}
[c]{ccl}%
\ln\mathbb{P}\left(  \widetilde{\mathbf{y}}|y_{0},d_{1},\mathbf{\theta}\right)
& = & {\sum\limits_{j=1}^{J}}{\sum\limits_{d\leq d_{j}^{\ast}-1}}%
H^{(j)}(d)\text{ }g_{\mathbf{\theta},1}(j,d)+{\sum\limits_{j=1}^{J}}\Big[%
%TCIMACRO{\tsum \limits_{d\geq d_{j}^{\ast}}}%
%BeginExpansion
{\textstyle\sum\limits_{d\geq d_{j}^{\ast}}}
%EndExpansion
H^{(j)}(d)\Big]g_{\theta,1}(j,d_{j}^{\ast})\\
&  & \\
& + &
%TCIMACRO{\tsum \limits_{k=1}^{J}}%
%BeginExpansion
{\textstyle\sum\limits_{k=1}^{J}}
%EndExpansion%
%TCIMACRO{\tsum \limits_{j=1}^{J}}%
%BeginExpansion
{\textstyle\sum\limits_{j=1}^{J}}
%EndExpansion
D^{(j,k)}\widetilde{\beta}_{y}(k,j)\\
&  & \\
& + & {\sum\limits_{j=1}^{J}}%
%TCIMACRO{\tsum \limits_{d\geq d_{j}^{\ast}-1}}%
%BeginExpansion
{\textstyle\sum\limits_{d\geq d_{j}^{\ast}-1}}
%EndExpansion
\Delta^{(j)}(d)g_{\theta,2}(j,d)+{\sum\limits_{j=1}^{J}}\Big[%
%TCIMACRO{\tsum \limits_{d\geq d_{j}^{\ast}}}%
%BeginExpansion
{\textstyle\sum\limits_{d\geq d_{j}^{\ast}}}
%EndExpansion
\Delta^{(j)}(d)\Big]g_{\theta,2}(j,d_{j}^{\ast})\\
&  & \\
& + & {\sum\limits_{j=1}^{J}}\Delta^{(j)}(d_{j}^{\ast})\text{ }(\gamma
(j,d_{j}^{\ast}-1)-\gamma(j,d_{j}^{\ast}))\qquad\blacksquare
\end{array}
\tag{A.28}%
\end{equation}

\bigskip

\noindent\textbf{Proof of Proposition 12. }It is clear that $\widehat
{\mathbb{P}}\left(  A_{n}\right)  \rightarrow_{p}\mathbb{P}_{0}\left(
A_{n}\right)  $ and $\widehat{\mathbb{P}}\left(  B_{n}\right)  \rightarrow
_{p}\mathbb{P}_{0}\left(  B_{n}\right)  $ such that the concentrated
likelihood function $N^{-1}\ell_{N}(d^{\ast})$ converges uniformly to the
function:%
\begin{equation}%
\begin{array}
[c]{ccl}%
\ell_{0}(d^{\ast}) & = & \sum\limits_{n=2}^{d^{\ast}}\mathbb{P}_{0}\left(
A_{n}\right)  \text{ }\ln\left[  \dfrac{\mathbb{P}_{0}\left(  A_{n}\right)
}{\mathbb{P}_{0}\left(  A_{n}\right)  +\mathbb{P}_{0}\left(  B_{n}\right)
}\right]  +\mathbb{P}_{0}\left(  B_{n}\right)  \text{ }\ln\left[
\dfrac{\mathbb{P}_{0}\left(  B_{n}\right)  }{\mathbb{P}_{0}\left(
A_{n}\right)  +\mathbb{P}_{0}\left(  B_{n}\right)  }\right] \\
&  & \\
& + & \sum\limits_{n=d^{\ast}+1}^{L}\mathbb{P}_{0}\left(  A_{n}\right)  \text{
}\ln\left[  \dfrac{1}{2}\right]  +\mathbb{P}_{0}\left(  B_{n}\right)  \text{
}\ln\left[  \dfrac{1}{2}\right]
\end{array}
\tag{A.29}%
\end{equation}

\textit{Lemma.} Consider the function $f(q)=p_{1}$ $\ln(q)+p_{2}$ $\ln(1-q)$
where $p_{1},p_{2},q\in(0,1)$. This function is uniquely maximized at
$q=p_{1}/[p_{1}+p_{2}]$.

Taking into account this Lemma, we have that for any value of $n$:%
\begin{align}
&  \mathbb{P}_{0}\left(  A_{n}\right)  \text{ }\ln\left[  \dfrac
{\mathbb{P}_{0}\left(  A_{n}\right)  }{\mathbb{P}_{0}\left(  A_{n}\right)
+\mathbb{P}_{0}\left(  B_{n}\right)  }\right]  +\mathbb{P}_{0}\left(
B_{n}\right)  \text{ }\ln\left[  \dfrac{\mathbb{P}_{0}\left(  B_{n}\right)
}{\mathbb{P}_{0}\left(  A_{n}\right)  +\mathbb{P}_{0}\left(  B_{n}\right)
}\right] \tag{A.30}\\
&  \geq\mathbb{P}_{0}\left(  A_{n}\right)  \text{ }\ln\left[  \dfrac{1}%
{2}\right]  +\mathbb{P}_{0}\left(  B_{n}\right)  \text{ }\ln\left[  \dfrac
{1}{2}\right] \nonumber
\end{align}
and the inequality is strict if and only if $\mathbb{P}_{0}\left(
A_{n}\right)  =\mathbb{P}_{0}\left(  B_{n}\right)  $. Given this result, it is
straightforward to show that: $\ell_{0}(d_{0}^{\ast})>\ell_{0}(d^{\ast})$ for
any $d^{\ast}<d_{0}^{\ast}$; and $\ell_{0}(d_{0}^{\ast})=\ell_{0}(d^{\ast})$
for any $d^{\ast}>d_{0}^{\ast}$.\qquad$\blacksquare$

\bigskip

\noindent\textbf{Proof of Proposition 13}. Let $n$ be a value of the parameter
$d^{\ast}$ different to the true value $d_{0}^{\ast}$. Given our
BIC\ function, we favor $\widehat{d_{N}^{\ast}}=n$ over $\widehat{d_{N}^{\ast
}}=d_{0}^{\ast}$ if and only if $BIC_{N}(n)>BIC_{N}(d_{0}^{\ast})$ and this is
equivalent to:%
\begin{equation}
2\left[  \ell_{N}(n)-\ell_{N}(d_{0}^{\ast})\right]  >\left[  n-d_{0}^{\ast
}\right]  \text{ }\ln(N) \tag{A.31}%
\end{equation}
We show below that, as $N\rightarrow\infty$, $\mathbb{P}(2\left[  \ell
_{N}(n)-\ell_{N}(d_{0}^{\ast})\right]  >\left[  n-d_{0}^{\ast}\right]  $
$\ln(N))\rightarrow0$, and therefore, $\mathbb{P}(\widehat{d_{N}^{\ast}}%
=d_{0}^{\ast})\rightarrow1$.

First, we show that $\mathbb{P}(\widehat{d_{N}^{\ast}}>d_{0}^{\ast
})\rightarrow0$ as $N\rightarrow\infty$. By definition,%
\begin{equation}
\mathbb{P}\left(  \widehat{d_{N}^{\ast}}>d_{0}^{\ast}\right)  =\mathbb{P}%
\left(  \exists n>d_{0}^{\ast}:2\left[  \ell_{N}(n)-\ell_{N}(d_{0}^{\ast
})\right]  >\left[  n-d_{0}^{\ast}\right]  \text{ }\ln(N)\right)  \tag{A.32}%
\end{equation}
Proposition 12 implies that, for any $n\geq d_{0}^{\ast}$, $N^{-1}\ell
_{N}(n)\rightarrow_{p}\ell_{0}(d_{0}^{\ast})$ and $2[\ell_{N}(n)-\ell
_{N}(d_{0}^{\ast})]\rightarrow_{d}$ $\chi_{n-d_{0}^{\ast}}^{2}=O_{p}(1)$.
Therefore, $\mathbb{P}\left(  \widehat{d_{N}^{\ast}}>d_{0}^{\ast}\right)  =$
$\mathbb{P}\left(  O_{p}(1)>\left[  n-d_{0}^{\ast}\right]  \text{ }%
\ln(N)\right)  $ that goes to zero as $N\rightarrow\infty$.

Now, we show that $\mathbb{P}(\widehat{d_{N}^{\ast}}<d_{0}^{\ast}%
)\rightarrow0$ as $N\rightarrow\infty$. We need to prove that, for any
$n<d_{0}^{\ast}$, the probability that $2\left[  \ell_{N}(d_{0}^{\ast}%
)-\ell_{N}(n)\right]  <\left[  d_{0}^{\ast}-n\right]  $ $\ln(N)$ goes to zero
as $N\rightarrow\infty$. We can write%
\begin{equation}
2\left[  \ell_{N}(d_{0}^{\ast})-\ell_{N}(n)\right]  =2\left[  \ell_{N}%
(d_{0}^{\ast})-\ell_{N}(d_{0}^{\ast}-1)\right]  +\sum_{j=n+1}^{d_{0}^{\ast}%
-1}2\left[  \ell_{N}(j)-\ell_{N}(j-1)\right]  \tag{A.33}%
\end{equation}
Since $\beta_{0}(d_{0}^{\ast})\neq0$, classical results imply that: (a) there
exist constants $\mathit{c}$ and $\mathcal{C}$ such that $\mathit{c}N$
$\leq2\left[  \ell_{N}(d_{0}^{\ast})-\ell_{N}(d_{0}^{\ast}-1)\right]
\leq\mathcal{C}N$; and (b) $\sum_{j=n+1}^{d_{0}^{\ast}-1}2\left[  \ell
_{N}(j)-\ell_{N}(j-1)\right]  =O_{p}(N)$ for all $n<d_{0}^{\ast}$, therefore
$\mathbb{P}(2\left[  \ell_{N}(d_{0}^{\ast})-\ell_{N}(n)\right]  <\left[
d_{0}^{\ast}-n\right]  $ $\ln(N))\rightarrow0$ as $N\rightarrow\infty$%
.\qquad$\blacksquare$

\newpage

\noindent\textbf{Appendix 2. Model with stochastic transition of the
endogenous state variables}

\medskip

Consider a model with the same structure as the model in Section 2 and
Assumption 1 but now the vector of endogenous state variables is
$\mathbf{x}_{t}=(x_{t}^{y},x_{t}^{d})$ and variables $x_{t}^{y}$ and
$x_{t}^{d}$ \textit{stochastic versions} of the variables $y_{t-1}$ and
$d_{t}$, respectively. We now describe precisely the stochastic process of
these variables.

The support of state variable $x_{t}^{y}$ is the choice set $\mathcal{Y}$, and
its transition rule is $x_{t+1}^{y}=f_{y}(y_{t},\xi_{t+1}^{y})$ where
$\xi_{t+1}^{y}$ is i.i.d. over time and independent of $\mathbf{x}_{t}$. The
support of state variable $x_{t}^{d}$ is the set of possible durations,
$\{1,2,...,\infty\}$, and its transition rule is $x_{t+1}^{d}=1\{y_{t}>0\}[$
$1\left\{  y_{t}=x_{t}^{y}\right\}  $ $x_{t}^{d}+1+\xi_{t+1}^{d}]$, where
$\xi_{t+1}^{d}$ has support $\{0,1,...,\infty\}$, and it is i.i.d. over time
and independent of $\mathbf{x}_{t}$. Importantly, the stochastic shocks
$\xi_{t+1}^{y}$ and $\xi_{t+1}^{d}$ are not known to the agent when she makes
her decision at period $t$. Note that this model becomes our model in the main
text when these shocks have a degenerate probability distribution with
$p(\xi_{t+1}^{y}=0)=p(\xi_{t+1}^{d}=0)=1$.

Assumption 1' below is simply an extension of our Assumption 1 to this
stochastic version of the model. We omit the exogenous state variables
$\mathbf{z}_{t}$ for notational simplicity.

\medskip

\noindent\textit{ASSUMPTION 1'. (A) The time horizon is infinite and }%
$\delta\in(0,1)$\textit{. (B) The utility function is }$\Pi_{t}(y)=\alpha
_{\mathbf{\theta}}\left(  y\right)  +1\{y=x_{t}^{y}\}$ $\beta_{d}\left(
y,x_{t}^{d}\right)  +$ $1\{y\neq x_{t}^{y}\}$ $\beta_{y}\left(  y,x_{t}%
^{y}\right)  +\varepsilon_{t}(y)$\textit{, and functions }$\alpha
_{\mathbf{\theta}}\left(  y\right)  $\textit{, }$\beta_{d}\left(  y,x_{t}%
^{d}\right)  $\textit{, and }$\beta_{y}(y,x_{t}^{y})$ \textit{are bounded. (C)
}$\beta_{y}(y,y)=0$\textit{, }$\beta_{d}\left(  0,x^{d}\right)  =0$\textit{.
(D) }$\{\varepsilon_{t}(y):y\in\mathcal{Y}\}$\textit{\ are }$i.i.d.$%
\textit{\ over }$(i,t,y)$\textit{\ with a extreme value type I distribution.}
\textit{(E) }$\mathbf{z}_{t}$\textit{\ has discrete and finite support
}$\mathcal{Z}$\textit{\ and follows a time-homogeneous Markov process. (F) The
probability distribution of }$\mathbf{\theta}$ \textit{conditional on}
$\{\mathbf{z}_{t},\mathbf{x}_{t}:t=1,2,...\}$\textit{\ is nonparametrically
specified and completely unrestricted. (G) }$x_{t}^{y}\in\mathcal{Y}$\textit{,
and} $x_{t+1}^{y}=f_{y}(y_{t},\xi_{t+1}^{y})$\textit{ where }$\xi_{t+1}^{y}$
\textit{is i.i.d. over time and independent of }$\mathbf{x}_{t}$\textit{;}
$x_{t}^{d}\in\{0,1,...,\infty\}$\textit{, and }$x_{t+1}^{d}=1\{y_{t}>0\}[$
$1\left\{  y_{t}=x_{t}^{y}\right\}  $ $x_{t}^{d}+1+\xi_{t+1}^{d}]$\textit{,
where }$\xi_{t+1}^{d}$\textit{ has support }$\{0,1,...,\infty\}$\textit{, and
it is i.i.d. over time and independent of }$\mathbf{x}_{t}$.$\qquad
\blacksquare$

\medskip

The model has the following integrated Bellman equation:%
\[
V_{\mathbf{\theta}}\left(  \mathbf{x}_{t}\right)  =\ln\left(  \sum
_{y\in\mathcal{Y}}\exp\left\{  \text{ }\alpha_{\mathbf{\theta}}\left(
y\right)  +\beta\left(  y,\mathbf{x}_{t}\right)  +\delta\text{ }%
\mathbb{E}_{\xi_{t+1}}\left[  V_{\mathbf{\theta}}\left(  f_{y}(y_{t},\xi
_{t+1}^{y})\text{, }1\left\{  y_{t}=x_{t}^{y}\right\}  x_{t}^{d}+1+\xi
_{t+1}^{d}\right)  \right]  \text{ }\right\}  \right)
\]
where $\mathbb{E}_{\xi_{t+1}}(.)$ the expectation over the distribution of
$(\xi_{t+1}^{y}$,$\xi_{t+1}^{d})$. Let $v_{\mathbf{\theta},t}$ be the
continuation value function $\delta$ $\mathbb{E}_{\xi_{t+1}}[V_{\mathbf{\theta
}}\left(  f_{y}(y_{t},\xi_{t+1}^{y})\text{, }1\left\{  y_{t}=x_{t}%
^{y}\right\}  x_{t}^{d}+\text{ }1+\xi_{t+1}^{d}\right)  ]$ . Under our
assumptions on the distribution of $(\xi_{t+1}^{y}$,$\xi_{t+1}^{d})$, the
continuation value function has very similar properties as in the model with a
deterministic transition of the endogenous state variables. More specifically,
(a)\ it depends only $y_{t}$ and $1\left\{  y_{t}=x_{t}^{y}\right\}  x_{t}%
^{d}+1$, i.e., $v_{\mathbf{\theta},t}=v_{\mathbf{\theta}}(y_{t},1\left\{
y_{t}=x_{t}^{y}\right\}  x_{t}^{d}+1)$; (b)\ If $y_{t}\neq x_{t}^{y}$, then
$v_{\mathbf{\theta},t}=v_{\mathbf{\theta}}(y_{t},1)$; (c)\ If $y_{t}=x_{t}%
^{y}$, then $v_{\mathbf{\theta},t}=v_{\mathbf{\theta}}(y_{t},x_{t}^{d}+1)$;
and (D) if $x_{t}^{d}\geq d_{y}^{\ast}-1$ and $y_{t}=x_{t}^{y}$, then
$v_{\mathbf{\theta},t}=v_{\mathbf{\theta}}(y_{t},d_{y}^{\ast})$.

\end{document}